\documentstyle[preprint,tighten,eqsecnum,aps,floats,psfig,epsfig,amssymb]{revtex}

\begin{document}
\draft
\title{
Critical behavior of O(2)$\otimes$O($N$) symmetric models
}
\author{Pasquale Calabrese,$^{1}$ Pietro Parruccini,$^2$ 
Andrea Pelissetto,$^3$ and Ettore Vicari$\,^4$ }
\address{$^1$ 
R.~Peierls Center for Theoretical Physics, University of Oxford, \\
1 Keble Road, Oxford OX1 3NP, United Kingdom.}
\address{$^2$ Dip. Chimica Applicata e Scienza dei Materiali (DICASM) 
dell' Universit\`a di Bologna, \\
Via Saragozza 8, I-40136 Bologna, Italy}
\address{$^3$ Dip. Fisica dell'Universit\`a di Roma ``La Sapienza"  and INFN, \\
P.le Moro 2, I-00185 Roma, Italy}
\address{$^4$
Dip. Fisica dell'Universit\`a di Pisa
and INFN, V. Buonarroti 2, I-56127 Pisa, Italy}
\address{
\bf e-mail: \rm 
{\tt calabres@df.unipi.it,}
{\tt parrucci@df.unipi.it,} 
{\tt Andrea.Pelissetto@roma1.infn.it},
{\tt vicari@df.unipi.it}
}

\date{\today}

\maketitle

\begin{abstract}
We investigate the controversial issue of the existence of
universality classes describing critical phenomena in
three-dimensional systems characterized by a matrix order
parameter with symmetry O(2)$\otimes$O($N$) and symmetry-breaking
pattern O(2)$\otimes$O($N$)$\rightarrow$O(2)$\otimes$O($N-2$).
Physical realizations of these systems are, for example, frustrated
spin models with noncollinear order.

Starting from the field-theoretical Landau-Ginzburg-Wilson
Hamiltonian, we consider the massless critical theory and the
minimal-subtraction scheme without $\epsilon$ expansion.  The
three-dimensional analysis of the corresponding five-loop series
shows the existence of a stable fixed point for $N=2$ and $N=3$,
confirming recent field-theoretical results based on a six-loop
expansion in the alternative zero-momentum renormalization scheme
defined in the massive disordered phase.

In addition, we report numerical Monte Carlo simulations of a class of
three-dimensional O(2)$\otimes$O(2)-symmetric lattice models.  The
results provide further support to the existence of the
O(2)$\otimes$O(2) universality class predicted by the
field-theoretical analyses.
\end{abstract}



\section{Introduction}
\label{intro}

Several interesting critical transitions are effectively described by
a matrix order parameter with symmetry O(2)$\otimes$O($N$) and
symmetry-breaking pattern
O(2)$\otimes$O($N$)$\rightarrow$O(2)$\otimes$O($N-2$).  This is the
case, for $N=2$ or $N=3$, of multicomponent frustrated magnetic
systems with noncollinear order, in which frustration may arise either
because of the special geometry of the lattice or from the competition
of different kinds of interactions.  Typical examples of systems of
the first type are stacked triangular antiferromagnets (STA's), in
which the magnetic ions are located at the sites of a stacked
triangular lattice.  Frustration due to the competition of
interactions is realized in helimagnets, in which a magnetic spiral is
formed along a certain direction of the lattice. The nature of the
magnetic transition in these materials has been the object of several
studies, see, e.g.,
Refs.~\cite{CP-97,DPVB-95,Kawamura-98,PV-rev,DMT-03} for reviews.  In
particular, the order of the transition is still controversial, with
several contradictory results both on theoretical and experimental
sides.

The Landau-Ginzburg-Wilson (LGW) theory with O(2)$\otimes$O($N$) symmetry
that is expected to describe these systems is given by
\cite{Kawamura-88}
\begin{equation}
{\cal H}_{\rm LGW}  = \int d^d x
 \Bigl\{ {1\over2}
      \sum_{a} \Bigl[ (\partial_\mu \phi_{a})^2 + r \phi_{a}^2 \Bigr]
+ {1\over 4!}u_0 \Bigl( \sum_a \phi_a^2\Bigr)^2
+ {1\over 4!}  v_0
\sum_{a,b} \Bigl[ ( \phi_a \cdot \phi_b)^2 - \phi_a^2\phi_b^2\Bigr]
             \Bigr\},
\label{LGWH}
\end{equation}
where $\phi_a$ ($a=1,2$) are $N$-component vectors.  The
symmetry-breaking pattern \cite{SBP}
O(2)$\otimes$O($N$)$\rightarrow$O(2)$\otimes$O($N-2$) is obtained by
requiring $v_0>0$.  Negative values of $v_0$ lead to a different
symmetry-breaking pattern: the ground state configurations have a
ferromagnetic or antiferromagnetic order and correspond to
O(2)$\otimes$O($N$)$\rightarrow{\mathbb{Z}}_2\otimes$O($N-1$).  For
$v_0<0$ the model is also of interest; it describes magnets with
sinusoidal spin structures~\cite{Mukamel-75,Kawamura-88} and, for
$N=3$, the superfluid transition of $^3$He
\cite{JLM-76,BLM-77,DPV-03}; see, e.g.,
Refs.~\cite{AS-94,Kawamura-98,PV-rev,Sachdev-02} for other
applications. Here, we will only focus on the case $v_0>0$ and thus
whenever we speak of an O(2)$\otimes$O($N$) universality class we
refer to the case in which the symmetry-breaking pattern is
O(2)$\otimes$O($N$)$\rightarrow$O(2)$\otimes$O($N-2$).

The O(2)$\otimes$O($N$) theory (\ref{LGWH}) has been much studied
using field-theoretical (FT) methods.  Different perturbative schemes
have been exploited, such as the $\epsilon\equiv 4-d$ expansion
\cite{WF-72} and the three-dimensional (3-$d$) massive zero-momentum
(MZM) renormalization scheme \cite{Parisi-80}.  A detailed discussion
of the scenario emerging from the $\epsilon$ expansion is presented in
Ref.~\cite{Kawamura-98}.  Near four dimensions a stable
O(2)$\otimes$O($N$) fixed point (FP) with $v_0 > 0$ is found
\cite{Kawamura-88,ASV-95,CP-03} only for large values of $N$, $N> N_c=
21.80 - 23.43 \epsilon + 7.09 \epsilon^2 + O(\epsilon^3)$.
Resummations of the $\epsilon$ expansion of $N_c$, known to
$O(\epsilon^4)$ \cite{CP-03}, suggest \cite{CP-03,PRV-01b} $N_c\approx
6$ in three dimensions.  Therefore, a {\em smooth extrapolation} of
the scenario around $d=4$ to $d=3$ would indicate that a new
O(2)$\otimes$O($N$) universality class does not exist for the
physically interesting cases $N=2$ and 3.  On the other hand, six-loop
calculations in the framework of the 3-$d$ MZM scheme provide a rather
robust evidence for the existence of a new stable FP for $N=2$ and
$N=3$ with attraction domain in the region $v_0>0$
\cite{PRV-01,CPS-02}.  This FP was found only in the analysis of
high-order series, starting at four loops, while earlier lower-order
calculations up to three loops \cite{AS-94} did not find it. According
to renormalization-group (RG) theory, the stable FP of the
O(2)$\otimes$O($N$) theory should describe the critical behavior of
3-$d$ systems undergoing continuous transitions
characterized by the symmetry-breaking pattern
O(2)$\otimes$O($N$)$\rightarrow$O(2)$\otimes$O($N-2$).  The main
problem of the calculations within the MZM scheme is the fact that,
for $N=2$ and 3, the O(2)$\otimes$O($N$) FP is found in a region of
quartic couplings in which the perturbative expansions are not Borel
summable.  Therefore, a Borel transformation only provides an
asymptotic expansion and convergence is not guaranteed, at variance
with the case of O($N$) theories in which the Borel summability of the
corresponding MZM expansion provides a solid theoretical basis for the
resummation methods.  In the case of the O(2)$\otimes$O($N$) theory
the reliability of the results concerning the new stable FP is
essentially verified {\em a posteriori} from their stability with
respect to the perturbative order.  The MZM expansions have also been
analyzed by using the pseudo-$\epsilon$ expansion method
\cite{CP-03,HID-04}.  No stable FP is found for $N=2$ and 3, but this
is not unexpected since this resummation method can only find FPs that
are already present at one loop, similarly to the $\epsilon$
expansion.  We finally mention that perturbative studies of the 
corresponding nonlinear $\sigma$ models near two dimensions have been
reported in Refs.~\cite{ADJ-90,DJ-96}.

We mention that there are other physically interesting cases in which
low-order $\epsilon$-expansion calculations fail to provide the
correct physical picture: for example, the Ginzburg-Landau model of
superconductors, in which a complex scalar field couples to a gauge
field.  Although $\epsilon$-expansion calculations do not find a
stable FP \cite{HLM-74}, thus predicting first-order transitions, it
is now well established (see, e.g., Refs.~\cite{kleinertbook,MCs})
that 3-$d$ systems described by the Ginzburg-Landau model
can also undergo a continuous transition---this implies the presence
of a stable FP in the 3-$d$ Ginzburg-Landau theory---in agreement with
experiments \cite{GN-94}.

The O(2)$\otimes$O($N$) theory has also been studied by exploiting an
alternative FT method based on the analysis of the RG flow of the
so-called effective average
action~\cite{BTW-00,Zumbach-94,TDM-00,TDM-03,DMT-03}.  This approach
does not rely on a perturbative expansion around the Gaussian FP and
it is therefore intrinsically nonperturbative.  However, the practical
implementation requires approximations and truncations of the average
effective action.  For this purpose, a derivative expansion of the
effective average action is usually performed.  The studies of the
O(2)$\otimes$O($N$) theory reported in the
literature~\cite{Zumbach-94,TDM-00,TDM-03,DMT-03}, based on the
zeroth- and first-order approximations, do not find evidence of stable
O(2)$\otimes$O($N$) FPs for $N=2$ and 3, in contradiction with the
perturbative MZM results.  This would imply that phase transitions
characterized by the symmetry-breaking pattern
O(2)$\otimes$O($N$)$\rightarrow$O(2)$\otimes$O($N-2$) with $N=2$ or 3
are always of first order.

The issue concerning the existence of O(2)$\otimes$O($N$) universality
classes is most important to understand the physics of STA's and of
magnets with helical order, because the absence of stable
O(2)$\otimes$O($N$) FPs implies that none of them can undergo a
continuous transition.  On the experimental side, experiments
\cite{CP-97,DPVB-95} have apparently observed continuous transitions
belonging to the O(2)$\otimes$O($N$) universality class. However, as
discussed in Ref.~\cite{DMT-03}, experimental results are not
consistent---STA's and helimagnets show a critical behavior with
apparently different exponents---and, in some cases, do not satisfy
general exponent inequalities, for instance $\gamma \le 2 \nu$ and
$2\beta \ge \nu$.

The most recent Monte Carlo (MC) simulations of the antiferromagnetic
XY Hamiltonian on a stacked triangular lattice have observed a
first-order transition \cite{PM-97,Itakura-03,PSDMT-04} with very
small latent heat.  Moreover, first-order transitions have been
observed in MC investigations~\cite{LS-00,PSDMT-04} of modified
lattice spin systems whose transitions are characterized by the same
symmetry-breaking pattern.  Therefore, MC simulations of the models
considered up to now do not support the existence of an
O(2)$\otimes$O(2) universality class.  On the other hand, MC
simulations of Heisenberg STA models, corresponding to $N=3$, give
results that are substantially consistent with a continuous
transition, see, e.g., Ref.~\cite{PS-03}.

We would like to stress that the existence of a universality class is
not contradicted by the observation of first-order transitions in some
systems sharing the same symmetry-breaking pattern.  The universality
class determines the critical behavior only if the system undergoes a
continuous transition. Instead, first-order transitions are expected
for systems that are outside the attraction domain of the stable
FP. This is evident in mean-field calculations and also within the FT
approach, in which some RG trajectories do not flow towards the stable
FP but run away to infinity.  Therefore, the above-mentioned MC
results for the XY STA models are still compatible with the hypothesis
of the existence of an O(2)$\otimes$O(2) universality class; XY STA
models may be simply outside the attraction domain of the stable FP.

In this paper we further investigate the existence of the
O(2)$\otimes$O($N$) universality class for XY ($N=2$) and Heisenberg
($N=3$) systems.  First, we consider an alternative 3-$d$ perturbative
approach, the so-called minimal-subtraction ($\overline{\rm MS}$)
scheme without $\epsilon$ expansion \cite{tHV-72,Dohm-85,SD-89}, for
which five-loop series have been recently computed in
Ref.~\cite{CP-03}.  This scheme is strictly related to the $\epsilon$
expansion, but, unlike it, no $\epsilon$ expansion is performed and
$\epsilon$ is set to the physical value $\epsilon=1$, providing a
3-$d$ scheme.  It works within the massless critical theory, thus
providing a nontrivial check of the results obtained within the MZM
scheme, which is defined in the massive disordered phase.  The
analysis of the corresponding five-loop expansions shows the existence
of an O(2)$\otimes$O($N$) FP for $N=2$ and 3, confirming the
conclusions of the analysis of the six-loop expansions within the MZM
scheme.  Concerning the critical exponents, the
analysis of the five-loop $\overline{\rm MS}$ series gives
$\nu=0.65(6)$ and $\eta=0.09(4)$ for $N=2$, and $\nu=0.63(5)$ and
$\eta=0.08(3)$ for $N=3$.  These results should be compared with the
estimates obtained from the six-loop MZM series, which are
$\nu=0.57(3)$ and $\eta=0.09(1)$ for $N=2$, and $\nu=0.55(3)$ and
$\eta=0.10(1)$ for $N=3$.  It is important to note that, although the
available $\overline{\rm MS}$ series have one order less, the
corresponding results are expected to be more reliable than the MZM
ones, because the $\overline{\rm MS}$ FPs are at the boundary of the
region in which the expansions are Borel summable, and not outside it
as in the MZM case.  We finally mention that the $\overline{\rm MS}$
scheme without $\epsilon$ expansion allows us to obtain
fixed-dimension results at any $d$. Thus, we are able to recover the
results of the $\epsilon$ expansion sufficiently close to four
dimensions and to obtain a full picture of the fate of the different
FPs as $d$ varies from four to three dimensions.

We also address numerically the question of identifying a
3-$d$ lattice model with symmetry O(2)$\otimes$O(2) and
with the expected symmetry-breaking pattern that shows a continuous
transition. This would conclusively show that the O(2)$\otimes$O(2)
universality class really exists. For this purpose we consider the
following lattice model
\begin{eqnarray}
&&{\cal H} = - \beta \sum_{x,\mu} 
\left( \varphi_x\cdot \varphi_{x+\mu} +
\psi_x\cdot \psi_{x+\mu} \right)
+ \sum_x \left( \varphi_x^2 + \psi_x^2\right)+
\nonumber \\
&& \qquad\qquad
+ A_4 \sum_x \left[ (\varphi_x^2-1)^2 + (\psi_x^2-1)^2\right]
+ 2 A_{22} \sum_x \varphi_x^2 \psi_x^2,
\label{HLi} 
\end{eqnarray}
where $\varphi$ and $\psi$ are two-component real variables.  The
Hamiltonian ${\cal H}$ describes two identical two-component
O(2)-symmetric lattice $\phi^4$ models coupled by an energy-energy
term.  By an appropriate change of variables, see Sec.~\ref{secmc},
one can show that model (\ref{HLi}) corresponds to a lattice
discretization of the Hamiltonian (\ref{LGWH}) for $N=2$ with $u_0\sim
(A_4 + A_{22})$ and $v_0 \sim (A_{22} - A_4)$.  When $A_{22}>A_4>0$
the critical behavior at the phase transition should be described by
the $N=2$ Hamiltonian (\ref{LGWH}) with $v_0>0$.  Therefore, a region
of continuous transitions in the quartic parameter space with
$A_{22}>A_4$ would imply the existence of the O(2)$\otimes$O(2)
universality class.  In order to investigate this point, we present MC
simulations for $A_4=1$ and several values of $A_{22}$.  The phase
diagram emerging from the simulations is characterized by a line of
first-order transitions extending from large values of $A_{22}$ down
to a tricritical point at $A_{22}=A_{22}^*>A_4= 1$, where the latent
heat vanishes, and, for $1 = A_4 < A_{22} < A_{22}^*$, by a line of continuous 
transitions that should belong to the O(2)$\otimes$O(2) universality class 
identified by the perturbative FT
approaches.  The possible extension of the first-order transition line
up to $A_{22} = A_4$, i.e. up to the 4-vector theory, is apparently
incompatible with the theoretically predicted behavior of the latent
heat near an O(4) tricritical point.

The paper is organized as follows.  In Sec.~\ref{MSbar} we present the
analysis of the five-loop $\overline{\rm MS}$ expansions, providing
evidence for the existence of a stable FP with attraction domain in
the region $v_0>0$, in the two- and three-component cases.  There, we
also show that for $d\rightarrow 4$ the results of the $\epsilon$
expansion are recovered.  In Sec.~\ref{sec:crossover} we discuss the
crossover behaviors predicted by the FT approach and their relation
with those that may be observed in realistic models.  In
Sec.~\ref{secmc} we report the results of the MC simulations for the
model defined by Hamiltonian (\ref{HLi}), determining its phase
diagram in the region of quartic parameters $A_{22}>A_4=1$. We
investigate the finite-size scaling (FSS) behavior using cubic
lattices of size $L\le 120$.  In Sec.~\ref{Conclusions} we report some
conclusive remarks.  In App.~\ref{appmsb} we provide some details on
the perturbative expansions in the $\overline{\rm MS}$ scheme.  In
App.~\ref{App:mediumrange} we discuss some properties of
O($M$)$\otimes$O($N$)-symmetric medium-range models.

\section{Analysis of the  five-loop minimal-subtraction series}
\label{MSbar}

\subsection{The minimal-subtraction scheme without $\epsilon$ expansion}
\label{sec2a}

The FT approach is based on the Hamiltonian (\ref{LGWH}).  In the
$\overline{\rm MS}$ scheme one considers the massless critical theory
in dimensional regularization and determines the RG functions from the
divergences appearing in the perturbative expansion of the correlation
functions \cite{tHV-72}.  In the standard $\epsilon$-expansion scheme
\cite{WF-72} the FPs, i.e. the common zeroes of the $\beta$-functions,
are determined perturbatively as expansions in powers of $\epsilon$,
while exponents are obtained by expanding the corresponding RG
functions, i.e. $\eta_{\phi,t}$ (see App.~\ref{appmsb}), computed at
the FP in powers of $\epsilon$.  The $\overline{\rm MS}$ scheme
without $\epsilon$ expansion \cite{SD-89} is strictly related. The RG
functions $\beta_{u,v}$ and $\eta_{\phi,t}$ are the $\overline{\rm
MS}$ functions. However, $\epsilon$ is no longer considered as a small
quantity but it is set to its physical value, i.e. in three dimensions
one simply sets $\epsilon = 1$. Then, one determines the FP values
$u^*$, $v^*$ from the common zeroes of the resummed $\beta$ functions.
Finally, critical exponents are determined by evaluating the resummed
RG functions $\eta_\phi$ and $\eta_t$ at $u^*$ and $v^*$.  Notice that
the FP values $u^*$ and $v^*$ are different from the FP values of the
renormalized quartic couplings of the MZM renormalization scheme,
since $u$ and $v$ indicate different quantities in the two schemes.

The $\overline{\rm MS}$ RG functions have been computed to five loops
in Ref.~\cite{CP-03}.  In App.~\ref{appmsb} we report the series for
$N=2$ and $N=3$.  We also consider the critical exponents associated
with the chiral degrees of freedom. They can be determined from the RG
dimension of the chiral operator \cite{Kawamura-88}
\begin{equation}
C_{ckdl}(x) = \phi_{ck}(x) \phi_{dl}(x) - \phi_{cl}(x) \phi_{dk}(x).
\label{chiralop}
\end{equation} 
We computed the $\overline{\rm MS}$ RG function $\eta_c(u,v)$ associated with 
the chiral operator $C_{ckdl}$ to five loops.
The series are reported in App.~\ref{appmsb}.

\subsection{The resummation}
\label{sec2b}

Since perturbative expansions are divergent, resummation methods must
be used to obtain meaningful results.  Given a generic quantity
$S(u,v)$ with perturbative expansion $S(u,v)= \sum_{ij} c_{ij} u^i
v^j$, we consider
\begin{equation}
S(x u,x v) = \sum_k s_k(u,v) x^k,
\label{seriesx}
\end{equation}
which must be evaluated at $x=1$. The expansion (\ref{seriesx}) in
powers of $x$ is resummed by using the conformal-mapping method
\cite{LZ-77,ZJbook} that exploits the knowledge of the large-order
behavior of the coefficients, generically given by
\begin{equation}
s_k(u,v) \sim k! \,[-A(u,v)]^{k}\,k^b\,\left[ 1 + O(k^{-1})\right].
\label{lobh}
\end{equation}
The quantity $A(u,v)$ is related to the singularity $t_s$ of the Borel
transform $B(t)$ that is nearest to the origin: $t_s=-1/A(u,v)$.  The
series is Borel summable for $x > 0$ if $B(t)$ does not have
singularities on the positive real axis, and, in particular, if
$A(u,v)>0$.  Using semiclassical arguments, one can argue that
\cite{PRV-01} the expansion is Borel summable when (see
App.~\ref{appmsb} for the precise definition of $u$ and $v$)
\begin{equation}
u \geq 0,\qquad u - \frac{1}{2}v \geq 0.
\label{brr}
\end{equation}
In  this region we have
\begin{equation}
A(u,v) = \frac{1}{2} \,{\rm Max} \left( u,u-v/2\right).
\label{afg}
\end{equation}
Under the additional assumption that the Borel-transform singularities
lie only in the negative axis, the conformal-mapping
method~\cite{LZ-77} turns the original expansion into a convergent one
in the region (\ref{brr}).  Outside, the expansion is not Borel
summable.  However, if the condition
\begin{equation}
u-\frac{1}{4}v >0
\label{brr2}
\end{equation}
holds, then the Borel-transform singularity closest to the origin is
still in the negative axis, and therefore the large-order behavior is
still given by Eq.~(\ref{lobh}) with $A(u,v)$ given by
Eq.~(\ref{afg}). Thus, by using this value of $A(u,v)$ and the
conformal-mapping method one still takes into account the leading
large-order behavior.  One may therefore hope to get an asymptotic
expansion with a milder behavior, which may still provide reliable
results.

We should mention that the $\overline{\rm MS}$ series are essentially
four-dimensional, so that they may be affected by renormalons that
make the expansion non-Borel summable for any $u$ and $v$, and are not
detected by a semiclassical analysis; see, e.g.,
Ref.~\cite{highorder}.  The same problem should also affect the
$\overline{\rm MS}$ series of O($N$) models.  However, the good
agreement between the results obtained from the FT analyses
\cite{SD-89} and those obtained by other methods indicates that
renormalon effects are either very small or absent (note that, as
shown in Ref.~\cite{BD-84}, this may occur in some renormalization
schemes).  For example, the analysis of the five-loop perturbative
series \cite{SD-89} gives $\nu=0.629(5)$ for the Ising model and
$\nu=0.667(5)$ for the XY model, that are in good agreement with the
most precise estimates obtained by lattice techniques, such as
$\nu=0.63012(16)$ \cite{CPRV-02} and $\nu=0.63020(12)$ \cite{DB-03}
for the Ising model, and $\nu=0.67155(27)$ \cite{CHPRV-01} for the XY
universality class.  On the basis of these results, we will assume
renormalon effects to be negligible in the analysis of the
two-variable series of the O(2)$\otimes$O($N$) theory.

\subsection{Three-dimensional analysis of the five-loop series for $N=2,3$}
\label{sec2c}

The RG flow of the theory is determined by the FPs.  Two FPs are
easily identified: the Gaussian FP, which is always unstable, and the
O($2N$) FP located along the $u$ axis.  The results of
Ref.~\cite{CPV-03,CPV-04nova} on the stability of the
three-dimensional O($M$)-symmetric FP under generic perturbations can
be used to prove that also the O($2N$) FP is unstable for any $N\geq
2$ \cite{footnotesp4}. Indeed, the Hamiltonian term $(\phi_a \cdot
\phi_b)^2 - \phi_a^2\phi_b^2$, which acts as a perturbation at the
O($2N$) FP, is a particular combination of quartic operators
transforming as the spin-0 and spin-4 representations of the O($2N$)
group, and any spin-4 quartic perturbation is relevant \cite{CPV-03}
at the O($M$) FP for $M\geq 3$, since its RG dimension $y_{4,4}$ is
positive for $M\geq 3$.  In particular, $y_{4,4}\approx 0.11$ at the
O(4) FP and $y_{4,4}\approx 0.27$ at the O(6) FP \cite{CPV-03}.  Note
that these values are rather small, especially in the O(4) case.  The
$u$-axis plays the role of a separatrix and thus the RG flow
corresponding to $v_0>0$ cannot cross the $u$-axis.  Therefore, since
models with the symmetry-breaking pattern
O(2)$\otimes$O($N$)$\to$O(2)$\otimes$O($N-2$) have $v_0 > 0$, the
relevant FPs lie in the region $v>0$.

The analyses of the six-loop series in the MZM scheme reported in
Refs.~\cite{PRV-01,CPS-02} provided a rather robust evidence for the
presence of a stable FP with attraction domain in the region $v_0>0$
for $N=2$ and $N=3$.  In the following this result will be confirmed
by the analysis of the $\overline{\rm MS}$ five-loop series.  In order
to investigate the RG flow in the region $v_0>0$, we apply essentially
the same analysis method of Refs.~\cite{PRV-01,CPV-00} (we refer to
these references for details).  We resum the perturbative series by
means of the conformal-mapping method and, in order to understand the
systematic errors, we vary two different parameters $b$ and $\alpha$
(see Ref.~\cite{CPV-00} for definitions).  We also apply this method
for those values of $u$ and $v$ for which the series are not Borel
summable but still satisfy $u - {1\over4} v > 0$.  As already
discussed, the conformal-mapping method should still provide
reasonable estimates since we are taking into account the leading
large-order behavior.

\begin{figure*}[t!]
\centerline{\includegraphics[width=12truecm,height=10truecm]{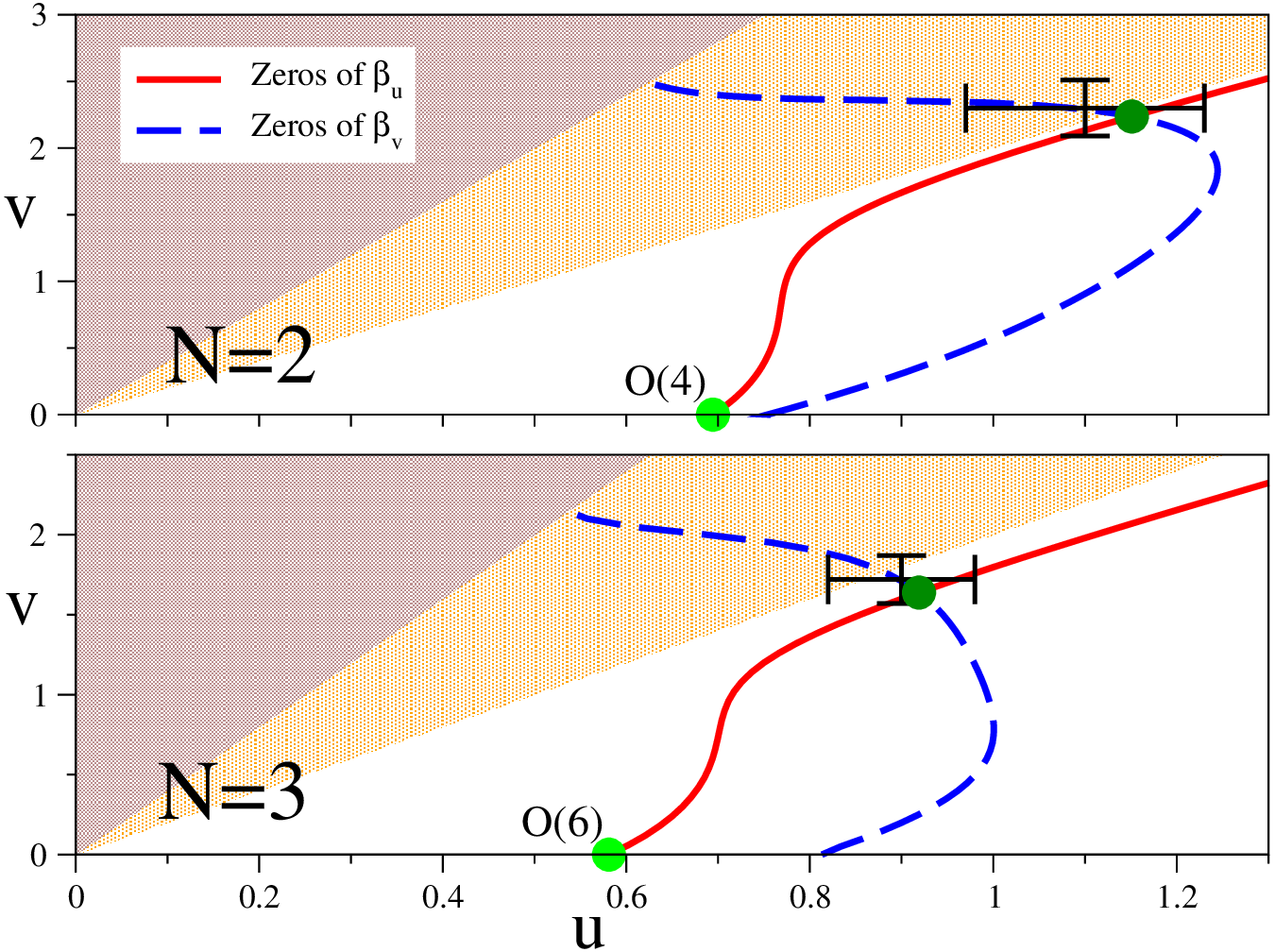}}
\caption{
Zeroes of the $\beta$-functions in the region $v>0$ for $N=2$ and $3$
for a particular approximant, see text.
The three colored regions correspond (from below) to: 
(1) $u - {1\over2} v > 0$ (domain in which the perturbative series are
Borel summable), (2) $u - {1\over4} v > 0$ and $u - {1\over2} v < 0$
(domain in which the perturbative series are not Borel summable but 
one can take into account the leading large-order behavior), (3) 
$u - {1\over4} v < 0$.
The full lines at the O(2)$\otimes$O($N$) FP 
show the final estimate (\ref{stfpms}) with its uncertainty.
}
\label{zeroes}
\end{figure*}

In order to find the zeroes of the $\beta$-functions we first resummed
the expansions of $B_u(u,v)$ and $B_v(u,v)$ defined in
Eq.~(\ref{Bdef}).  More precisely, we considered the functions
$R_{u,v}(u,v,x)\equiv B_{u,v}(ux,vx)/x^2$.  For each function
$R_{u,v}$ we considered several approximants corresponding to
different values of the resummation parameters $\alpha$ and $b$, see,
e.g., Refs.~\cite{CPV-00,PRV-01} for details.  In Fig.~\ref{zeroes} we
show the common zeroes of the $\beta$-functions in the region $v>0$.
The figure is obtained by using a single approximant, the one with
$\alpha_u=\alpha_v=1$, $b_u=b_v=10$, but others give qualitatively
similar results.  A common zero $(u^*, v^*)$ with $v^*>0$ is clearly
observed at $u^*\approx 1.1$, $v^*\approx 2.3$ for $N=2$, and at
$u^*\approx 0.9$, $v^*\approx 1.7$ for $N=3$.  In order to give an
estimate of the FP, we considered resummations of $B_u(u,v)$ and
$B_v(u,v)$ with parameters $\alpha_u$, $b_u$, $\alpha_v$, and $b_v$,
assuming integer values in the range $-1\leq \alpha_{u,v} \leq 3$ and
$4\leq b_{u,v}\leq 16$.  Most combinations, approximately 90\% for
$N=2$ and 97\% for $N=3$, have a common zero in the region $v>0$
(these percentages increase if we only consider 
approximants with $\alpha_u=\alpha_v$ and $b_u=b_v$, becoming
approximately 94\% for $N=2$ and 99\% for $N=3$).  We take the average
of the results as final estimate, obtaining
\begin{eqnarray}
&u^*=1.10(13), \quad v^*=2.30(21) \quad &{\rm for}\quad N=2,
\nonumber \\
&u^*=0.90(8), \quad v^*=1.72(15) \quad &{\rm for}\quad N=3.
\label{stfpms}
\end{eqnarray}
The errors are related to the variation of the results with respect to
changes of the resummation parameters $\alpha_u$, $b_u$, $\alpha_v$,
and $b_v$ in the considered range of values, and correspond to one
standard deviation. As a check, we also tried a different method.  We
determined optimal values of $\alpha$ and $b$ by minimizing the
difference between the results of the four- and five-loop resummations
of the functions $B_{u,v}$ (independently) close to the
O(2)$\otimes$O($N$) FP. The results are consistent with those reported
in Eq.~(\ref{stfpms}).  Notice that in the case $N=3$, since
$v^*/u^*\approx 1.9$, the FP is substantially within the region in
which the perturbative expansions should be Borel summable, while for
$N=2$, since $v^*/u^*\approx 2.1$, the FP is slightly above its
boundary $v/u=2$. Therefore, Borel resummations are expected to be
effective.  In this respect the $\overline{\rm MS}$ scheme seems to
behave better than the MZM scheme, in which the FPs are in the
non-Borel summable region \cite{PRV-01}, although still in the region
in which the conformal-mapping resummation method should be able to
take into account the leading large-order behavior.  The analysis of
the stability matrix shows that the FP is stable, i.e. its eigenvalues
have positive real part.  Most approximants give complex eigenvalues,
supporting the hypothesis that the FP is a focus, as discussed in
Ref.~\cite{CPS-02}. We obtain $\omega = 1.0(5)\pm i 0.8(5)$ for $N =
2$ and $\omega=0.9(4)\pm i 0.7(3)$ for $N = 3$, in rough agreement
with the MZM scheme results \cite{CPS-02}.  We finally mention that
consistent results are obtained by resumming the series using the
Pad\'e-Borel technique, which does not exploit the knowledge of the
large-order behavior of the series.

Critical exponents are obtained by evaluating the RG functions
$\eta_{\phi}$ and $\eta_t$ or appropriate combinations at the FP. We found
\begin{equation}
\nu=0.65(2+4),\quad \eta=0.09(2+2), \quad \gamma=1.24(3+8),  
              \quad \phi_c=1.42(6+10)
\end{equation}
for $N=2$, and
\begin{equation}
\nu=0.63(1+4),\quad \eta=0.08(2+1), \quad \gamma=1.20(1+7), 
              \quad \phi_c=1.35(2+7) 
\end{equation}
for $N=3$.  The errors are reported as the sum of two terms, related
respectively to the dependence on $b$ and $\alpha$ and to the
uncertainty of the FP coordinates.  For comparison we report the
corresponding results obtained from the analysis of the six-loop
series in the MZM scheme \cite{PRV-01,PRV-02}: $\nu=0.57(3)$,
$\eta=0.09(1)$ and $\phi_c=1.43(4)$ for $N=2$, and $\nu=0.55(3)$,
$\eta=0.10(1)$ and $\phi_c=1.27(4)$ for $N=3$.  We note that the
$\overline{\rm MS}$ estimates of $\nu$ and $\gamma$ are larger than
those obtained in the MZM scheme, but still substantially compatible
with them taking into account their relatively large errors.  We
stress again that this comparison represents a nontrivial consistency
check since the two schemes are quite different: in the MZM scheme one
works in the massive high-temperature phase, while in the
$\overline{\rm MS}$ scheme one considers the massless critical theory.

\begin{figure*}[t!]
\centerline{\includegraphics[width=12truecm]{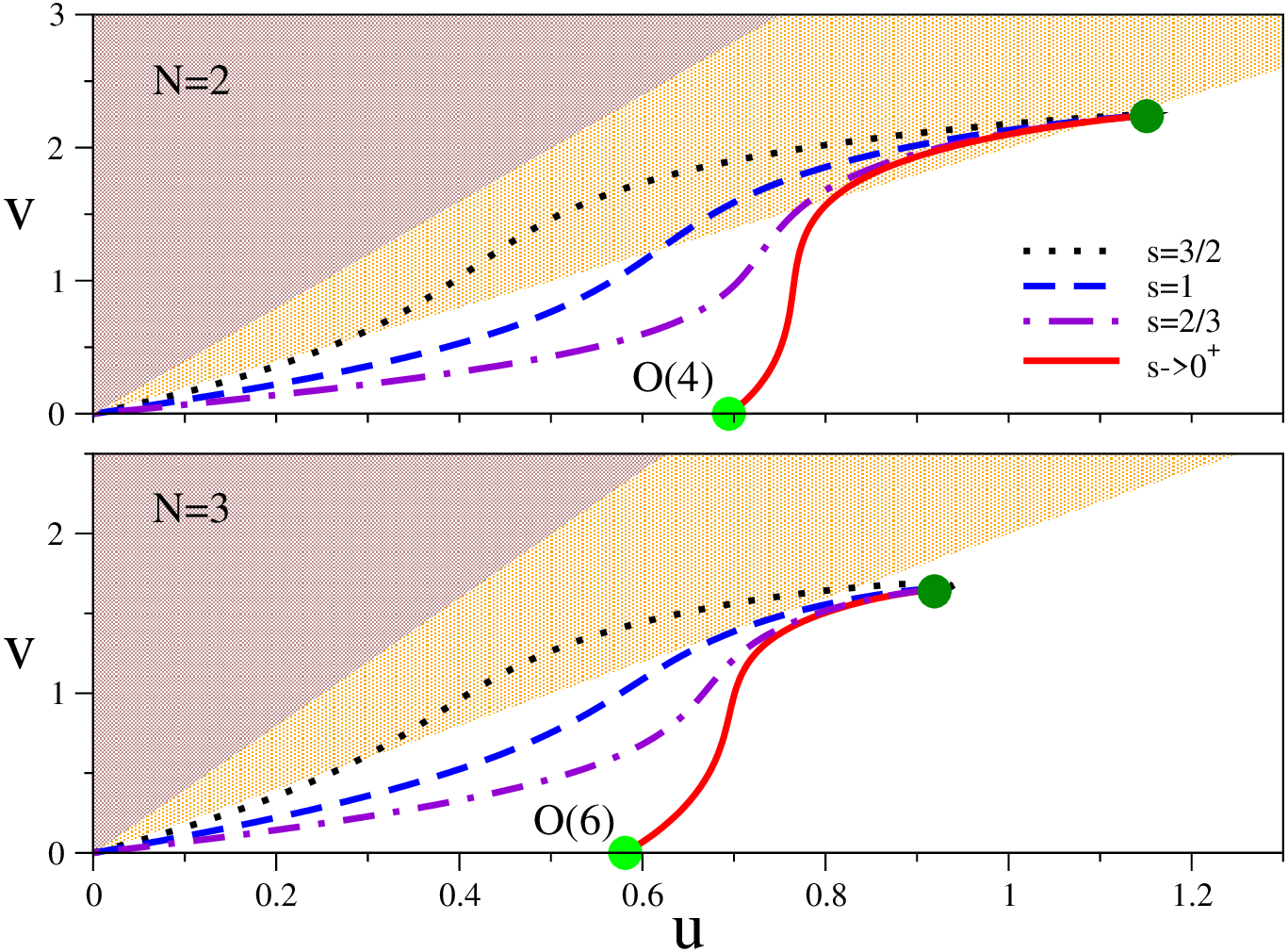}}
\caption{
The RG flow in the quartic-coupling plane for $N=2,3$ in the 
$\overline{\rm MS}$ scheme for $d=3$.
}
\label{rgflow}
\end{figure*}

Finally, we computed the RG flow, in order to determine the attraction
domain of the stable FP.  We refer the reader to Ref.~\cite{CPPV-03}
for the relevant definitions.  In Fig.~\ref{rgflow} we show the RG
flow in the quartic-coupling $u,v$ plane corresponding to different
values of the ratio $s\equiv v_0/u_0$ for $v_0>0$.  All trajectories
corresponding to $s\lesssim 3/2$ belong to the region $u - {1\over 2}
v \gtrsim 0$, in which the resummation should be reliable, and appear
to be attracted by the stable FP.

\subsection{Results for generic values of $N$ and dimension $d$}
\label{sec2d}

The fixed-dimension $\overline{\rm MS}$ scheme allows us to obtain
fixed-dimension results for any dimension $d$.  Since in three
dimensions and for $N=2$, 3 this scheme provides results that are
substantially different from those of the strictly related $\epsilon$
expansion, it is interesting to compare the two perturbative methods
for generic values of $d$ and $N$.

\begin{figure*}[t!]
\centerline{\includegraphics[width=12truecm]{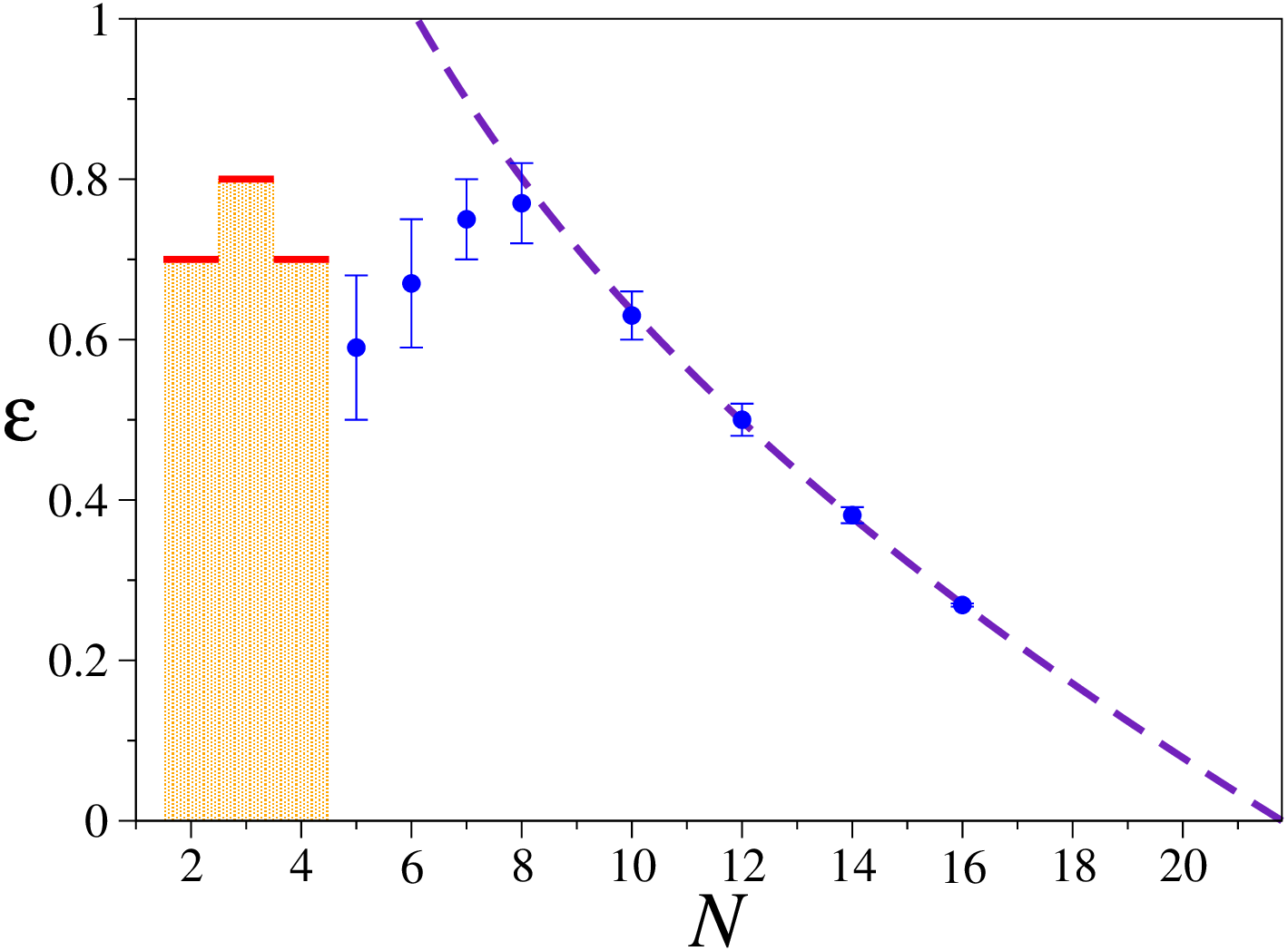}}
\caption{
Results for $\epsilon_c(N)$ as obtained from the fixed-dimension
$\overline{\rm MS}$ analysis and from the $\epsilon$ expansion (dashed line).
The thick segments at $N=2,3,4$ represent conservative upper bounds
on $\epsilon_c(N)$.
}
\label{en}
\end{figure*}

Using the five-loop series of the $\overline{\rm MS}$
$\beta$-functions, we investigate the presence of FPs in the region
$0\leq v < 4u$, where resummations seem to be under control, for
generic $d$ and $N$.  In any $d$, or $\epsilon\equiv 4-d$, and for
sufficiently large values of $N$, we find a stable O(2)$\otimes$O($N$)
FP.  If we decrease $N$ at fixed $\epsilon$, for $\epsilon$ smaller
than a critical value $\epsilon_{c,\rm max}$, we find a value
$N_c(\epsilon)$ such that, for $N=N_c(\epsilon)$, the stable FP
disappears.  In Fig.~\ref{en} we plot the results for the inverse
quantity $\epsilon_c(N)\equiv 4-d_c(N)$, where $d_c(N)$ represents the
dimension below which one finds a stable FP in the region $v>0$ at
fixed $N$.  This quantity may be estimated by averaging the values of
the largest dimension $d$ (smallest value of $\epsilon$) for which
each pair of approximants of the $\beta$-functions (we use the same
set as in the three-dimensional analysis reported in Sec.~\ref{sec2c})
has a stable O(2)$\otimes$O($N$) FP.  The reported error corresponds
to one standard deviation. Unfortunately, for $4 \lesssim N\lesssim 5$
the stable FP moves outside the region $u - {1\over4} v > 0$, in which
we are able to resum reliably the perturbative series (if this
condition is satisfied we can take into account the leading
large-order behavior).  Therefore, for $N\le 4$ we are unable to
compute $\epsilon_c(N)$.  In this case we can compute a conservative
upper bound by finding the smallest value of $\epsilon$ such that at
least 95\% of the approximants still present a stable FP in the region
$0<v<4u$.  The bounds corresponding to $N=2$, 3, and 4 are represented
by thick segments in Fig.~\ref{en}.  There, we also compare these
results with the curve obtained by resumming the $O(\epsilon^4)$
expansion \cite{CP-03} of $N_c(\epsilon)$ (we actually report the
curve obtained by resumming $1/N_c(\epsilon)$).  A nice agreement is
observed for sufficiently large values of $N$, down to $N\approx
8$. For smaller values of $N$ the fixed-dimension $\overline{\rm MS}$
results differ from the $\epsilon$-expansion curve.  In particular,
unlike the $\epsilon$ expansion, the fixed-dimension $\overline{\rm
MS}$ series provides estimate for $\epsilon_c(2)$ and $\epsilon_c(3)$
that are definitely smaller than one, leading to the bounds
$\epsilon_c(2)<0.7$ and $\epsilon_c(3)<0.8$.  As Fig.~\ref{en} shows,
the results for $\epsilon_c(N)$ are nonmonotonic, with a maximum value
$\epsilon_{c,{\rm max}}\approx 0.8$ for $N\approx 8$.  Thus, for
$\epsilon \lesssim \epsilon_{c,{\rm max}}$ there exists another
limiting value of $N$, $N_{c2}(\epsilon)$, such that for
$N_{c2}(\epsilon)<N<N_c(\epsilon)$ no stable FP exists in the region
$v\geq 0$, while for $N<N_{c2}(\epsilon)$ the O(2)$\otimes$O($N$) FP
is again present.  Note that while for $N>N_c(\epsilon)$ the stable FP
has real stability eigenvalues, for $N<N_{c2}(\epsilon)$ it is a
stable focus, i.e., the stability eigenvalues are complex with
positive real part.

The $\overline{\rm MS}$ results are qualitatively consistent with the
MZM ones, although the estimate of $\epsilon_{c,\rm max}$,
$\epsilon_{c,\rm max} \approx 0.8$, apparently contradicts the
conclusions of Ref.~\cite{PRV-01}. Indeed, from the analysis of the
MZM scheme expansions, Ref.~\cite{PRV-01} did not find a clear
evidence of stable FPs for $5 \lesssim N \lesssim 7$, which would
imply that $\epsilon_{c,\rm max} \gtrsim 1$. This conclusion was also
in contrast with the MC results of Ref.~\cite{LSDASD-00} that
apparently found a continuous transition for $N=6$.  On the basis of
the present analysis we are now inclined to believe that this may be only
a resummation problem, in some sense connected to the fact that in the
extended space $(N,d)$, $N=6$ and $d=3$ is close to the line $d =
d_c(N)$.

Although in the three-dimensional $\overline{\rm MS}$ calculation,
$N=6$ does not represent a special point for the existence of the FP,
such a value still plays a special role. In $d=3$ the FP is
topologically different for small and large values of $N$ since the
stability eigenvalues are real for large $N$, while for $N=2$ and $3$
they are complex. Therefore, there should be a value $N_{\rm eq}$ that
separates the two behaviors.  We shall show that $N_{\rm eq} \approx
6$.

\begin{table}
\squeezetable
\caption{
Results obtained from the analysis of the five-loop series in the
$\overline{\rm MS}$ perturbative scheme.  No values of $\omega_{1,2}$
are reported for $N=6$: in this case it is not clear whether the
eigenvalues of the stability matrix are complex or real.  }
\begin{tabular}{cllcclll}
\multicolumn{1}{c}{$N$}& 
\multicolumn{1}{c}{$u^*$} & 
\multicolumn{1}{c}{$v^*$} & 
\multicolumn{1}{c}{$\omega_1 \qquad \omega_2$ }&
\multicolumn{1}{c}{$\nu$}&
\multicolumn{1}{c}{$\gamma$}&
\multicolumn{1}{c}{$\eta$}\\
\hline
2 & 1.10(13) & 2.30(21) &  \multicolumn{1}{c}{1.0(5) $\pm i$0.8(5)}
 & 0.65(6) & 1.24(11) & 0.09(4) \\
3 & 0.90(8)  & 1.72(15) &  \multicolumn{1}{c}{0.9(4) $\pm i$0.7(3)}
 & 0.63(5) & 1.20(8)  & 0.08(3) \\
4 & 0.74(3)  & 1.29(8)  &  \multicolumn{1}{c}{0.7(2)  $\pm i$0.4(2)}
 & 0.64(4) & 1.24(6)  & 0.073(10) \\
5 & 0.63(3)  & 1.04(7)  &  \multicolumn{1}{c}{0.7(2)  $\pm i$0.3(2)}
 & 0.64(4) & 1.25(6)  & 0.061(10) \\
6 & 0.56(4)  & 0.86(7)  & &  0.66(4) & 1.29(8)  & 
0.052(14) \\
7 & 0.51(5)  & 0.73(5)  &  \multicolumn{1}{c}{0.8(2)$\qquad$0.5(2)} & 0.68(4) & 
1.34(9) & 0.047(15) \\
8 & 0.47(4)  & 0.64(4)  &  \multicolumn{1}{c}{0.8(2)$\qquad$0.5(2)} & 0.70(5) & 
1.37(10)  & 0.042(10) \\
10 & 0.41(4) & 0.51(1)  &  \multicolumn{1}{c}{0.9(2)$\qquad$0.5(2)} & 0.74(5) & 
1.46(11)  & 0.036(8) \\
16 & 0.300(14) & 0.334(8)& 
    \multicolumn{1}{c}{$\hphantom{40}$0.9(2)$\qquad$0.74(12)} & 0.82(4) 
& 1.62(8)  & 0.025(4) \\
\end{tabular}
\label{tab1}
\end{table}

For large values of $N$ the stability eigenvalues are real. As $N$
decreases the difference between $\omega_1$ and $\omega_2$ decreases
and for $N=N_{\rm eq}$ we have $\omega_1= \omega_2$. Then, for $N <
N_{\rm eq}$ the eigenvalues become complex and the FP is a focus.  As
it can be seen from the results reported in Table~\ref{tab1}, $N_{\rm
eq} \approx 6$. In this case, $\approx$ 50\% of the approximants give
real estimates for $\omega_1$ and $\omega_2$, while $\approx$ 50\%
give complex estimates with a small imaginary part. In all cases the
real part satisfies $0.3<\hbox{{$\Re$}{\rm e}}\, \omega_i < 0.8$.

Beside the stability eigenvalues in Table~\ref{tab1} we also report
the critical exponents and the FP coordinates for several values of
$N$.  These results are in agreement with the MZM estimates of
Ref.~\cite{CPS-02}.  We also note that the $\overline{\rm MS}$ results
for $N=6$ are in substantial agreemeent with the MC results of
Ref.~\cite{LSDASD-00}, $\nu=0.700(11)$ and $\gamma =1.383(36)$, and
with the nonperturbative RG results of Ref.~\cite{DMT-03}, $\nu\approx
0.707$ and $\gamma \approx 1.377$.

\section{Crossover behavior} \label{sec:crossover}

\subsection{Effective exponents} \label{sec:effexp}

The perturbative analysis presented in Sec.~\ref{MSbar} as well as the
analyses in the MZM scheme of Ref.~\cite{PRV-01} predict the presence
of a stable FP for the physically interesting cases $N=2$ and $N=3$.
However, this FP has a quite unusual feature: the stability
eigenvalues are apparenly complex with positive real part
\cite{PRV-01,CPS-02}. In this Section we wish to understand the
consequences on experimental and numerical determinations of the
critical exponents and, in general, of RG-invariant quantities.

The presence of complex stability eigenvalues changes the approach to
criticality. If ${\cal O}$ is a generic critical quantity we expect
close to the critical point
\begin{equation}
{\cal O} \approx C \xi^{\sigma} \left[
1 + a \xi^{-\omega_R} \cos(\omega_I \log \xi + b)\right],
\label{subleading}
\end{equation}
where $\xi$ is the correlation length and the stability eigenvalues
are written as $\omega_R \pm i \omega_I$. Scaling corrections
oscillate and the approach to the asymptotic behavior is nonmonotonic.

In order to characterize the behavior of critical quantities outside
the critical point, it is useful to introduce effective exponents.
From the susceptibility $\chi$ and the correlation length $\xi$ one
can define the effective exponents
\begin{equation}
\nu_{\rm eff}(t) \equiv -
{\partial {\rm ln} \xi \over \partial {\rm ln} t},
\qquad
\gamma_{\rm eff}(t) \equiv -
{\partial {\rm ln} \chi \over \partial {\rm ln} t},
\qquad
\eta_{\rm eff}(t) \equiv 2 - 
{\partial {\rm ln} \chi\over \partial {\rm ln} \xi},
\label{effexp}
\end{equation}
where $t\equiv (T-T_c)/T_c$ is the reduced temperature.  One can
easily check that $\eta_{\rm eff} = 2 - \gamma_{\rm eff}/\nu_{\rm
eff}$.  The effective exponents are not universal and depend on the
specific model. Nonetheless, it is usually assumed (but there are
notable exceptions; for instance, the 3-$d$ Ising model
and the corresponding scalar $\phi^4$ theory behave differently near
the critical point \cite{LF-90,Nickel-91,Schaefer-94}) that the
qualitative features are similar in all models belonging to the same
universality class. For this reason, in the following we shall compute
the effective exponents in the FT model.  We shall present numerical
results for $N=2$, in order to be able to compare them with the MC
results of Sec.~\ref{secmc}. For $N=3$ effective exponents are
qualitatively similar.

\begin{figure*}[t!]
\centerline{\includegraphics[width=12truecm]{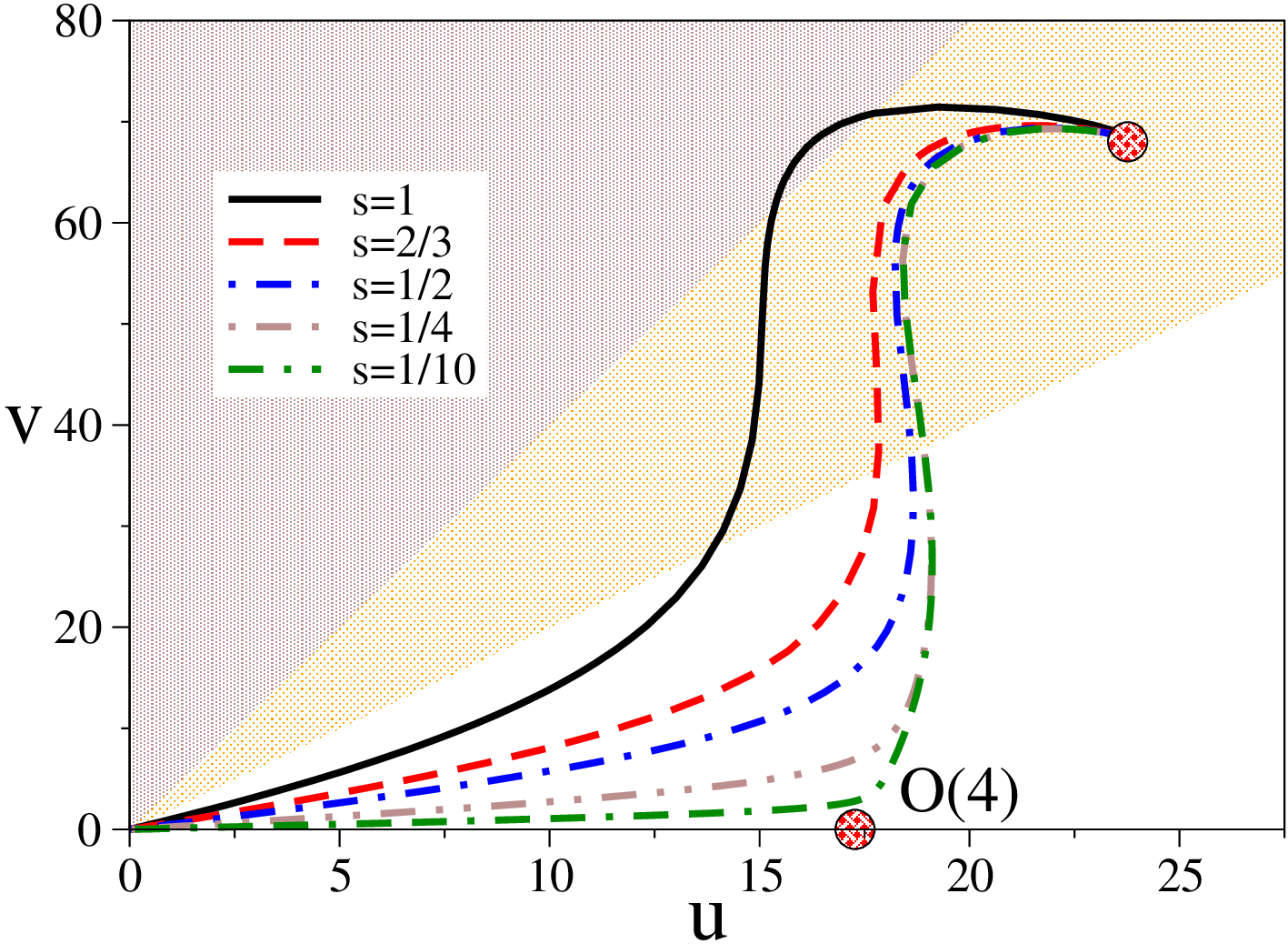}}
\caption{
RG flow in the MZM scheme for several values of $s$ and $N = 2$.
The O(2)$\otimes$O(2) FP corresponds to $u^* = 23.9(1.3)$, $v^* = 68.7(2.5)$ 
(Ref.~\protect\cite{PRV-01}).
}
\label{flowMZM}
\end{figure*}

We shall consider the MZM scheme since all necessary formulas have
already been presented in Ref.~\cite{CPPV-03}, although the same
analysis could have been done in the $\overline{\rm MS}$ scheme by
generalizing to the present case the results of Ref.~\cite{KSD-90}. If
$u$ and $v$ are the zero-momentum renormalized couplings normalized so
that $u\approx u_0/m$ and $v\approx v_0/m$ at tree level
\cite{norm-U-V}, RG trajectories are determined by solving the
differential equations
\begin{eqnarray}
&&-\lambda {d u\over d\lambda} = \beta_u(u(\lambda),v(\lambda)),\nonumber\\
&&-\lambda {d v\over d\lambda} = \beta_v(u(\lambda),v(\lambda)),
\label{rgfloweq}
\end{eqnarray}
where $\lambda\in [0,\infty)$, with the initial conditions
\begin{eqnarray} 
&&u(0) = v(0) = 0 ,\nonumber \\
&& \left. {d u\over d\lambda} \right|_{\lambda=0} = 1 ,\qquad
\left. {d v\over d\lambda} \right|_{\lambda=0} = s, \label{ini-rgflow}
\end{eqnarray}
where $s \equiv {v_0/ u_0}$ parametrizes the different models. The 
results of Ref.~\cite{CPPV-03} allow us to derive general scaling formulas 
for the rescaled $\widetilde{\chi} \equiv \chi u_0^2$ and
$\widetilde{\xi} \equiv \xi u_0$, where $\chi$ and $\xi$ 
are respectively the susceptibility and the second-moment correlation length.
In particular, if $\tau \equiv r - r_c$ is the reduced temperature and 
$\widetilde{\tau} \equiv \tau/u_0^2$, we have 
\begin{equation}
\widetilde{\chi} = F_\chi(\widetilde{\tau},s), \qquad \qquad
\widetilde{\xi} = F_\xi(\widetilde{\tau},s).
\label{def-Ffunctions}
\end{equation}
The functions $F_\chi$ and $F_\xi$ can be expressed in terms of RG
functions---in the present case they are known to six loops---and can
be computed rather accurately, as we shall show below.

\begin{figure}[t!]
\centerline{\psfig{width=10truecm,angle=0,file=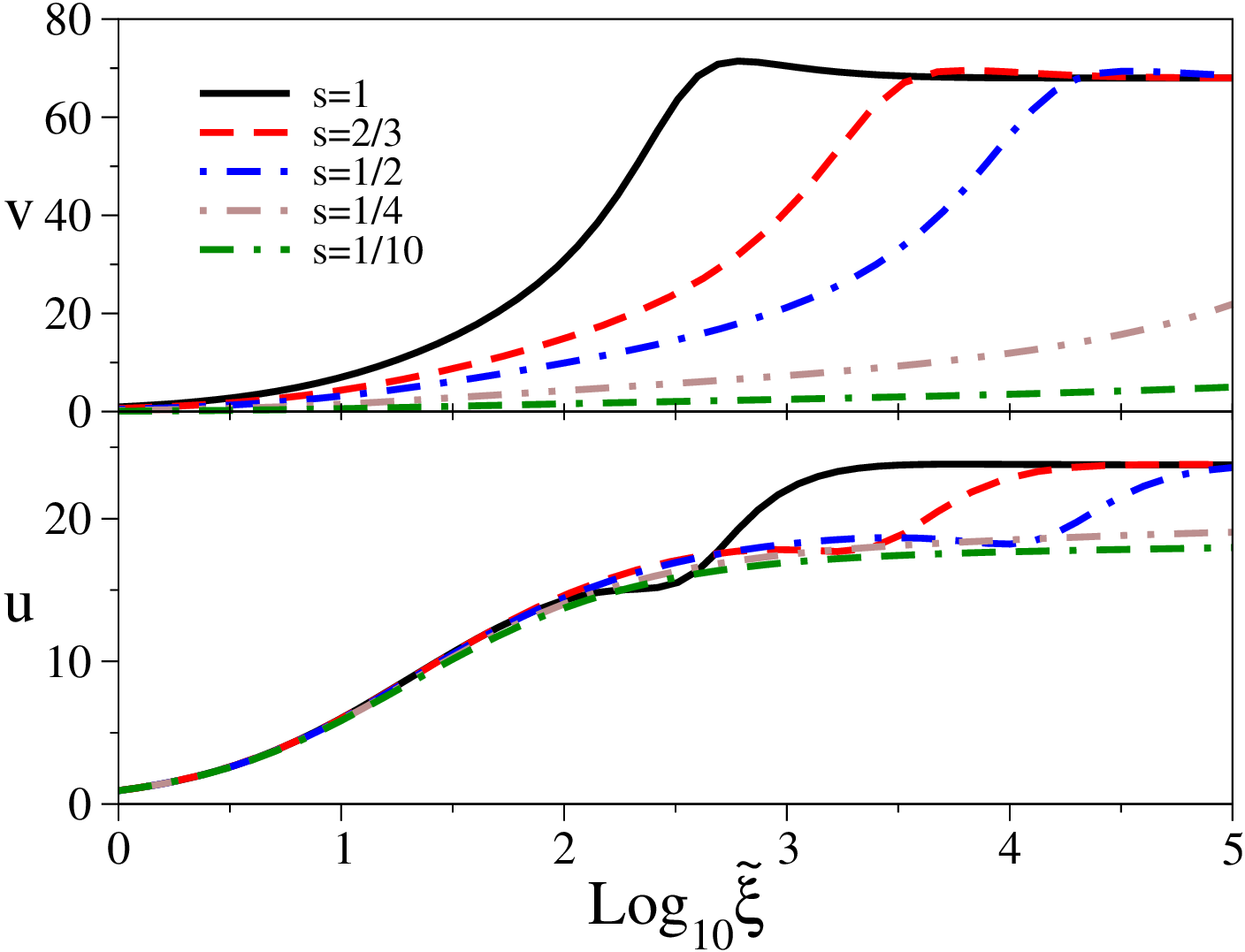}}
\caption{
The four-point couplings $u$ and $v$ as a function of $\widetilde{\xi}$ 
for several values of $s$ and $N = 2$. For $\widetilde{\xi}\to\infty$,
$u$ and $v$ converge to $u^* = 23.9(1.3)$ and $v^* = 68.7(2.5)$ 
(Ref.~\protect\cite{PRV-01}).
}
\label{UV}
\end{figure}

\begin{figure}[tb!]
\vspace{6mm}
\centerline{\psfig{width=10truecm,angle=0,file=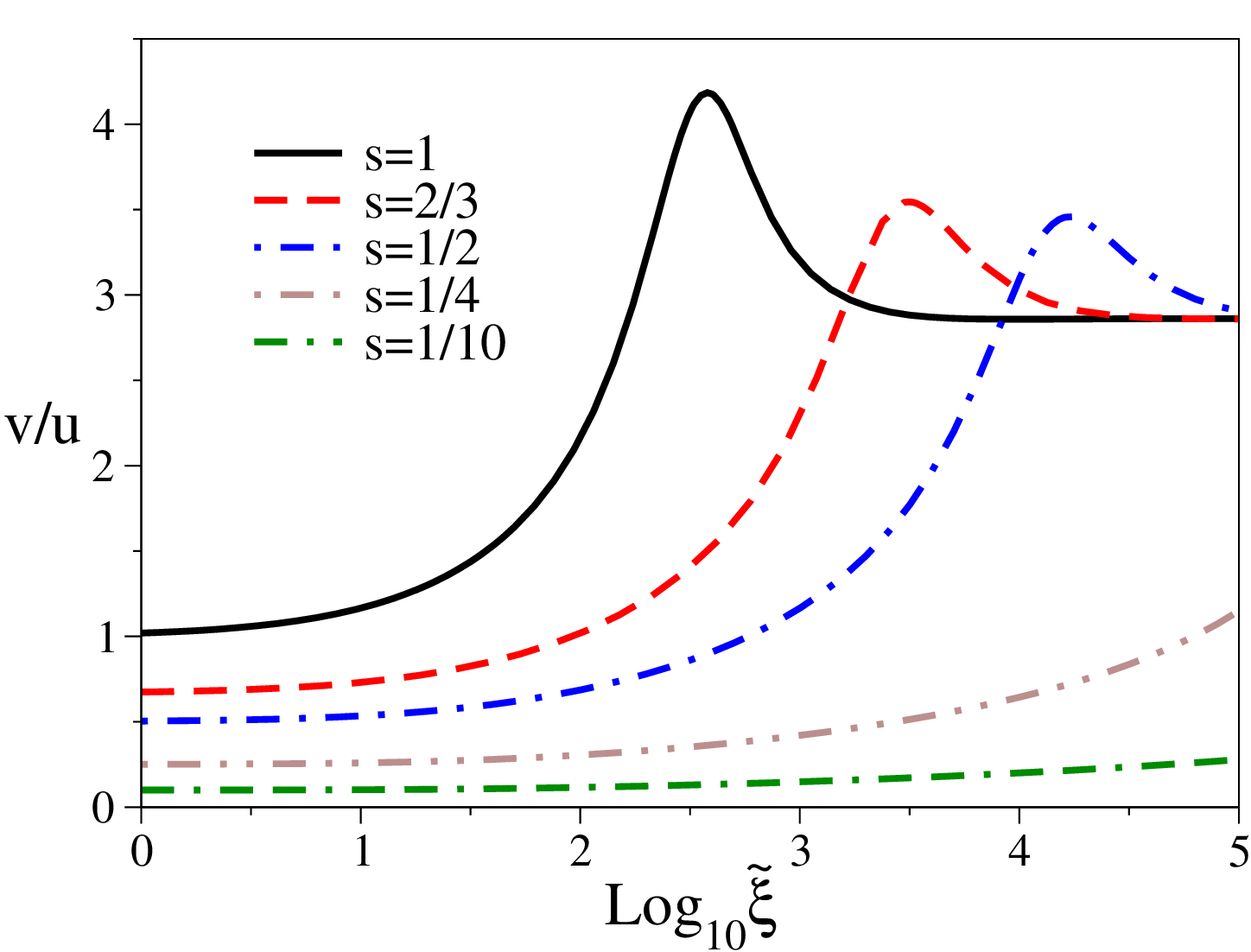}}
\caption{
The ratio $v/u$ of the four-point couplings $u$ and $v$ as a function of 
$\widetilde{\xi}$ for several values of $s$ and $N = 2$.
Asymptotically, the ratio converges to $v^*/u^*=2.9(2)$ 
(Ref.~\protect\cite{PRV-01}). 
}
\label{usuv}
\end{figure}

In Fig.~\ref{flowMZM} we show the RG trajectories for several values
of $s$ with $0 < s \le 1$ \cite{foot-d2}.  For larger values of $s$
trajectories run in the region $u - {1\over4} v < 0$, where we are not
able to resum the perturbative series.  Correspondingly, in
Fig.~\ref{UV} we report the behavior of the four-point couplings $u$
and $v$ as a function of $\widetilde{\xi}$.  The corresponding FP
values \cite{PRV-01,norm-U-V} are $u^* = 23.9 (1.3)$ and $v^* =
68.7(2.5)$.  Considering first $v(\widetilde{\xi})$, it is interesting
to observe that oscillations are significant only for $s \approx
1$. For smaller values of $s$, $v(\widetilde{\xi})$ increases
essentially monotonically with $\widetilde{\xi}$. More peculiar is the
behavior of $u(\widetilde{\xi})$.  Indeed, for all $s \lesssim 2/3$,
$u$ flattens first at a value around 18 and then suddenly 
increases towards the asymptotic value. This is due to the presence of
the unstable O(4) FP that gives rise to strong crossover effects, even
when $s$ is as large as 2/3. Indeed, the plateau observed in $u$
corresponds to the FP value of $u$ in the O(4) theory \cite{Ustar-o4},
$u^*_{O(4)} \approx 17.4$.  Thus, unless $s \gtrsim 1$, the flow first
feels the presence of the O(4) FP, so that $u \approx u^*_{O(4)}$ and
then goes towards the O(2)$\otimes$O(2) FP.  In Fig.~\ref{usuv} we
also show the ratio $v/u$. Note that for small $s$ such a ratio is
very small, while for $s \gtrsim 1/2$ the behavior is nonmonotonic
with a pronounced peak.

\begin{figure}[t!]
\vspace{8truemm}
\centerline{\psfig{width=10truecm,angle=0,file=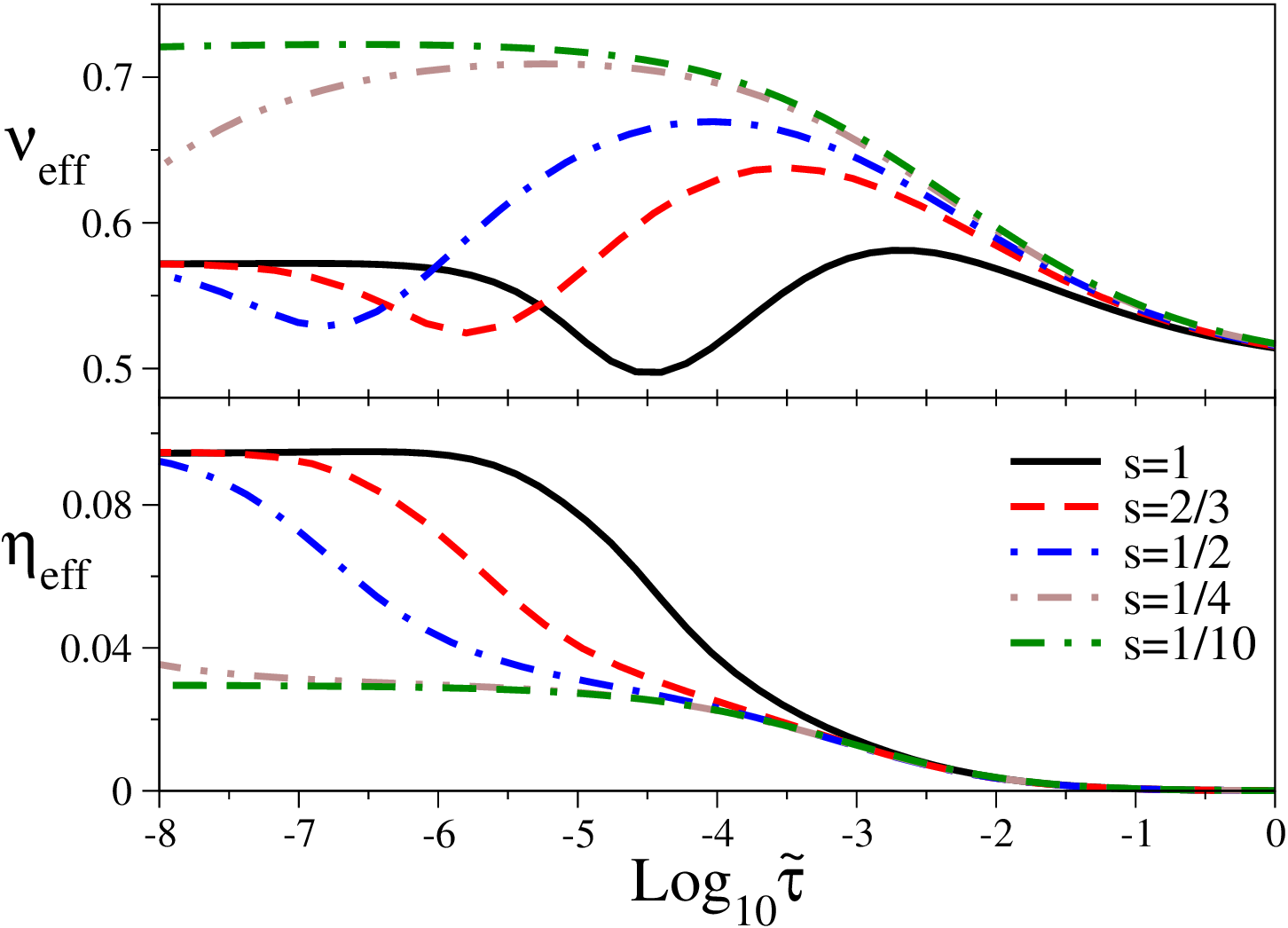}}
\caption{
Effective exponents as a function of the rescaled reduced temperature 
$\widetilde{\tau}$ for several values of $s$ and $N = 2$.  
}
\label{Fig:effexp}
\end{figure}

Finally, we determine the effective exponents. In
Fig.~\ref{Fig:effexp} we report the effective exponents $\nu_{\rm
eff}$ and $\eta_{\rm eff}$ as a function of the rescaled reduced
temperature $\tilde{\tau}$.  The exponent $\nu_{\rm eff}$ shows quite
large oscillations, especially for small $s$. They are not only due to
the complex stability eigenvalues but also to crossover effects
related to the presence of the O(4) FP. As we already remarked above,
for small $s$ the trajectories are close to the O(4) FP and thus
$\nu_{\rm eff}$ is close to $\nu_{O(4)}$ (the best available estimate
is $\nu_{O(4)} = 0.749(2)$, Ref.~\cite{Hasenbusch-01}). For instance,
for $s=1/4$ (resp. $s=1/2$) the maximum value of $\nu_{\rm eff}$ is
0.71 (resp. 0.67).  For $s$ close to 1, crossover effects are less
relevant and $\nu_{\rm eff}$ does not increase much. However, in this
case there is a large downward oscillation.  The exponent $\nu_{\rm
eff}$ decreases below the asymptotic value and it may be even less
than 0.5: for $s=1$ the minimum value of $\nu_{\rm eff}$ is 0.49 and
it is expected to further decrease if $s$ increases.  These
oscillations show how difficult is the determination of the critical
exponents: extrapolations may provide completely incorrect
estimates. The effective exponent $\gamma_{\rm eff}$ has a behavior
similar to that of $\nu_{\rm eff}$. On the other hand, $\eta_{\rm
eff}$ shows an approximately monotonic behavior without detectable
oscillations, although the crossover effects due to the presence of
the O(4) FP ($\eta_{O(4)} = 0.0365(10)$, Ref.~\cite{Hasenbusch-01})
are clearly visible for $s \lesssim 1/4$.

\subsection{Crossover behavior in lattice systems} \label{sec:mediumrange}

In Sec.~\ref{sec:effexp} we computed the FT crossover curves.  It is
of course of interest to relate them to the results obtained in
lattice models and in experimental systems. Strictly speaking, the
mapping cannot go beyond the leading correction term appearing in
Eq.~(\ref{subleading}) (see the discussion in Sec. IV.A of
Ref.~\cite{CPPV-03}). In some cases even the leading critical behavior
cannot be reproduced \cite{LF-90,Nickel-91,Schaefer-94}: this happens
in the nearest-neighbor Ising model and in the lattice self-avoiding
walk.  However, there are limiting cases in which the FT results {\em
exactly} describe the lattice model: this is the case of the critical
crossover limit in weakly coupled lattice models and in medium-range
models \cite{crossover1,PRV-lr,crossover2}. Consider, for instance, a
$d$-dimensional hypercubic lattice and the lattice discretization of
the FT Hamiltonian (\ref{LGWH}),
\begin{equation}
{\cal H} = - {\beta\over2}
    \sum_{x,y} J(x - y) \sum_{a}
    \varphi_{x,a} \cdot \varphi_{y,a} +
    \sum_x V(\varphi_x),
\label{H-MR1}
\end{equation}
where the sums over $x$ and $y$ are extended over all lattice points,
$J(x)$ is a generic short-range coupling,  and 
\begin{equation}
V(\varphi) = r \sum_a \varphi_a^2 + {U_0\over 4!} (\sum_a \varphi_a^2)^2 +
    {V_0\over 4!} \sum_{ab} \left[
        (\varphi_{a}\cdot \varphi_{b})^2 - \varphi_a^2 \varphi_b^2
     \right].
\label{lattice-potential}
\end{equation}
The parameter $r$ is irrelevant and can be made equal to $\pm1$ by
changing the normalization of the fields.

The first interesting case corresponds to weakly coupled theories in
which $r > 0$ and $U_0,V_0 \to 0$.  Let $\beta_c(U_0,V_0)$ be the
critical point for given $U_0$ and $V_0$ and let $t$ be the reduced
temperature. Then, consider the limit $t\to 0$, $U_0,V_0\to 0$ keeping
fixed $s_L\equiv V_0/U_0$ and $\widetilde{t}\equiv
t/U_0^{2/(4-d)}$. In this limit
\begin{eqnarray}
\chi(\beta,U_0,V_0) U_0^2 \to \mu_\chi F_\chi(a \widetilde{t},s_L),
\nonumber \\
\xi(\beta,U_0,V_0) U_0 \to \mu_\xi F_\xi(a \widetilde{t},s_L),
\end{eqnarray}
where $F_\chi(t,s)$ and $F_\xi(t,s)$ are exactly the FT functions
defined in Eq.~(\ref{def-Ffunctions}). The constants $\mu_\chi$,
$\mu_\xi$, and $a$ can be easily computed by comparing the
perturbative expansions (at one loop) for the continuum and the
lattice model. The additive mass renormalization---it requires a
nonperturbative matching, see Ref.~\cite{PRV-lr}---also fixes the
first terms of the expansion of $\beta_c(U_0,V_0)$ in powers of $U_0$
and $V_0$.

The second interesting case corresponds to medium-range models.  In
this case we assume that the coupling $J(x)$ depends on a parameter
$\rho$.  For instance, one may take
\begin{equation}
J_\rho(x) = \cases{1 & $\qquad$ if $|x| \le \rho$, \cr
                0 & $\qquad$ otherwise.}
\label{Coupling-J}
\end{equation}
This specific form is not necessary for the discussion that will be
presented below, and indeed one can consider more general families of
couplings, as discussed in Sec. 3 of Ref.~\cite{PRV-lr}. The relevant
property is that $J_\rho(x)$ couples all lattice points for $\rho\to
\infty$, i.e., that for $\rho\to \infty$ one recovers a mean-field
theory. The interaction range is characterized by $R$ defined by
\begin{equation}
   R^2 = {1\over 2d} {\sum_x x^2 J_\rho(x)\over \sum_x J_\rho(x)}.
\label{def-R2}
\end{equation} 
These models are called medium-range models and admit an
interesting scaling limit called critical crossover limit
\cite{crossover1,crossover2}. If $\beta_c(R)$ is
the critical temperature as a function of $R$ (here $U_0$ and $V_0$
are fixed and do not play any role in the limit), then for
$R\to \infty$, $t\equiv (\beta_c(R) - \beta)/\beta_c(R) \to 0$ at
fixed $\widetilde{t} \equiv R^{2d/(4 - d)} t$, critical quantities show a
scaling behavior. For instance, the susceptibility $\chi(\beta,R)$ and
the correlation length $\xi(\beta,R)$ scale as
\begin{eqnarray}
  \widetilde{\chi} &\equiv& \chi(\beta,R) R^{-2d/(4 - d)} \approx
    f_\chi(\widetilde{t}),
\nonumber \\
   \widetilde{\xi} &\equiv& \xi(\beta,R)  R^{-4/(4 - d)}  \approx
    f_\xi(\widetilde{t}).
\label{MR-criticalscaling}
\end{eqnarray}
The functions $f_\chi(\widetilde{t})$ and $f_\xi(\widetilde{t})$ are
directly related to the crossover functions $F_\chi(\widetilde{\tau},s)$ and
$F_\xi(\widetilde{\tau},s)$ computed in field theory, 
cf.~Eq.~(\ref{def-Ffunctions}). Indeed,
\begin{eqnarray}
    f_\chi(\widetilde{t}) &=& \mu_\chi F_\chi(a \widetilde{t}, s),
\nonumber \\
    f_\xi(\widetilde{t})  &=& \mu_\xi F_\xi(a \widetilde{t}, s) ,
\label{relations-CCL}
\end{eqnarray} 
where $\mu_\chi$, $\mu_\xi$, $a$, and $s$ are nonuniversal constants
that depend on the model \cite{PRV-lr,crossover2}. Therefore, the FT crossover
functions are expected to describe accurately the crossover behavior
for large $R$: in practice, numerical simulations show that for $\rho
\approx 3$ one already obtains a good agreement.  All constants
appearing in Eq.~(\ref{relations-CCL}) can be exactly computed by
performing a one-loop calculation.  The relevant formulae are reported
in App.~\ref{App:mediumrange}.

\section{Numerical results from Monte Carlo simulations}
\label{secmc}

\subsection{The lattice model}
\label{secmc1}

In order to investigate the existence of the O(2)$\otimes$O(2)
universality class by numerical MC simulations, we considered a simple
cubic lattice $L^3$ and the following Hamiltonian:
\begin{eqnarray}
&&{\cal H} = - \beta \sum_{x,\mu} 
\left( \varphi_x\cdot \varphi_{x+\mu} +
\psi_x\cdot \psi_{x+\mu} \right)
+ \sum_x \left( \varphi_x^2 + \psi_x^2\right)+
\label{HL} \\
&& \qquad\qquad
+ A_4 \sum_x \left[ (\varphi_x^2-1)^2 + (\psi_x^2-1)^2\right]
+ 2 A_{22} \sum_x \varphi_x^2 \psi_x^2,
\nonumber
\end{eqnarray}
where $\varphi$ and $\psi$ are two-component
real variables.
The Hamiltonian ${\cal H} $
describes two identical two-component lattice $\phi^4$ models coupled
by an energy-energy term. Note that
if $A_{22}=A_4$ the symmetry is enlarged to O(4) and
we have the standard four-component lattice $\phi^4$ model.
By applying the trasformation
\begin{equation}
\phi_{11} = {\varphi_{1}-\psi_{2}\over \sqrt{2}},\quad
\phi_{12}  = - {\varphi_{2}-\psi_{1}\over \sqrt{2}},\quad 
\phi_{21} = {\varphi_{2}+\psi_{1}\over \sqrt{2}},\quad
\phi_{22} = {\varphi_{1}+\psi_{2}\over \sqrt{2}},
\label{transf}
\end{equation}
one can easily see that model (\ref{HL}) 
corresponds to the Hamiltonian (\ref{H-MR1}) with nearest-neighbor coupling 
$J$ and  potential (\ref{lattice-potential}) with
\begin{eqnarray}
U_0 &=& 12 (A_{22} + A_4), \nonumber \\
V_0 &=& 24 (A_{22} - A_4), \nonumber \\
r &=& 1 - 2 A_4. 
\label{parHL-parMR}
\end{eqnarray}
Therefore, model (\ref{HL}) is a lattice discretization of the
O(2)$\otimes$O(2) Hamiltonian (\ref{LGWH}).  According to the FT
results presented in Sec.~\ref{MSbar}, continuous transitions in
models with $A_{22}>A_4$ should be controlled by the O(2)$\otimes$O(2)
FP.  For $A_{22}=A_4$ the symmetry is enlarged to O(4) and the
transition is controlled by the O(4) FP.  If $A_{22}<A_4$ continuous
transitions should belong to the XY universality class, because the
O(2)$\otimes$O(2) theory has a stable XY FP with attraction domain in
the region $v_0<0$, see, e.g., Ref.~\cite{PV-rev}.

\subsection{Monte Carlo simulations}
\label{MCsim}

We present the results of MC simulations for several values of the
quartic Hamiltonian parameters. We set $A_4=1$ and vary $A_{22}$,
considering $A_{22}=3$, 11/4, 5/2, 9/4, 2, 5/3, and 7/5.  If we define
\begin{equation}
s_L\equiv {V_0\over U_0} = 2{ A_{22}-A_4 \over A_{22}+A_4 },
\label{wpar}
\end{equation}
they correspond to $s_L=1$, 14/15, 6/7, 10/13, 2/3, 1/2, and 1/3 respectively.
Note that for $A_{22}>A_4>0$ we have $0<s_L<2$.

We simulate model (\ref{HL}) by using two different types of local
moves: (1) a Metropolis update in which $\varphi_x$ and $\psi_x$ are
both varied by adding a random term to each component in such a way to
obtain a 50\% acceptance; (2) an O(4) update \cite{microcanonical} in
which $\varphi_x$ and $\psi_x$ are both changed keeping fixed the
O(4)-symmetric part of the Hamiltonian, while the O(4)-breaking term
$2 (A_{22}-A_4) \sum_x \varphi_x^2 \psi_x^2$ is taken into account by
performing a standard Metropolis acceptance test (the acceptance of
this move is rather large, varying from approximately 78\% for
$A_{22}=3$ to 94\% for $A_{22}=7/5$).  In the simulations we use a
mixed algorithm in which we performed an O(4) sweep and a standard
Metropolis sweep with probability 1/4 and 3/4 respectively. A rough
investigation of the autocorrelation times shows that this is an
optimal combination.  The mixed algorithm is substantially faster (for
$s_L=1/3$ the autocorrelation time of the magnetic susceptibility
$\chi$ decreases by approximately a factor of 10) than the algorithm
in which only the Metropolis update is used.

We perform a FSS study using lattices with $16\le L \le 120$ for
values of $\beta$ close to $\beta_c$.  The integrated autocorrelation
time $\tau_\chi$ of the magnetic susceptibility (estimated by using
the blocking method) increases approximately as $\tau_\chi\approx c
L^2$ at $\beta_c$ with $c\approx 0.2$ for $A_{22}=7/5$ and $c\approx
0.5$ for $A_{22}=5/3$, where the time unit is an update of all spin
variables. For larger values of $A_{22}$ the transition becomes of
first order and the dynamics becomes very slow as $L$ increases.  The
large autocorrelation time, i.e. the difficulty of the updating
algorithm to provide independent configurations, represents the main
limitation to the study of the critical behavior at $\beta_c$ and for
large volumes.  For each value of $\beta$ we typically performed runs
of a few million iterations for the smallest values of $L$, and of
20-40 million iterations for the largest lattice sizes.  The total CPU
time was approximately 5 CPU years of a single 64-bit Opteron 246
(2Ghz) processor.

\subsection{Definitions and notations}
\label{definitions}

In order to investigate the phase diagram, it is useful to study the FSS
of quantities related to the energy 
\begin{equation}
E \equiv {1\over V} \langle {\cal H} \rangle
\label{edef}
\end{equation}
($V\equiv L^3$ is the volume),
such as the specific heat 
\begin{equation}
\quad C \equiv {1\over V} \left( \langle {\cal H}^2 \rangle -
\langle {\cal H} \rangle^2 \right),
\label{chdef}
\end{equation}
and the energy cumulant \cite{CLB-86}  
\begin{equation}
B_E \equiv 
1 - { \langle {\cal H}^4 \rangle\over 3 \langle {\cal H}^2 \rangle^2}.
\label{CLB-cumulant}
\end{equation}
We also define a quantity $M$ related to the magnetization:
\begin{equation}
M \equiv \langle \; \sqrt{\mu^2_\varphi + \mu^2_\psi}\; \rangle,
\label{|M|-def}
\end{equation}
where 
\begin{equation}
\vec{\mu}_{\varphi} \equiv  {1\over V} \sum_x \vec{\varphi}_x ,
\qquad \vec{\mu}_{\psi} \equiv  {1\over V} \sum_x \vec{\psi}_x .
\label{def-mu}
\end{equation}
The two-point correlation function $G(x)$ is defined as
\begin{equation}
G(x) \equiv  \langle \varphi_0 \cdot \varphi_x\;+\;\psi_0 \cdot \psi_x \rangle.
\end{equation}
The corresponding susceptibility $\chi$ and second-moment
correlation length $\xi$ are given by
\begin{equation}
\chi \equiv  \sum_x G(x), \qquad \qquad 
\xi^2 \equiv  {1\over 4 \sin^2 (q_{\rm min}/2)} 
{\widetilde{G}(0) - \widetilde{G}(q)\over \widetilde{G}(q)},
\end{equation}
where $\widetilde{G}(q)$ is the Fourier transform of $G(x)$, $q = (q_{\rm min},0,0)$,
and $q_{\rm min} = 2 \pi/L$. The finite-volume definition of $\xi$ 
is not unique. The definition used here has the advantage of a fast convergence
to the infinite-volume limit \cite{CP-98}.

In our FSS study we shall consider three RG-invariant ratios
\cite{Binder-81}
\begin{eqnarray}
l_\xi &\equiv & {\xi/L}, 
\label{lxidef} \\
B_1   &\equiv & { \langle ( \mu_\varphi^2 + \mu_\psi^2 )^2 \rangle\over
\langle \mu_\varphi^2 + \mu_\psi^2 \rangle^2 }, 
\label{b1def}\\
B_2   &\equiv & 
 { \langle  \mu_\varphi^2 \mu_\psi^2 - (\mu_\varphi\cdot\mu_\psi)^2 
       \rangle\over \langle \mu_\varphi^2 + \mu_\psi^2 \rangle^2 }.
\label{b2def}
\end{eqnarray}
Note that $B_1$ (resp. $B_2$) is equal to 3/2 (resp. 1/8) 
at $\beta=0$ and to 1 (resp. 0) at $\beta=\infty$.

\subsection{First-order transitions: summary of theoretical results}
\label{foth}

In the case of a first-order transition the probability distributions
of the energy and of the magnetization are expected to show a double
peak for large values of $L$.  Therefore, as a first indication, one
usually looks for a double peak in the distribution of the energy and
of the magnetization.  However, as discussed in the literature, see,
e.g., Refs.~\cite{Ape-90,Billoire-95} and references therein, the
observation of a double peak in the distribution of the energy for a
few finite values of $L$ is not sufficient to conclude that the
transition is a first-order one.  For instance, in the two-dimensional
Potts model with $q=3$ and $q=4$\cite{FMOU-90,Mc-90}, double-peak
distributions are observed for relatively large lattice sizes even if
the transition is known to be continuous.  In order to identify
definitely a first-order transition, it is necessary to perform a more
careful analysis of the large-$L$ scaling properties of the
distributions or, equivalently, of the specific heat, the energy
cumulant, and the Binder cumulants, see, e.g.,
Refs.~\cite{CLB-86,VRSB-93,LK-91}.

The difference of the two maximum values $E_+$ and $E_-$ of the
energy-density distribution gives the latent heat.  Alternative
estimates of the latent heat can be obtained from the lattice-size
scaling of the specific heat $C$ and of the energy cumulant $B_E$.
According to the phenomenological theory \cite{CLB-86} of first-order
transitions based on the two-Gaussian Ansatz, for a lattice of size
$L$ there exists a value $\beta_{\rm max}$ of $\beta$ where $C$ has a
maximum, $C_{\rm max}$, and
\begin{equation}
\beta_{\rm max}-\beta_c=O(1/V),\qquad
C_{\rm max} = V\left[ {1\over 4} \Delta_h^2 + O(1/V)\right],
\label{cmaxsc}
\end{equation}
where $\Delta_h$ is the (rescaled) latent heat
\begin{equation}
\Delta_h\equiv E_+ - E_-.
\label{latentheat}
\end{equation}
Note that, since the temperature parameter $\beta$ is included in the
Hamiltonian (\ref{HL}), $\Delta_h$ should be identified with the
dimensionless ratio between the latent heat and the critical
temperature.  We recall that in the case of a continuous transition
one expects
\begin{equation}
\beta_{\rm max}-\beta_c\approx a L^{-1/\nu},\qquad
C_{\rm max} \approx  b L^{\alpha/\nu} + c.
\end{equation}
The energy cumulant $B_E$ can also be used to identify first-order transitions.
Indeed, a careful analysis \cite{CLB-86} shows that 
there is a value $\beta_{\rm min}$ where 
$B_E$ has a minimum, $B_{E,{\rm min}}$, and which 
is related to the latent heat. The phenomenological theory gives
\cite{BLMGIP-90}
\begin{equation}
\beta_{\rm min}-\beta_c=O(1/V),\qquad
B_{E,{\rm min}} = {2\over 3} 
\left[ 1 - {1\over 2} \Delta^2 - {1\over 8} \Delta^4 \right]
+ O(1/V),
\label{limbe}
\end{equation}
where 
\begin{equation}
\Delta\equiv {E_{+}-E_{-}\over \sqrt{E_{+}E_{-}} } .
\label{deltadef}
\end{equation}
In continuous transitions $\Delta = 0$---there is only one peak in the
energy distribution---and the infinite-volume limit of $B_{E,{\rm
min}}$ is trivial: $\lim_{L\rightarrow \infty} B_{E,{\rm min}}=2/3$.

As discussed in Ref.~\cite{VRSB-93}, the distribution of the order
parameter is also expected to show two peaks at $M_+$ and $M_-$, $M_-
< M_+$, with $M_- \to 0$ as $L\to \infty$ since in the
high-temperature phase there is no spontaneous magnetization.  The
phenomenological theory predicts that the Binder parameter can still
be used to identify the critical point [the analysis shows that
$\beta_{\rm cross}(L_1,L_2) - \beta_c\sim \min(L_1,L_2)^{-2d}$, where
$\beta_{\rm cross}(L_1,L_2)$ is the value of $\beta$ at which
$B_1(L_1)=B_1(L_2)$]. Moreover, it predicts
\begin{equation}
{d B_1/ d\beta}|_{\beta=\beta_c} \approx c L^{d},
\label{db1sc}
\end{equation}
for sufficiently large lattice sizes, where $d$ is the space dimension.
More generally,
close to $\beta_c$ the phenomenological theory predicts 
\cite{VRSB-93,LK-91}  
\begin{equation}
    B_1(\beta,L) \approx f[(\beta - \beta_c) L^d].
\label{scalingB1-FOT}
\end{equation}
Such a relation is valid only sufficiently close to $\beta_c$ since
the scaling function $f(x)$ diverges for $x = x_{\rm peak}$ where $
x_{\rm peak}$ is related to the position of the peak present in $B_1$
for $L\to \infty$. Ref.~\cite{VRSB-93} indeed shows that
$B_1(\beta,L)$ at fixed $L$ has a maximum $B_{1,\rm max}(L)$ at $\beta
= \beta_{\rm peak}(L) < \beta_c$ with
\begin{equation}
B_{1,\rm max} \sim L^d, \qquad\qquad
\beta_{\rm peak} - \beta_c \approx c_p L^{-d}.
\label{B1peak}
\end{equation}
Thus, for $(\beta - \beta_c) L^d$ close to $(\beta_{\rm peak} -
\beta_c) L^d = x_{\rm peak} = c_p$ the scaling behavior
(\ref{scalingB1-FOT}) breaks down [$f(x_{\rm peak})$ diverges] and
subleading terms of order $L^{-d}$ must be included.  In the region in
which Eq.~(\ref{scalingB1-FOT}) holds, we have ${dB_1/ d\beta} \approx
L^d f'[(\beta - \beta_c) L^d]$. By using Eq.~(\ref{scalingB1-FOT}) we
can express $(\beta - \beta_c) L^d$ in terms of $B_1$ obtaining the
relation
\begin{equation}
    {dB_1\over d\beta} = L^d \hat{f}(B_1),
\label{db1vsB1}
\end{equation}
with a suitable scaling function $\hat{f}(x)$. In practice this means that 
$L^{-d} {dB_1/d\beta}$ converges to a  universal function of $B_1$ as 
$L\to \infty$.

It should be noted that the phenomenological theory has been developed
for systems with discrete symmetry group, for instance for the Potts
model \cite{CLB-86,VRSB-93}.  In our case the symmetry is continuous;
thus, one may wonder if it can be also applied to the present case.
The numerical results that we will present below in
Sec.~\ref{phasedia} show that no changes are needed and that all
predictions hold irrespective of the symmetry group. This can be
understood on the basis of a simple argument.  Imagine we introduce a
magnetic field $H$ in the model.  The first-order transition should be
robust with respect to $H$, since we expect here the transition to be
temperature driven. In other words, we expect the behavior to be
unchanged if $H$ is switched on, as long as $H$ is small.  For
$H\not=0$ we have a discrete system, thus all previous scalings
apply.  If the behavior is continuous in $H$, all results also apply
for $H=0$. Note that, for $H\not=0$, $B_2$ is expected to be
noncritical, since $B_2$ vanishes in a system magnetized in a specific
fixed direction.  Thus, $B_2$ should have no discontinuity at the
transition and its derivative at ${\beta=\beta_c}$ should be
finite, at variance with the behavior of $B_1$, cf.~Eq.~(\ref{db1sc}).
This fact will also be verified by the MC results that we shall
present in Sec.~\ref{phasedia}.

\begin{figure}[tb!]
\centerline{\psfig{width=10truecm,angle=0,file=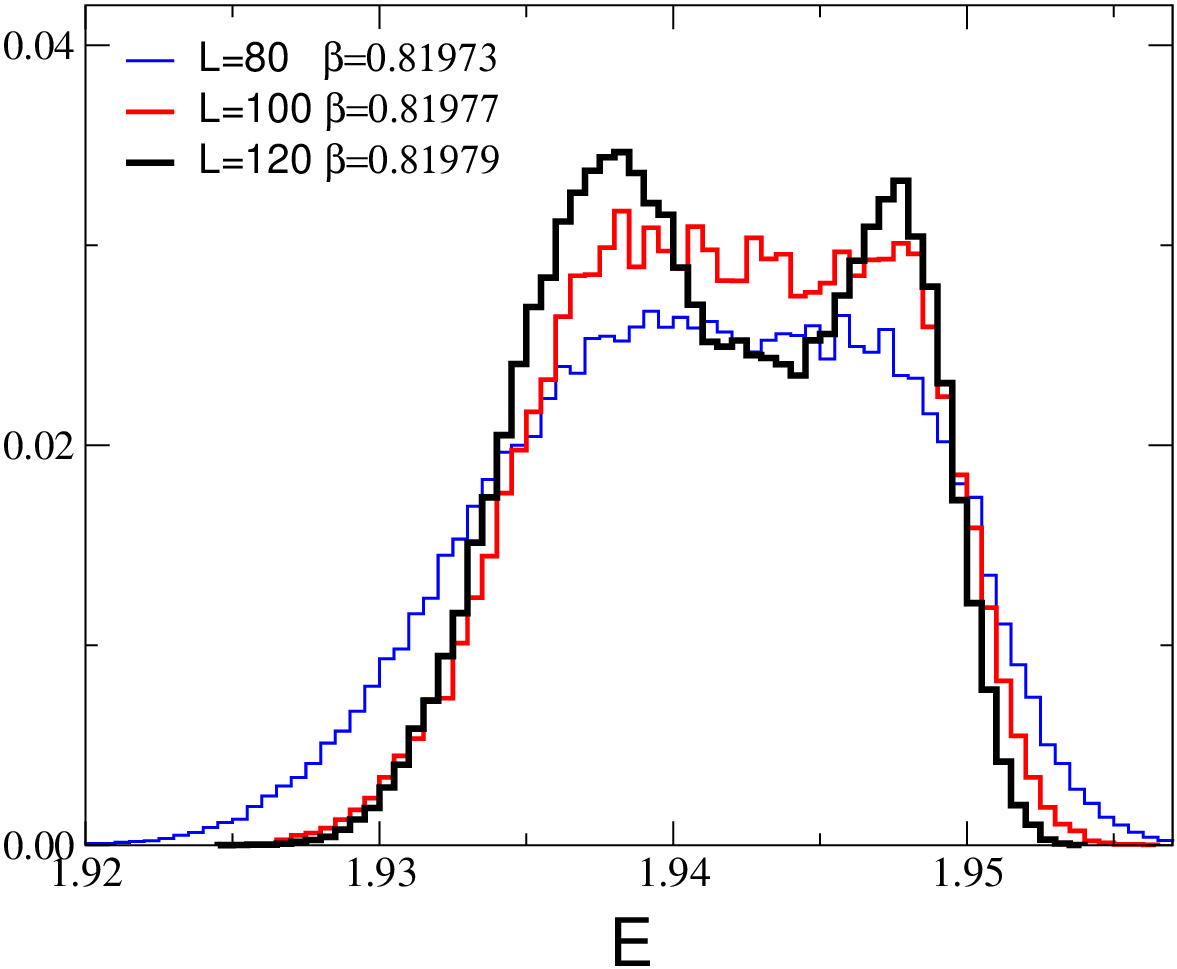}}
\vspace{15mm}
\centerline{\psfig{width=10truecm,angle=0,file=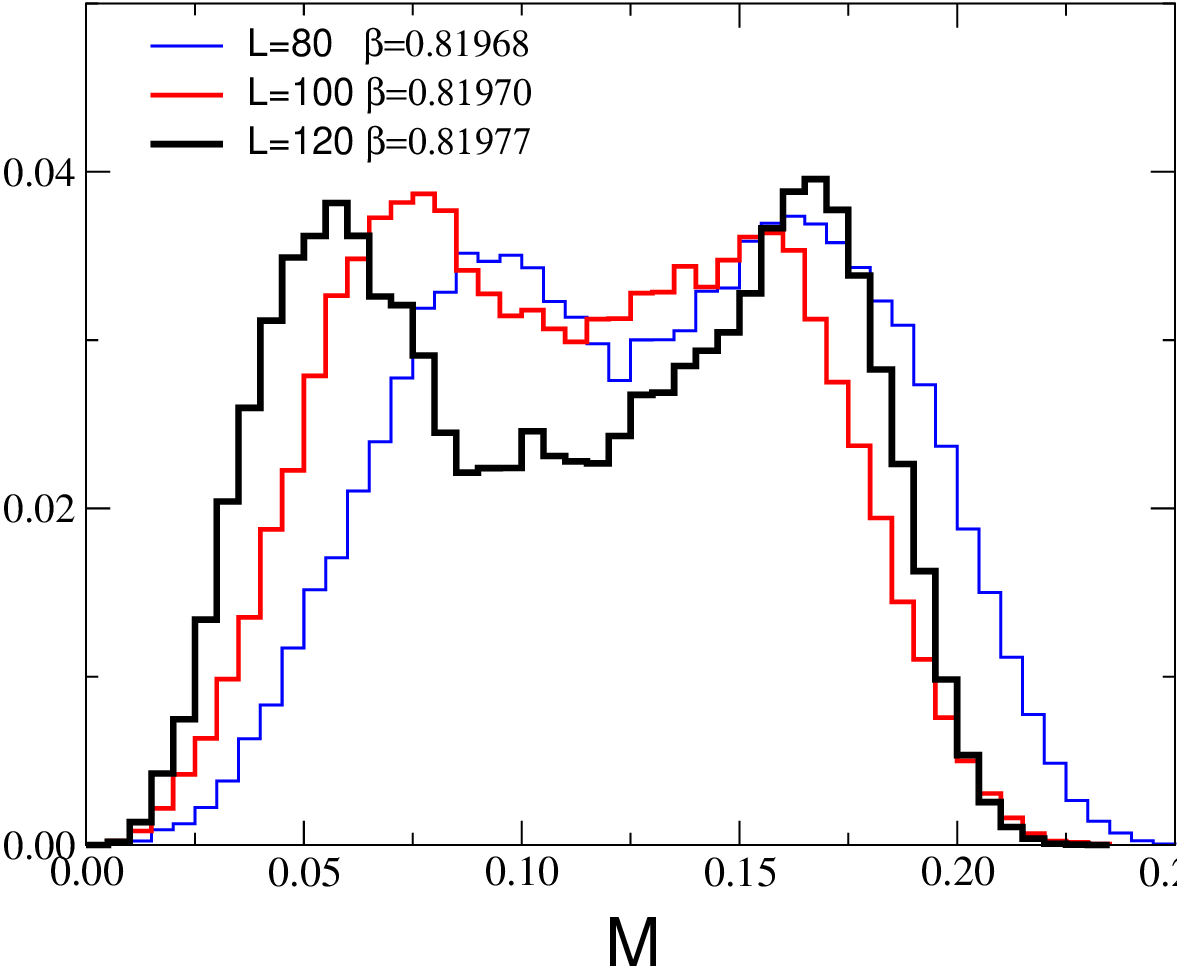}}
\caption{
Hystograms of the energy density (above) and of the 
magnetization (below) for $A_{22} = 2$ and lattice sizes $L=80$, 100, 120. 
For each $L$ we report the data for the value of $\beta$ at which 
the two peaks have approximately the same height. 
These results have been obtained by reweighting the MC data 
at $\beta=0.89170$, 0.89175, 0.81980. 
}
\label{hye}
\end{figure}

\subsection{The phase diagram for $A_{22}>A_4=1$}
\label{phasedia}

In this subsection we investigate the phase diagram of the lattice model
(\ref{HL}) for $A_{22}>A_4=1$ with the purpose of identifying the 
regions in which the model shows a first-order or a second-order 
phase transition.

\begin{figure}[tb]
\centerline{\psfig{width=12truecm,angle=0,file=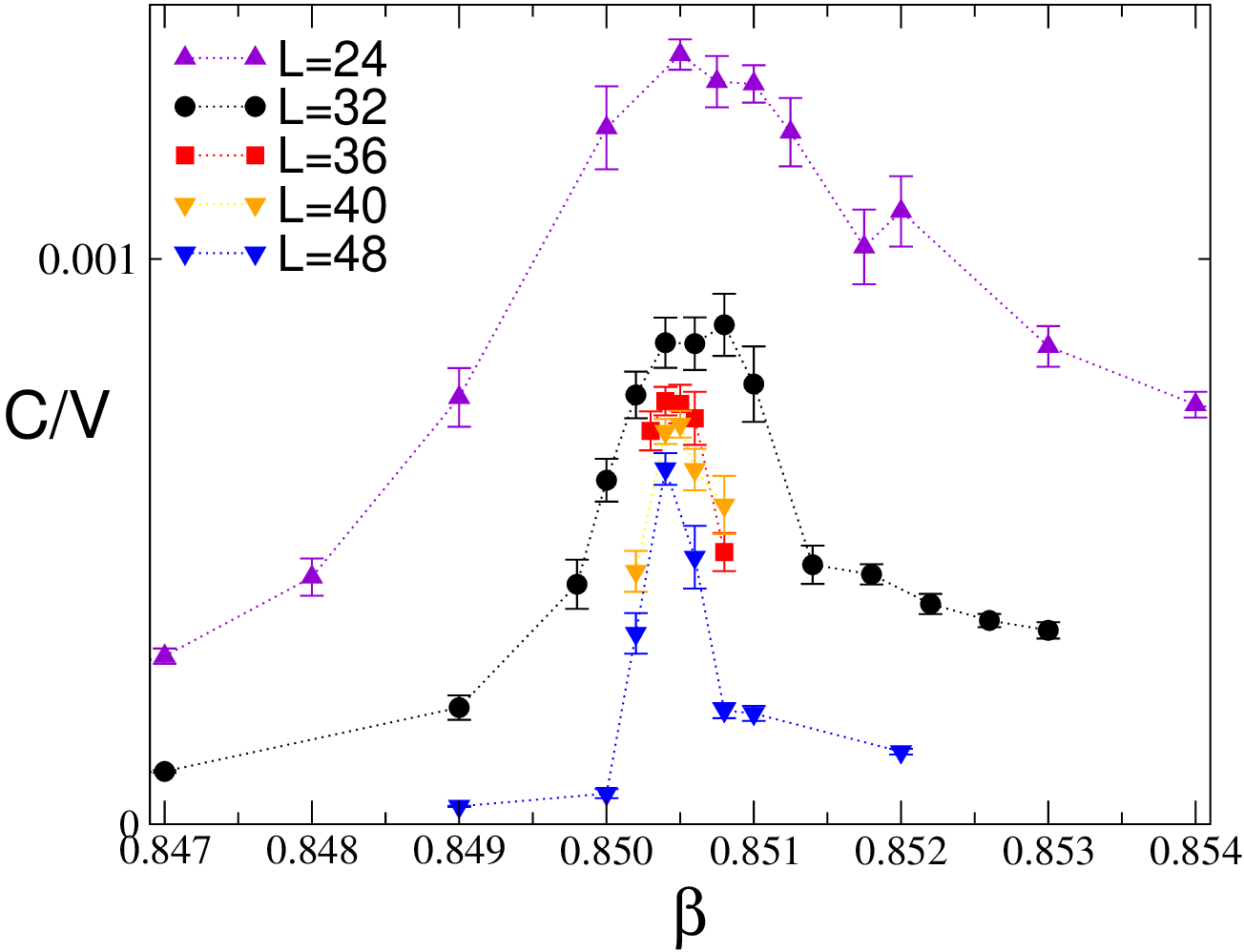}}
\vspace{2mm}
\caption{
Specific heat for $A_{22}=5/2$ and
several values of $L$. 
}
\label{cv5o2}
\end{figure}

\begin{figure}[tb]
\centerline{\psfig{width=12truecm,angle=0,file=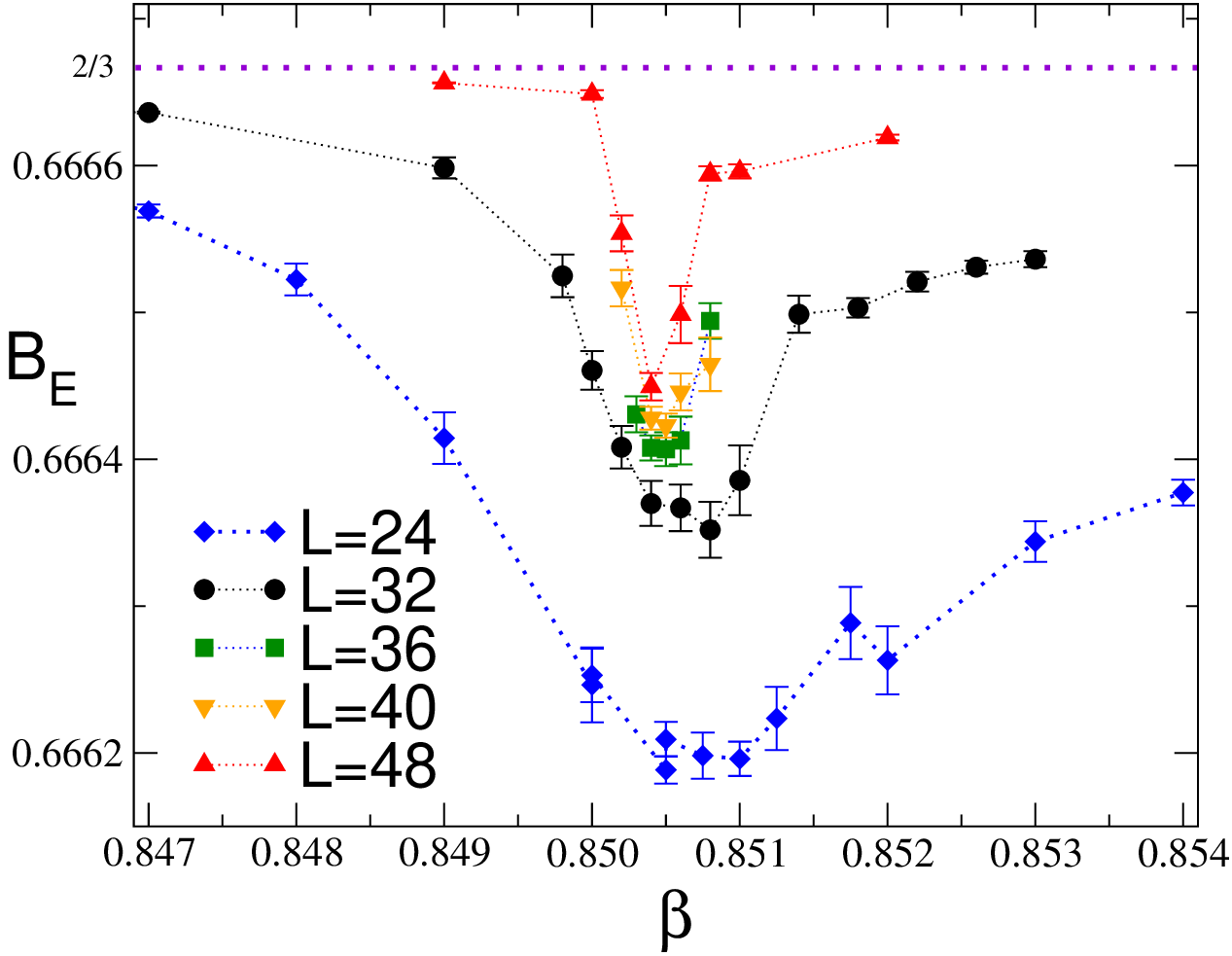}}
\vspace{2mm}
\caption{Energy cumulant $B_E$ for $A_{22}=5/2$ and
several values of $L$.
}
\label{be5o2}
\end{figure}

The distributions of $E$ and $M$ show two peaks for $A_{22} \ge 2$
when $L$ is large enough.  For $A_{22}=3$ two peaks are already
observed for $L = 16$, while for $A_{22} = 2$ two peaks are observed
only for $L\gtrsim 100$, cf. Fig.~\ref{hye}. Thus, the model with $A_4
= 1$ has apparently a relatively strong first-order transition for
$A_{22} \gtrsim 3$ that gradually weakens as $A_{22}$ is decreased.
In order to check that we are really in the presence of a first-order
transition, we check the scaling behavior of the different observables
for $L\to\infty$. The predictions for $C$ and $B_E$ are well
verified. For instance, in Fig.~\ref{cv5o2} we show $C/V$ for
$A_{22}=5/2$ for several values of $L$.  In agreement with
Eq.~(\ref{cmaxsc}) $C_{\rm max}/V$ has a finite limit as
$L\to\infty$. In Fig.~\ref{be5o2} we show the energy cumulant $B_E$
for the same value of $A_{22}$; $B_{E,\rm min}$ is different from 2/3,
confirming the first-order nature of the transition.  From $B_{E,\rm
min}$ we can determine the latent heat.  For each $L$ we determine
$\Delta_L$, which is obtained from $B_{E,\rm min}(L)$ by using
Eq.~(\ref{limbe}) and neglecting the $1/V$ corrections. The latent
heat is obtained by extrapolating $\Delta_L$ assuming $1/V$
corrections.

\begin{figure}[tb]
\centerline{\psfig{width=12truecm,angle=0,file=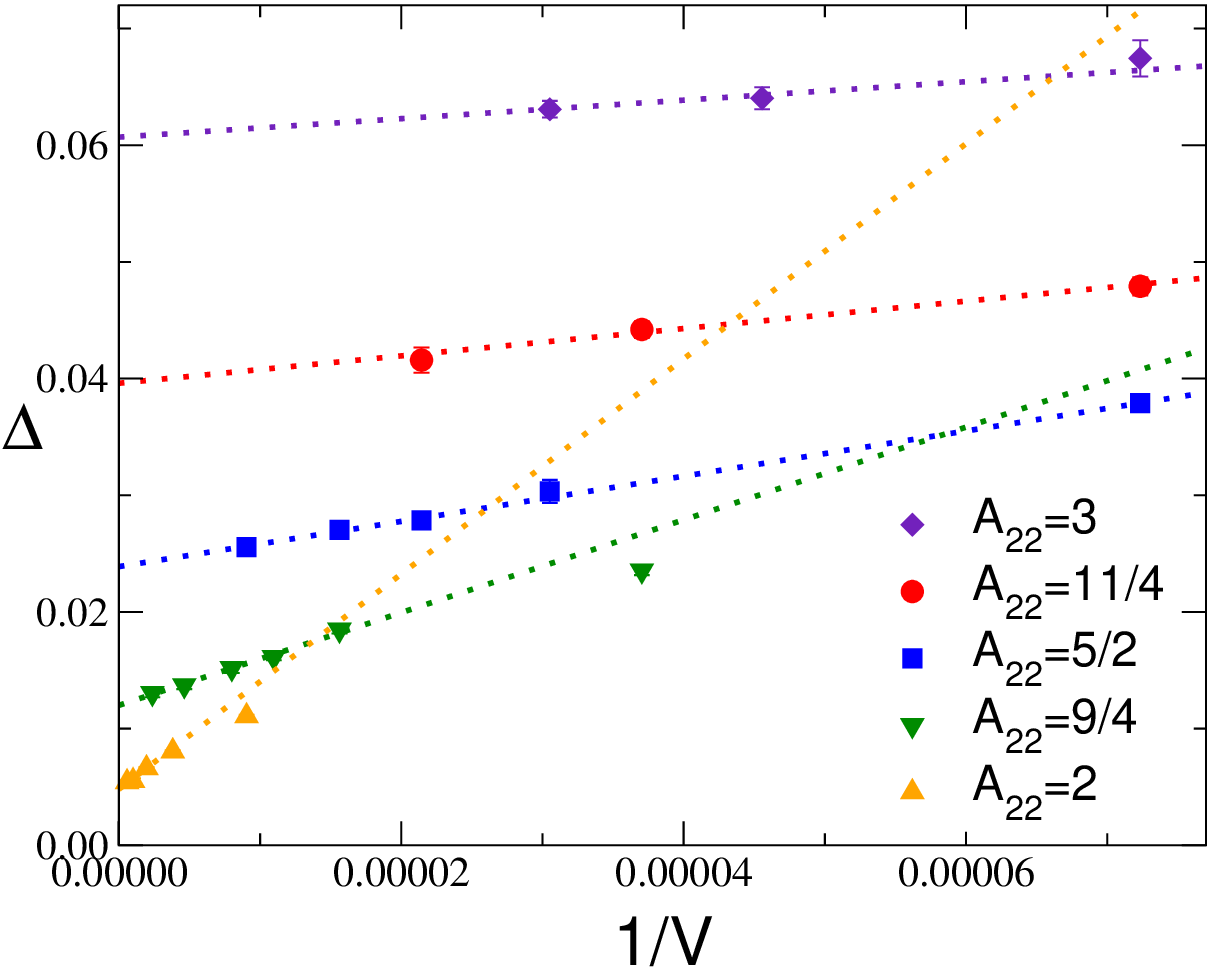}}
\vspace{2mm}
\caption{Values of $\Delta_L$ for $A_{22} = 2$, 9/4, 5/2, 11/4, 3,
as obtained from $B_{E,{\rm min}}$ by using Eq.~(\ref{limbe}). 
The lines represent the infinite-volume extrapolations 
assuming a $1/V$ asymptotic correction. 
}
\label{LHextra}
\end{figure}

Estimates of $\Delta_L$ for several values of $A_{22}$ and $L$ are
reported in Fig.~\ref{LHextra}. They show the expected $1/V$ behavior
and allow a precise determination of the latent heat in the
infinite-volume limit.  The results of the extrapolations are reported
in Table~\ref{bdtab}.  They are in perfect agreement with those
obtained from the maximum of the specific heat, using
Eq.~(\ref{cmaxsc}), and from the position of the peaks in the energy
distributions.

We have also performed MC simulations for $A_{22} = 5/3$ and 7/5 on
lattices of size $L\le 120$, without observing evidences for
first-order transitions. The hystograms of $E$ and $M$ do not show any
evidence of two peaks and are not significantly broad, as for instance
the distribution of $E$ for $L=80$ and $A_{22} = 2$,
cf. Fig.~\ref{hye}, which would indicate the onset of two peaks.  In
the cases $A_{22}=5/3,7/5$, $C$ and $B_E$ on lattices of size $L\le
120$ do not show the behavior predicted by the phenomenological theory
of first-order transitions \cite{CLB-86}.  In particular, $B_{E,\rm
min}$ continuously increases towards 2/3 and we can only put upper
bounds on $\Delta_h$. For $A_{22} = 5/3$ we obtain for instance
$\Delta_h < 0.005$.  A more stringent bound is suggested by the width
of the energy distribution at $\beta_c$, i.e. $\Delta_h < 0.003$.

\begin{table}
\caption{
For the model with $A_4=1$ and several values of $A_{22}$
we report $s_L$, $\beta_c$, the average energy density $E_c\equiv (E_++E_-)/2$,
$\Delta$, cf. Eq.~(\ref{deltadef}), and the latent heat $\Delta_h$.
For $\beta_c$ and $E_c$ all reported digits are exact.
}
\begin{tabular}{ccllll}
\multicolumn{1}{c}{$A_{22}$}& 
\multicolumn{1}{c}{$s_L$}& 
\multicolumn{1}{c}{$\beta_c$} & 
\multicolumn{1}{c}{$E_c$} & 
\multicolumn{1}{c}{$\Delta$}&
\multicolumn{1}{c}{$\Delta_h$} \\
\hline
3    & 1     &  0.8733 &  1.98 & 0.0607(6) &  0.1198(12) \\
11/4 & 14/15 &  0.8627 &  1.98 & 0.0396(12) & 0.0783(24) \\
5/2  &   6/7 &  0.8504 &  1.97 & 0.0239(4)  & 0.0471(8) \\ 
9/4  & 10/13 &  0.8364 &  1.96 & 0.0120(4) & 0.0235(8) \\ 
2    &   2/3 &  0.8198 &  1.94 & 0.0048(2)  & 0.0093(4) \\ 
5/3  &   1/2 &  0.7927 &  1.92 & $<$ 0.0015 & $<$ 0.003 \\
\end{tabular}
\label{bdtab}
\end{table}

Let us now discuss the behavior of the variables related to the
magnetization.  In particular, we focus on the derivatives with
respect to $\beta$ of $B_1$ and $B_2$.  Predictions (\ref{B1peak}) and
(\ref{db1vsB1}) are well verified by our data with $A_{22} \ge 2$. We
observe the presence of a peak in $B_1$ at fixed $L$ that becomes
sharper as $L$ increases and we also verify that far from this peak
Eq.~(\ref{db1vsB1}) holds. Of course, corrections increase as $A_{22}$
decreases, as expected. On the other hand, $B_1$ is monotonically
decreasing with $\beta$ at fixed $L$ for $A_{22} = 7/5$ and
$A_{22}=5/3$, providing no evidence that the transition is of first
order for these two values of $A_{22}$.

We have repeated the same analysis for $B_2$. Its behavior looks quite
different with respect to $B_1$. First of all, we observe a monotonic
behavior without peaks for all the considered values of $A_{22}$,
including the largest ones for which the peak in $B_1$ is rather
sharp.  Moreover, the data are reasonably well described by assuming
\begin{equation}
    {dB_2\over d\beta} = \hat{f}(B_1),
\end{equation}
suggesting that $B_2$ does not have a jump at the transition, as 
expected on the basis of the argument presented at the end of Sec.~\ref{foth}.
A similar analysis can also be performed for $l_\xi$ for which we have no 
prediction. Our data are roughly consistent with a behavior of the form
\begin{equation}
    {dl_\xi\over d\beta} = L^\theta \hat{f}(B_1)
\end{equation}
with $3.2 \lesssim \theta \lesssim 3.7$.  It is quite difficult to
intepret such a value of $\theta$ that could well be $d$ or $d+1$
depending on the size of the corrections.

\begin{figure}[tb]
\centerline{\psfig{width=12truecm,angle=0,file=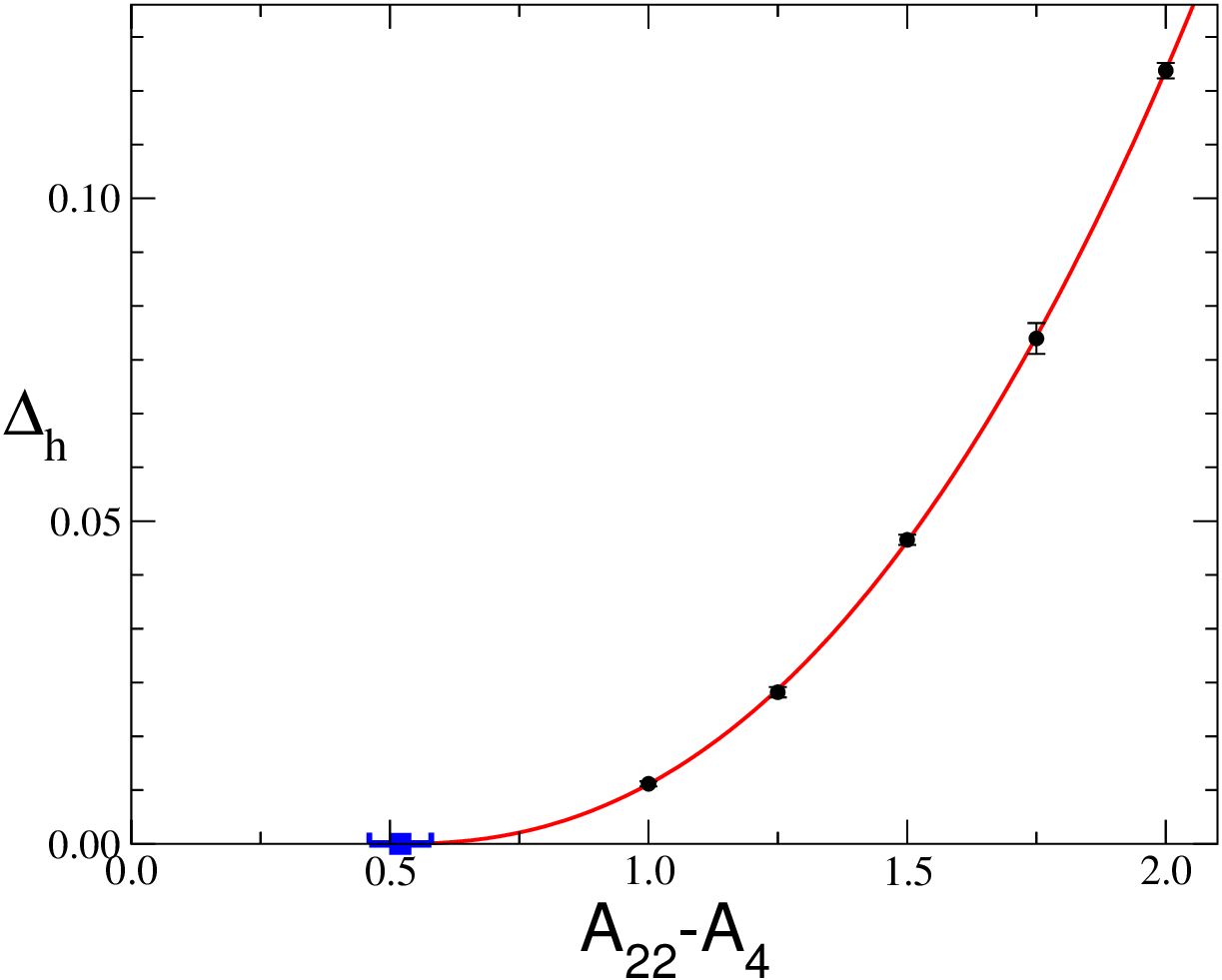}}
\vspace{2mm}
\caption{Latent heat $\Delta_h$ versus $A_{22}$.
The line corresponds to the best fit, $\Delta_h = 0.049(A_{22} - 1.52)^{2.29}$. 
We also report the estimate of $A_{22}^*$ with the corresponding error: 
$A_{22}^* = 1.52(6)$.
}
\label{LHw}
\end{figure}

The results for $\Delta_h$ reported in Fig.~\ref{LHw} suggest
therefore a phase diagram characterized by a line of first-order
transitions extending from large values of $A_{22}$ down to $A_{22}^*$
where $\Delta_h$ vanishes. In order to compute $A_{22}^*$, we should
first discuss the expected behavior of $\Delta_h$ close to the
tricritical point $A_{22}^*$ (see, e.g., Ref.~\cite{Sarbach}).  We
consider a generic model depending on two parameters $\beta$ and $g$
with a tricritical point at $g^*$, $\beta^*_c\equiv \beta_c(g^*)$. The
critical behavior can be parametrized in terms of two linear scaling
fields $u_1$ and $u_2$ with RG dimensions $y_1$ and $y_2$, satisfying
$y_1 > y_2$.  In the absence of any symmetry, the linear scaling
fields are combinations of $(g - g^*)$ and of $(\beta - \beta_c^*)$.
The scaling field $u_1$ is completely defined, while $u_2$ can be
arbitrarily chosen as long as it is independent of $u_1$. Therefore,
we write
\begin{eqnarray}
   u_1 &=& a_1 (\beta - \beta_c^*) + a_2 (g - g^*),\nonumber \\
   u_2 &=& g - g^*.
\label{scfield}
\end{eqnarray}
The first-order transition line is characterized by the equation $u_2
= c_1 u^{y_2/y_1}_1$ with, by a proper choice of the scaling field,
$u_1 > 0$. Analogously, the second-order transition line is given by
$u_2 = c_2 (s u_1)^{y_2/y_1}$ where $s$ may be either 1 or $-1$ and
$|c_1|\not= |c_2|$. Note that both lines are tangent to $u_1 = 0$, but
that the nonanalytic deviations are parametrized differently.

The free energy can be written as 
\begin{equation}
F \approx F_{\rm reg}(\beta,g) + |u_2|^{d/y_2} 
        F_{\rm sing,\pm} (u_1 |u_2|^{-y_1/y_2}),
\label{rgeq}
\end{equation}
where $F_{\rm sing,+}$ (resp. $F_{\rm sing,-}$) applies to the case
$u_2 > 0$ (resp. $u_2 < 0$), and $F_{\rm reg}(\beta,g)$ is a regular
function.  By hypothesis, assuming $c_1 > 0$ without loss of generality, 
$F_{\rm sing,+}(x)$ is continuous but has a discontinuous
derivative at the first-order transition line, i.e. for $x =
c_1^{-y_1/y_2}$.  It follows
\begin{equation}
 E_+ - E_- \approx a_2 |u_2|^{(d-y_1)/y_2} \Delta F_{\rm sing,+}' 
\sim  |g - g^*|^{(d-y_1)/y_2}.
\label{deltabeh}
\end{equation}
In the model we consider $g = A_{22}$, so that we predict
that sufficiently close to the tricritical point $A_{22}^*$ 
\begin{equation}
\Delta_h = \Delta_0 (A_{22} - A_{22}^*)^\theta, 
\end{equation}
where $\theta$ is an exponent 
defined by the tricritical theory. If we fit our data for the latent heat 
with this expression we obtain 
\begin{eqnarray} 
A_{22}^* = 1.52(6), \qquad \qquad
      \Delta_0 = 0.049(7), \qquad \qquad \theta = 2.29(15),
\label{stimaA22star}
\end{eqnarray}
with a $\chi^2$ per degree of freedom (DOF) of 0.24. The fit is stable
with respect to the number of points that are included: If the point
with $A_{22} = 3$ is not included we obtain $A_{22}^* = 1.50(12)$ and
$\theta = 2.35(36)$, in full agreement with the result reported
above. We have also investigated the possibility that $A_{22}^* = A_4
= 1$, which would imply the absence of critical transitions in the
O(2)$\otimes$O(2) universality class. A fit of the data for the latent
heat of the form $\Delta_h = \Delta_0(A_{22} - 1)^\theta$ gives
$\theta = 3.49(4)$ with $\chi^2/$DOF $ \approx 9$. If the point with
$A_{22} = 3$ is not included we obtain $\theta = 3.87(4)$
($\chi^2/$DOF $ \approx 11$), while if the two points with largest
$A_{22}$ are discarded we obtain $\theta = 4.00(3)$ ($\chi^2/$DOF $
\approx 9$).  These fits are significantly worse than that presented
above. Moreover, the estimate of $\theta$ does not agree with the
theoretical prediction that can be obtained assuming the tricritical
point to be at $A_{22}=1$.  Indeed, if $A_{22}^* = 1$ the tricritical
theory coincides with the O(4) theory.  There are two relevant
perturbations, the thermal perturbation with RG dimension $1/\nu$ and
the perturbation that breaks the O(4) symmetry down to
O(2)$\otimes$O(2), which is associated in field theory with the spin-4
quartic operator, with RG dimension $\phi/\nu$.  For the O(4)
universality class, $\alpha = -0.247(6)$ (Ref.~\cite{Hasenbusch-01})
and $\phi = 0.08(1)$ (Ref.~\cite{CPV-03}).  Since $\phi< 1$ in this
case, we have $y_1 = 1/\nu$ and $y_2 = \phi/\nu$, so that we should
have
\cite{nota-Sarbach}
\begin{equation}
\Delta_h \sim  (A_{22} - 1)^{(1-\alpha)/\phi} \sim (A_{22} - 1)^{16\pm2},
\end{equation}
which is not compatible with the MC results.  Thus, our results are
incompatible with $A_{22}^* = 1$, unless the limiting behavior sets in
for even smaller values of $\Delta_h$. This seems unlikely, since for
$A_{22}\lesssim 3$ we are already in a region in which $\Delta_h\ll
1$.  Thus, our simulations provide a nice evidence in favor of the
existence of a tricritical point at $A_{22}^*>1$ separating the
first-order transition line from a continuous transition line for
$1<A_{22}<A_{22}^*$.  Notice that a phase
transition is always expected due to the qualitative difference of the
minimum of the free energy in the high- and low-temperature regions.

The estimate (\ref{stimaA22star}) implies that for $A_{22} = 5/3$ one
should observe a first-order transition. As we have already discussed,
our data do not show any evidence for that. This is hardly surprising,
since the fit of the estimates of $\Delta_h$ predicts $\Delta_h =
0.8(6)\cdot 10^{-3}$ for $A_{22} = 5/3$, which is smaller than the
bound $\Delta_h < 3\cdot 10^{-3}$ obtained on lattices $L\le 100$.
{\em A posteriori}, however, one can convince oneself of the
first-order nature of the transition by considering the estimated
values of $\nu$. For instance, the derivative of the Binder parameter
$B_1$ with respect to $\beta$ is expected to scale at the critical
point as $B_1' \sim L^d$ for a first-order transition.  In our
analyses of the data with $A_{22} \ge 2$, we find that, if one
estimates $B_1'$ with $B_1'\sim L^\theta$, the effective $\theta$
rapidly increases towards $d$ as the considered set of sizes
increases, even when a double peak in the distributions is not yet
evident.  For $A_{22} = 5/3$ if we use the data with $L \ge 60$ we
obtain $1/\theta = 0.47(2)$, while including only the data with $L \ge
80$ gives $1/\theta = 0.42(4)$: the exponent $\theta$ is increasing
towards the expected value $\theta = d = 3$. At a second-order
transition one expects $B_1' \sim L^{1/\nu}$, and thus the previous
results would give a quite small value for $\nu$, definitely
incompatible with the FT predictions. Therefore, even if we do not
have a direct evidence, the data for $A_{22} = 5/3$ are better
explained by a very weak first-order transition than by a second-order
one.

\begin{figure}[tb]
\centerline{\psfig{width=12truecm,angle=0,file=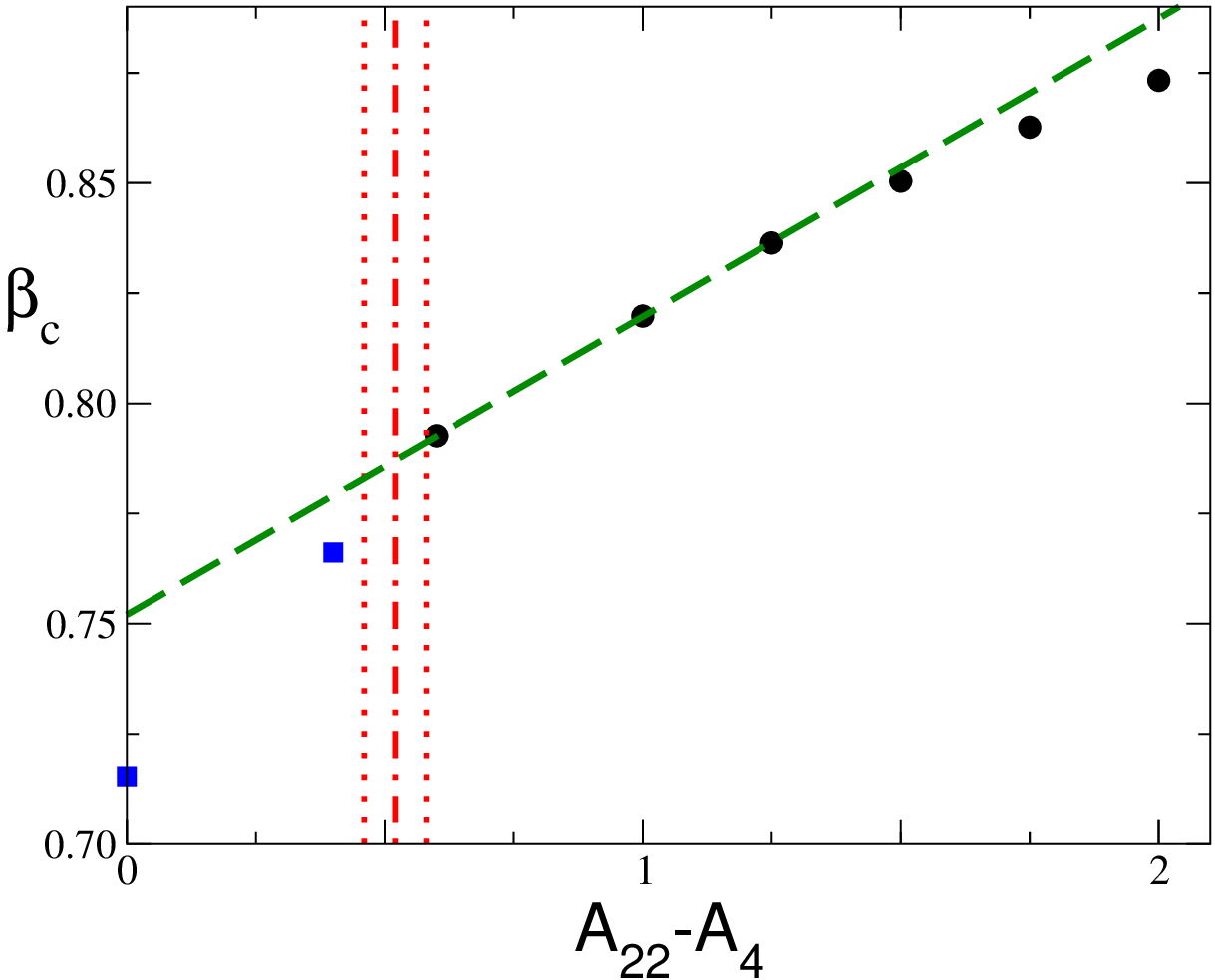}}
\vspace{2mm}
\caption{The inverse critical temperature $\beta_c$ versus $A_{22}$.
The vertical lines indicate the estimate of $A_{22}^*$ with its error.
The data for $A_{22}>A_{22}^*=1.52(6)$ are indicated by circles,
while the ones for $A_{22}< A_{22}^*$ are indicated by
a square. Errors are much smaller than the size of the symbols.
The dashed line shows the linear fit of the three data 
for $A_{22}>A_{22}^*$ that are closest to $A_{22}^*$.
}
\label{betacplot}
\end{figure}

The presence of a tricritical point at $A_{22}^*=1.52(6)$ is further
supported by the results for $\beta_c$ reported in Table~\ref{bdtab}
and plotted in Fig.~\ref{betacplot} versus $A_{22}-A_4$.  Equation
(\ref{scfield}) implies that, sufficiently close to the tricritical
point $A_{22}^*$, the values of $\beta_c$ along the first-order
transition line behave as
\begin{equation}
\beta_c - \beta_c^* \approx  -{a_2\over a_1} ( A_{22} - A_{22}^*) + 
      {1\over a_1} \left[ {1\over c_1} ( A_{22} - A_{22}^*) \right]^{y_1/y_2},
\label{betac1}
\end{equation}
while, along the second-order transition line, we have similarly
\begin{equation}
\beta_c - \beta_c^* \approx  -{a_2\over a_1} ( A_{22} - A_{22}^*) + 
    {1\over s a_1} 
    \left[ {1\over c_2} ( A_{22} - A_{22}^*) \right]^{y_1/y_2},
\label{betac2}
\end{equation}
Equations (\ref{betac1}) and (\ref{betac2}) strictly hold only if
$y_1/y_2 < 2$, otherwise one should also include additional integer
powers $( A_{22} - A_{22}^*)^n$, $n < y_1/y_2$, with coefficients that
are identical for the first-order and second-order transition line.

The linear dependence of $\beta_c$ on $A_{22}$ for $A_{22}\gtrsim
A_{22}^*$ is clearly observed in Fig.~\ref{betacplot}.  A linear fit
of the three points that are closest to $A_{22}^*$ and satisfy
$A_{22}>A_{22}^*$ gives $a_2/a_1\approx -0.068$ with a reasonable
$\chi^2$.  Note that the deviations from a straight line are very
small, indicating that the corrections, and in particular, the
nonanalytic ones, are tiny.  Fig.~\ref{betacplot} also shows the value
of $\beta_c$ for $A_{22}=7/5<A_{22}^*$, i.e. $\beta_c=0.76615(15)$
which will be determined in the next subsection, and for the O(4)
$\phi^4$ model obtained by setting $A_{22}=A_4=1$, which is given by
$\beta_c=0.7154(1)$ \cite{scres}.  Both values, and in particular the
one for $A_{22}=7/5$, differ significantly from the linear
extrapolation of the data for $A_{22}>A_{22}^*$.  This behavior of
$\beta_c$ as a function of $A_{22}$ is naturally explained by the
presence of a tricritical point at $A_{22}^*=1.52(6)$, and it is
another evidence of the fact that the point $A_{22} = 7/5$ belongs to
the second-order transition region. Indeed, in this scenario the
nonanalytic corrections are different ($|c_2| \not= |c_1|$) from those
observed in the first-order transition region.  Thus, the observed
value for $A_{22}=7/5$ can be explained by the presence of nonanalytic
corrections to the linear behavior that are substantially larger
($|c_2| \ll |c_1|$) than those observed on the side of the first-order
transition line.  The data shown in Fig.~\ref{betacplot} can hardly be
explained by assuming $A_{22}^*=1$, i.e.  a first-order transition
line extending down to the O(4) point $A_{22}=A_4=1$. Indeed, in this
case the linear behavior should extend down to the O(4) point, but
this is clearly contradicted by the values of $\beta_c$ obtained for
$A_{22} = 7/5$ and $A_{22}=1$.  Note also that the same RG arguments
leading to Eqs.~(\ref{scfield}), (\ref{rgeq}) and (\ref{betac1}) tell us
that the values of $\beta_c$ along the continuous transition line must
approach linearly the O(4) point $A_{22}=A_4$, but with a slope
different from the one observed at the tricritical point
$A_{22}^*=1.56(2)$.

In conclusion we have shown that, in the quartic parameter space with
$A_{22}>A_4$, there is a region in which the transition is continuous
and therefore belongs to the O(2)$\otimes$O(2) universality class
controlled by the O(2)$\otimes$O(2) FP found in the FT approach of
Sec.~\ref{MSbar}.

Finally, we should note that in the FT studies the basin of attraction
of the O(2)$\otimes$O(2) FP includes all theories with $s < s^*$ and
$s^* \gtrsim 1$, while here $s^*_L = 0.41(4)$. As we have already
discussed before, in spite of the similar definition, $s_L$ should not
be identified with $s$ and we have $s_L\to s$ only for a weakly
coupled theory, i.e. for $A_4\to0$ and $A_{22}\to 0$.  Here, $A_4 =
1$, so that we are far from this limiting case.  In any case, the FT
results imply that $s^*_L$ increases if $A_{4}$ decreases. The
critical value $s^*_L$ is also expected to increase if longer-range
interactions are added.  Indeed, as shown in
App.~\ref{App:mediumrange}, we expect $0.71 \lesssim s^*_L < 1.23$ for
medium-range models with $A_4 = 1$.

\subsection{Determination of the critical quantities for $A_{22} = 7/5$}
\label{secmcfs}

In this Section we analyze the results for $A_{22} = 7/5$. On the
basis of the analysis presented in the previous Section, in this case
the transition should be of second order.  In the FSS limit we expect
the following behavior
\begin{eqnarray}
  && R \approx f_R[(\beta - \beta_c)L^{1/\nu}],
\label{scalingR1} \\
  && \chi \approx L^{2-\eta} f_\chi[(\beta - \beta_c)L^{1/\nu}],
\label{scalingchi1} \\
  && M \approx L^{-\beta/\nu} f_M[(\beta - \beta_c)L^{1/\nu}],
\label{scalingM1} 
\end{eqnarray}
where $R$ is any RG invariant quantity (we will take $R$ to be 
$B_1$, $B_2$, or $l_\xi$). These scaling forms are valid for 
$\beta\to\beta_c$, $L\to \infty$ at fixed argument $(\beta - \beta_c)L^{1/\nu}$.
From Eq.~(\ref{scalingR1}) we obtain moreover 
\begin{equation}
{dR\over d\beta} \approx L^{1/\nu} f_R'[(\beta - \beta_c)L^{1/\nu}].
\label{scalingdRdb1} 
\end{equation}
Eqs.~(\ref{scalingchi1}), (\ref{scalingM1}), and (\ref{scalingdRdb1})
can be used to determine $\eta$, $\beta/\nu$, and $\nu$. It is also possible 
to avoid the use of  the two unknown quantities, $\beta_c$ and $\nu$. 
We use Eq.~(\ref{scalingR1}) to express
$(\beta - \beta_c)L^{1/\nu}$ in terms of $R$, and rewrite all
equations as
\begin{eqnarray}
&& \chi \approx L^{2-\eta} g_{\chi,R}(R) ,
\label{scalingchi2} \\
&& M \approx L^{-\beta/\nu} g_{M,R}(R) ,
\label{scalingM2} \\
&& {dR_1\over d\beta} \approx L^{1/\nu} g_{R_1,R_2}(R_2) ,
\label{scalingdRdb2} 
\end{eqnarray}
where $R$ is a RG-invariant quantity.

\begin{figure}[!tb]
\centerline{\psfig{width=9truecm,height=6.7truecm,angle=0,file=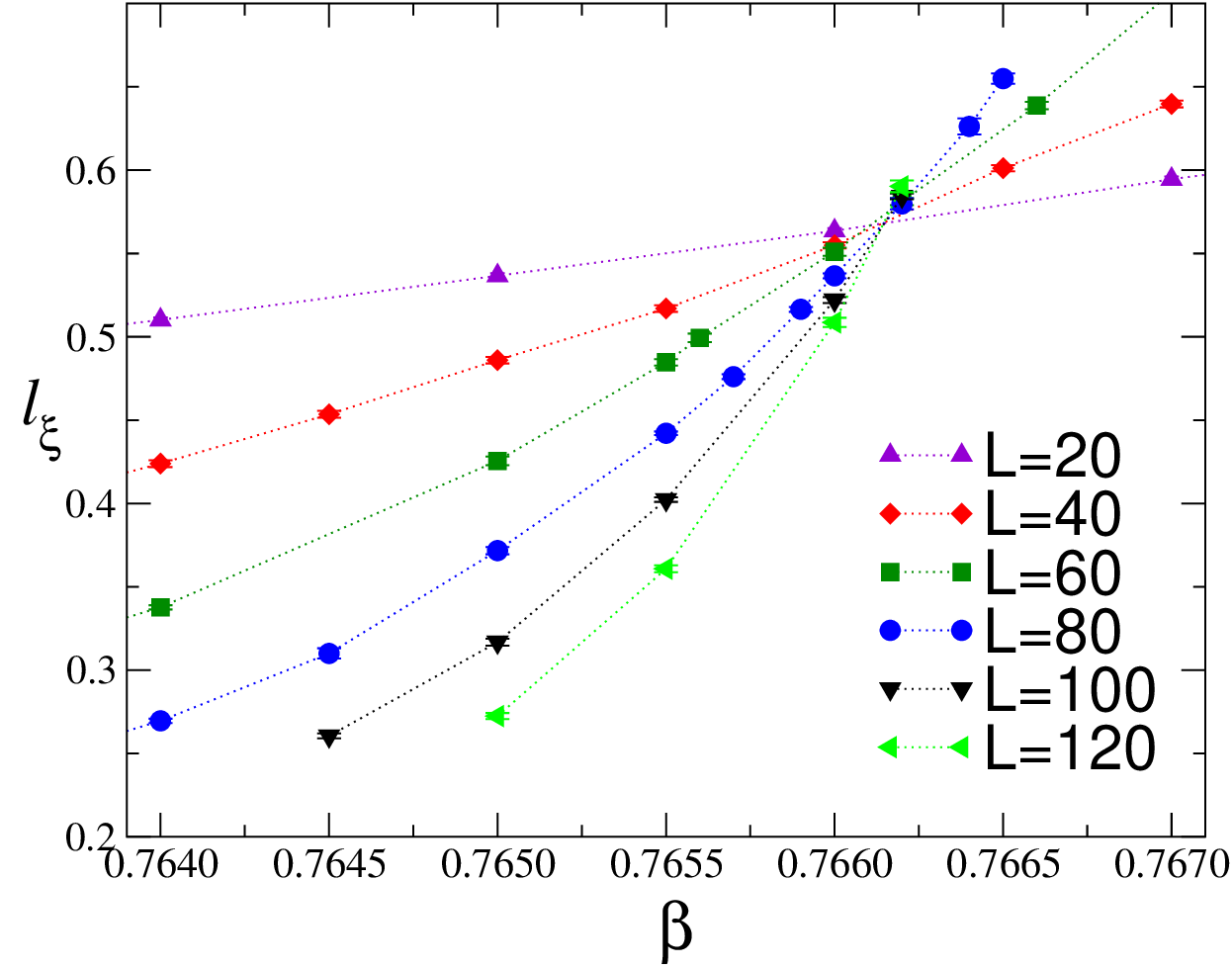}}
\vspace{6mm}
\centerline{\psfig{width=9truecm,height=6.7truecm,angle=0,file=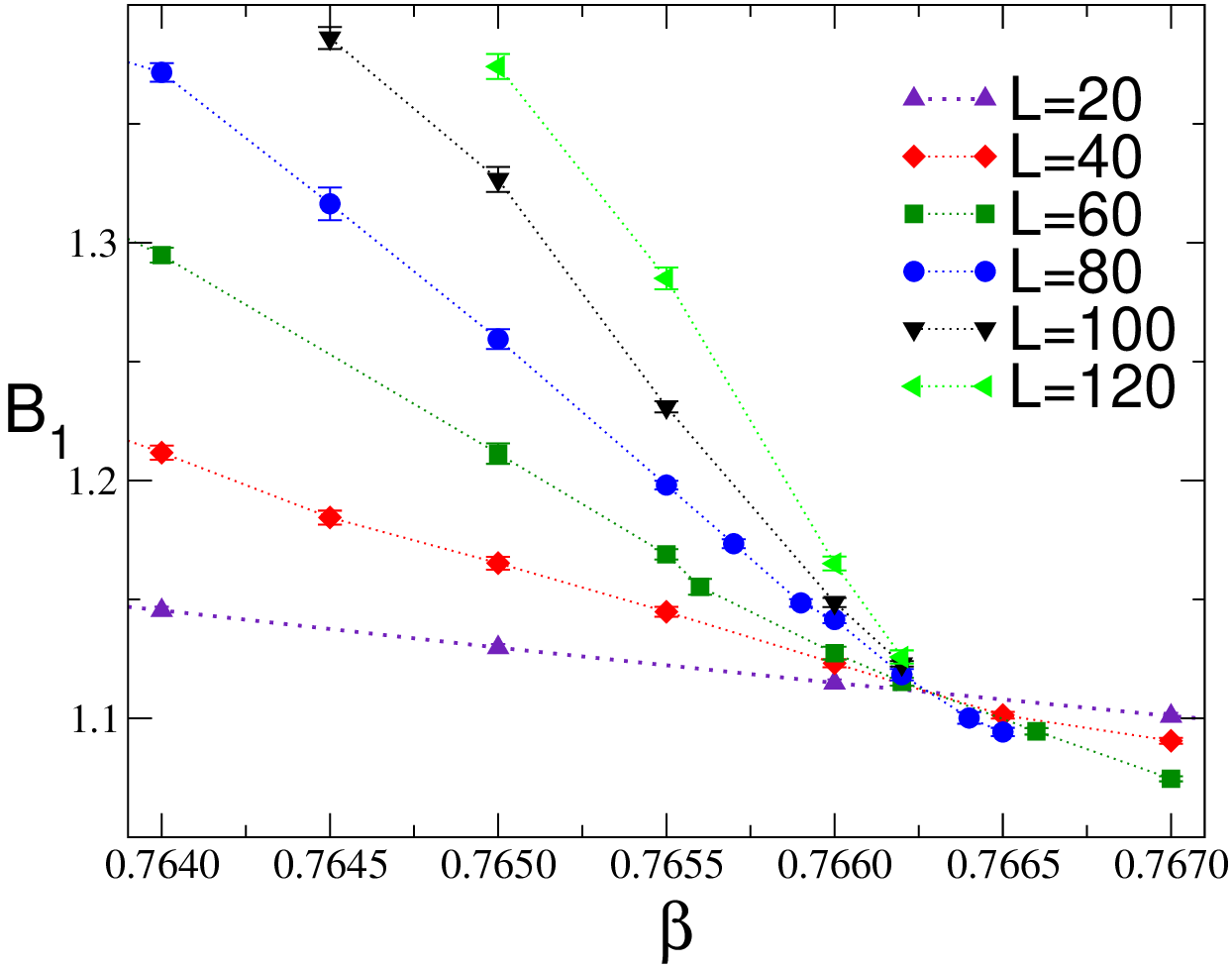}}
\vspace{6mm}
\centerline{\psfig{width=9truecm,height=6.7truecm,angle=0,file=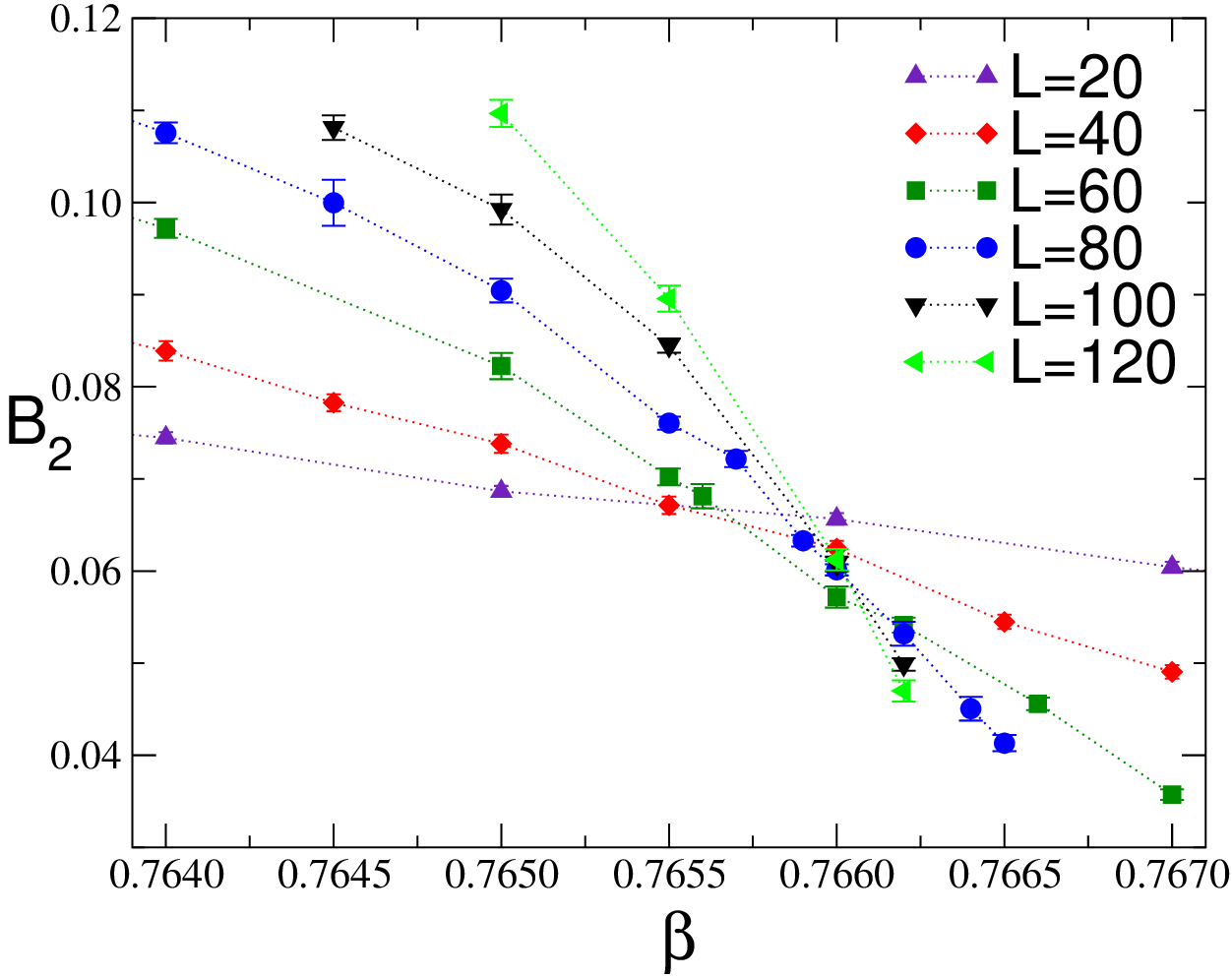}}
\caption{Plots of $l_\xi$, $B_1$, and $B_2$ for $A_{22} = 7/5$.
The lines are drawn to guide the eye.  
}
\label{cross}
\end{figure}

Let us begin by performing a direct analysis of $B_1$, $B_2$, and
$l_\xi$.  We fit our results corresponding to $20 \le L \le 120$ and
$0.755 \le \beta \le 0.767$ (see Fig.~\ref{cross}) by using
Eq.~(\ref{scalingR1}).  For this purpose we must somehow parametrize
the scaling function $f_R(x)$. We use a simple polynomial expression,
writing
\begin{equation}
     f_R(x) = \sum_{n=0}^p a_n x^n.
\end{equation}
The order $p$ is chosen in the following way. For a given set of data
we perform a nonlinear fit, increasing each time $p$ until the
$\chi^2$ changes by approximately 1 by going from $p$ to $p+1$. Of
course, one should also worry about scaling corrections and crossover
effects. In order to detect them we perform the fit several times,
each time including only the data corresponding to lattice sizes $L$
larger than some value $L_{\rm min}$. The corresponding results are
reported in Table~\ref{tab-anaR}.

\begin{table}
\caption{Analysis of the RG-invariant quantities $R$, $R=B_1$, $B_2$, and 
$l_\xi$, by using Eq.~(\protect\ref{scalingR1}). 
We report $\nu$, $\beta_c$, and $R^* = R(\beta_c)$.}
\label{tab-anaR}
\begin{tabular}{llclll}
\multicolumn{1}{c}{$R$}& 
\multicolumn{1}{c}{$L_{\rm min}$}& 
\multicolumn{1}{c}{$\chi^2$/DOF} & 
\multicolumn{1}{c}{$\nu$} & 
\multicolumn{1}{c}{$\beta_c$}&
\multicolumn{1}{c}{$R^*$} \\
\hline\hline 
$B_1$
 & 20 & 78/70
   & 0.728(4) & 0.766284(8) & 1.1111(6)
 \\
 & 40 & 63/54
   & 0.716(7) & 0.766278(11) & 1.1121(9)
 \\
 & 60 & 40/38
   & 0.711(14)& 0.766281(17) & 1.1121(16)
 \\
 & 80 & 24/20
   & 0.665(27)& 0.766260(25) & 1.1148(30)
 \\
\hline
$B_2$
 & 20 & 305/69
   & 0.662(5) & 0.765890(7) & 0.0643(2)
 \\
 & 40 & 92/54
   & 0.639(9) & 0.765992(10) & 0.0604(4)
 \\
 & 60 & 51/38
   & 0.650(17)& 0.766044(15) & 0.0580(7)
 \\
 & 80 & 25/21
   & 0.587(33)& 0.766038(19) & 0.0585(9)
 \\
\hline
$l_\xi$
 & 20 & 117/71
   & 0.687(2) & 0.766153(3) & 0.5692(6)
 \\
 & 40 & 61/55
   & 0.692(2) & 0.766168(5) & 0.5726(10)
 \\
 & 60 & 29/37
   & 0.694(5) & 0.766181(8) & 0.5770(19)
 \\
 & 80 & 9/18
   & 0.673(13)& 0.766168(12) & 0.5741(39)
 \\
\end{tabular}
\end{table}

The quality of the fits of $B_1$ and $l_\xi$ is reasonably good
($\chi^2$/DOF $\approx 1$), while for $B_2$ one should certainly
discard the results with $L_{\rm min} = 20$ and 40. As far as the
estimates of $\nu$ are concerned, we observe in all cases a systematic
drift. The analyses of $B_1$ and $l_\xi$ give first $\nu \approx $
0.69 - 0.71 and then, by using only the data with $L_{\rm min} = 80$,
one obtains $\nu \approx 0.66$-0.67. On the other hand, fits of $B_2$
give first $\nu \approx 0.66$ and then, for $L_{\rm min} = 80$, $\nu
\approx 0.59$, although with a large error of $\pm 0.03$. Clearly the
data are affected by large scaling corrections and, apparently, even
on these relatively large lattices one is not able to obtain a precise
estimate of $\nu$, but only an upper bound $\nu \lesssim 0.67$. In any
case, if we assume that the observed discrepancies give a reasonable
estimate of the systematic error, the results with $L_{\rm min} = 80$
give
\begin{equation}
    \nu = 0.63(7),
\end{equation} 
which includes the estimates from $B_1$, $B_2$, and $l_\xi$ with their
errors. This is fully compatible with the FT estimates. 

The results obtained from the analysis can be interpreted in terms of
a crossover due to the presence of a nearby O(4) FP. Indeed, the
observed behavior rensembles quite closely what is observed in field
theory for, say, $s = 1/4$. The effective exponent $\nu$ is first
close to the O(4) value and then decreases towards its asymptotic
value. This interpretation is somehow supported by the observed values
of $B_1^*$ and $l_\xi^*$, where $R^* = R(\beta_c)$.  They are close to
the corresponding O(4) ones
\cite{Hasenbusch-01}: $l^*_\xi \approx 0.547$, $B_1^* = 1.0945(2)$. 
Only for $B_2$ do we observe a relatively large difference since in
the O(4) case $B_2^* = {1\over12} B_1^* = 0.09121(2)$. This may
explain why the estimates of $\nu$ from $B_2$ are those that most
differ from the O(4) ones.  Of course, one may not exclude that the
asymptotic values $B_1^*$ and $l_\xi^*$ are close to the O(4)
estimates.

As a check we have also considered the derivatives of $B_1$, $B_2$,
and $l_\xi$ and we have used Eq.~(\ref{scalingdRdb2}). We find that in
all cases the best fit (smallest $\chi^2$/DOF) is obtained by taking
$R_1 = R_2$ with results that are (not surprisingly) fully compatible
with those obtained in the previous analysis. Fits with $R_1\not= R_2$
are somewhat worse but always show the same pattern. While for small
values of $L_{\rm min}$, $\nu$ varies between 0.64 and 0.70, depending
on the choice of $R_1$ and $R_2$, for $L_{\rm min} = 80$ the analyses
indicate a smaller value, fully compatible with the result reported
above.

\begin{figure}[!tb]
\centerline{
       \psfig{width=10truecm,height=7truecm,angle=0,file=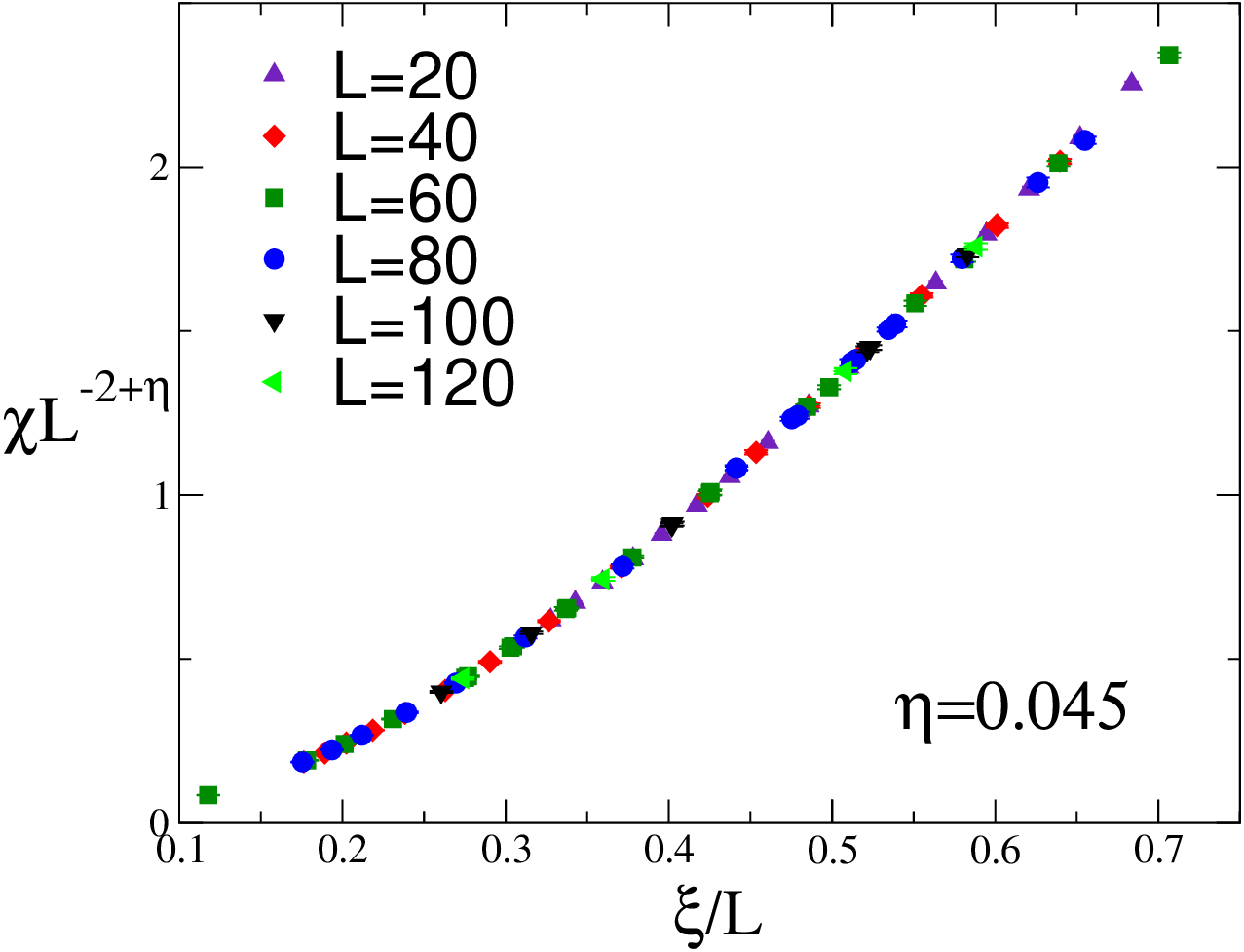}}
\vspace{12mm}
\centerline{\psfig{width=10truecm,height=7truecm,angle=0,file=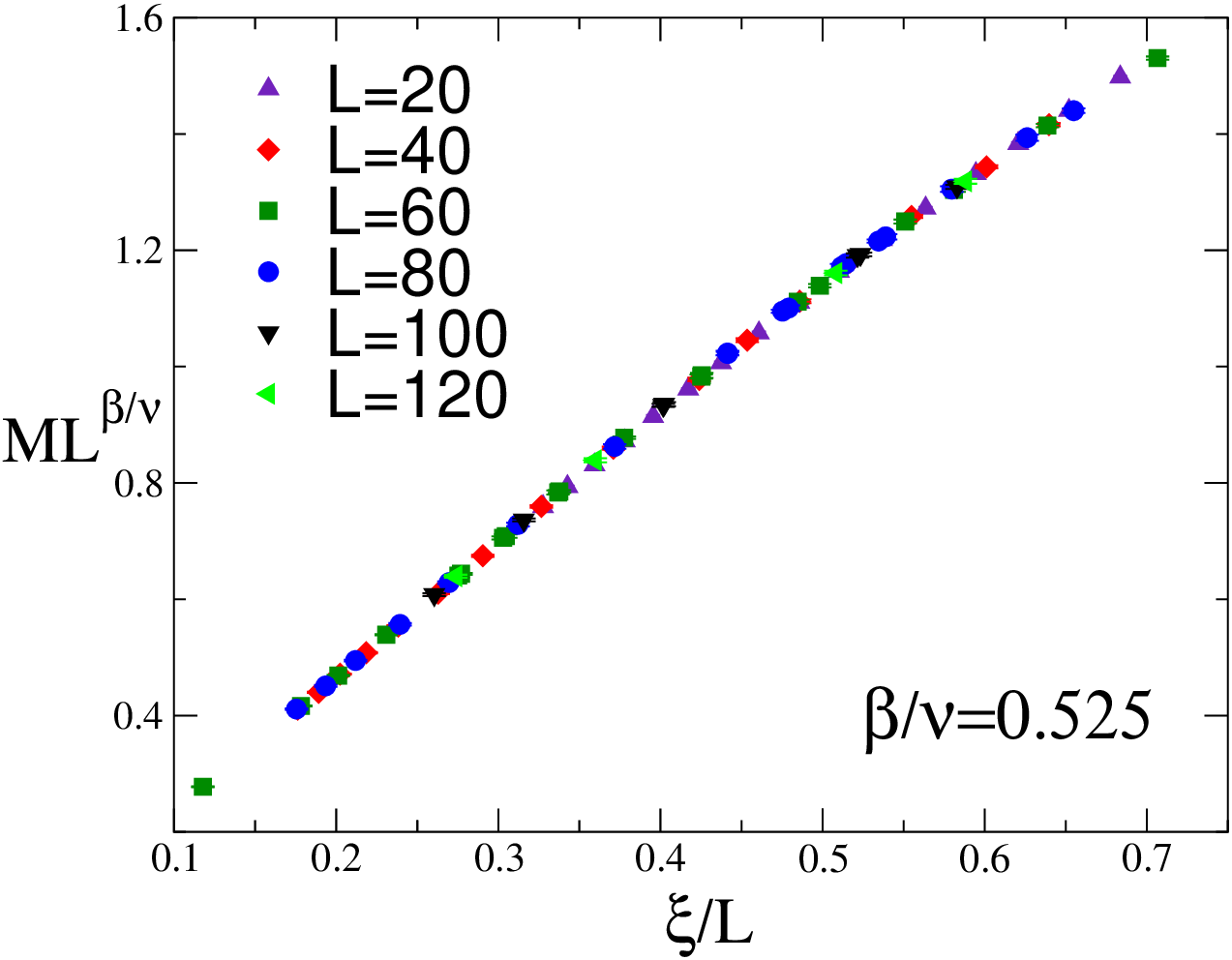}}
\caption{Plots of $\chi/L^{2-\eta}$ (above) and of $M L^{\beta/\nu}$ (below)
vs $\xi/L$. Here $\eta = 0.045$ and $\beta/\nu = 0.525$. 
}
\label{scalingchiM}
\end{figure}

The analyses of $B_1$, $B_2$, and $l_\xi$ also provide estimates of
$\beta_c$.  There is a clear upward trend in the estimates obtained
from $B_2$, as it can be also understood from Fig.~\ref{cross}: the
value at which the lines $B_2(\beta,L)$ at fixed $L$ cross moves
significantly towards higher values of $\beta$ as $L$ increases. The
estimates obtained from $B_1$ and $l_\xi$ are apparently stable, but
not compatible within the tiny statistical errors indicating that
there are strong (compared to the statistical errors) crossover
effects, as observed for $\nu$.  If we assume that the discrepancies
among the estimates obtained from the analyses of the three
RG-invariant quantities give a reasonable estimate of the systematic
error, an estimate of $\beta_c$ that includes all results is
\begin{equation}
\beta_c = 0.76615(15).
\end{equation}
We finally compute $\eta$ and $\beta/\nu$ from the analysis of
$\chi$ and $M$. We have performed the analyses by using
Eqs.~(\ref{scalingchi2}) and (\ref{scalingM2}).  The analyses using $R
= l_\xi$ are well behaved and show little dependence on $L_{\rm min}$,
leading to the estimates
\begin{eqnarray}
\eta = 0.045(10), \qquad \qquad {\beta\over \nu} = 0.0525(10). 
\label{est-eta-beta}
\end{eqnarray}
The corresponding scaling plots are reported in
Fig.~\ref{scalingchiM}.  On the other hand we observe systematic
deviations if we use $R = B_1$ or $B_2$. The goodness of the fit,
$\chi^2$/DOF, is a factor of ten larger than for the analysis with $R
= l_\xi$, and the estimates vary significantly, between $-0.1$ and 0.2
($\eta$) and 0.45 and 0.6 ($\beta/\nu$).

The results (\ref{est-eta-beta}) satisfy the scaling relation
$2\beta/\nu = 1 + \eta$ quite precisely. They can be compared with the
FT results: $\eta = 0.09(1)$ (MZM scheme, Ref.~\cite{PRV-01}) and
$\eta = 0.09(4)$ ($\overline{\rm MS}$ scheme).  Although larger, the
$\overline{\rm MS}$ estimate is compatible within error bars. A
discrepancy is observed for the MZM result, whose error might have
been underestimated (after all, the MZM series are not Borel
summable). Of course, it is also possible that scaling corrections
play an important role, as it is the case in the analyses of $B_1$,
$B_2$, and $l_\xi$. In this case, we expect the MC result to be
influenced by the presence of the nearby O(4) FP. Such an
interpretation is supported by the fact that the estimated $\eta$ is
close to the O(4) result $\eta = 0.0365(10)$
(Ref.~\cite{Hasenbusch-01}).

\begin{figure}[!tb]
\centerline{
     \psfig{width=10truecm,height=7truecm,angle=0,file=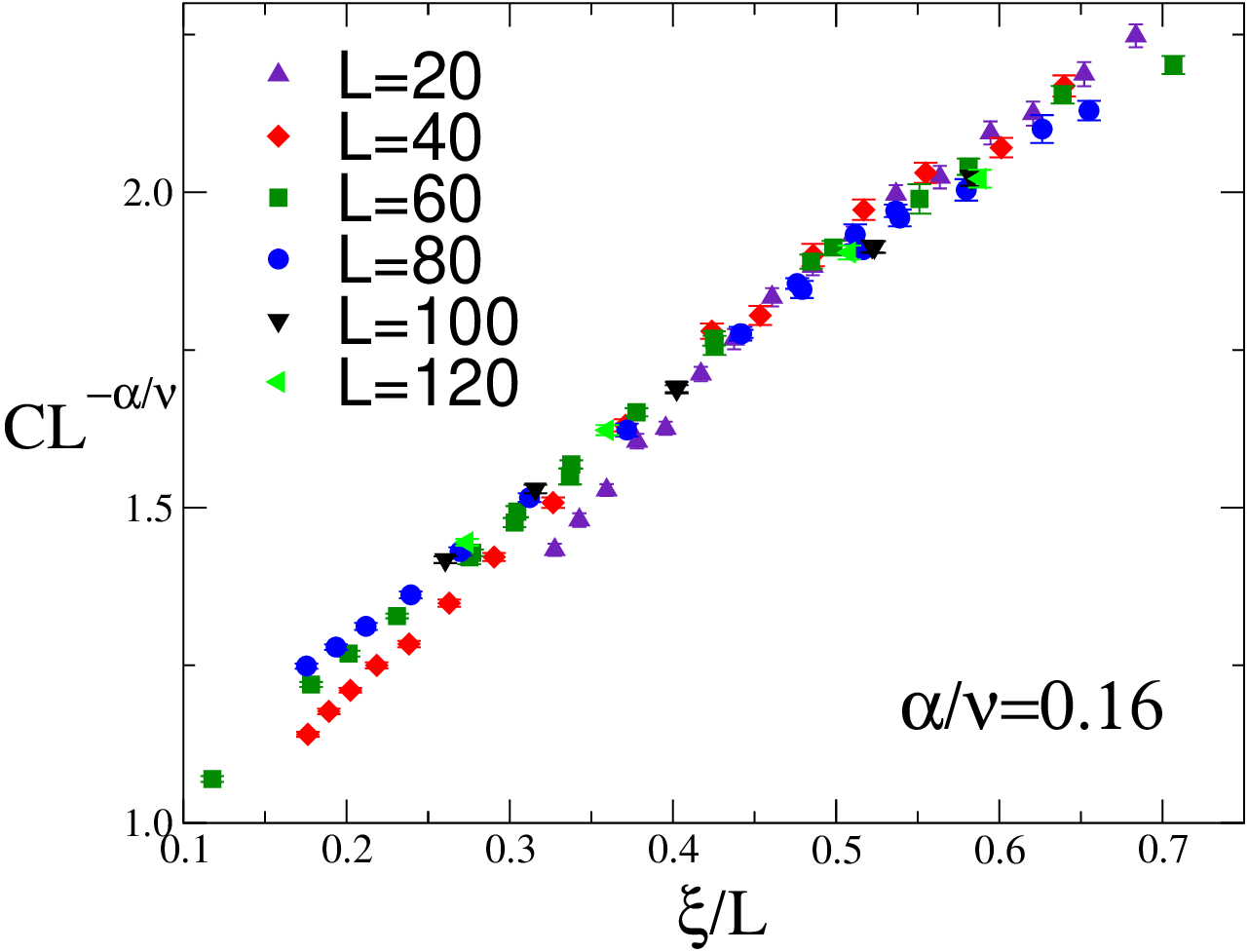}}
\vspace{3mm}
\caption{Plot of $C/L^{\alpha/\nu}$ 
vs $\xi/L$. Here $\alpha/\nu = 0.16$. 
}
\label{plot-CH}
\end{figure}

Finally, we analyzed the specific heat by using 
\begin{equation}
C = L^{\alpha/\nu} f(R),
\end{equation}
which is valid as long as $\alpha > 0$.  We do not expect this fit to
be very precise since we are neglecting the analytic contribution that
gives rise to corrections of order $L^{-\alpha/\nu}$, which are
expected to be sizeable since $\alpha$ is small. All fits have a poor
$\chi^2$. Only the fit with $R = l_\xi$ and $L_{\rm min} = 80$ has
$\chi^2$/DOF of order one.  Fits using $l_\xi$ give estimates of
$\alpha/\nu$ that decrease as $L_{\rm min}$ increases, varying between
0.20 and 0.16.  Fits using $B_1$ show the same decreasing trend with
$0.19\lesssim \alpha/\nu\lesssim 0.20$. Fits using $B_2$---they have a
very large $\chi^2$, $\chi^2$/DOF = 14 for $L_{\rm min} = 80$---show a
more erratic behavior with $L_{\rm min}$ and give $0.14\lesssim
\alpha/\nu\lesssim 0.18$. We quote as final result that obtained by
using $l_\xi$ and $L_{\rm min} = 80$:
\begin{equation}
{\alpha\over\nu} = 0.16(3).
\label{est-asun}
\end{equation}
The error has been chosen such that it includes all estimates with
$L_{\rm min} = 80$.  Using hyperscaling Eq.~(\ref{est-asun}) gives
$\nu = 0.63(1)$, in good agreement with the result reported above, and
$\alpha = 0.10(2)$. A scaling plot is reported in Fig.~\ref{plot-CH}.

The very large $\chi^2$/DOF of the above-reported analysis is probably due
the fact that the analytic contribution to the specific heat is neglected.
We have thus performed a second set of analyses using 
\begin{equation}
C = L^{\alpha/\nu} f_1(R) + f_2(\beta)
\end{equation}
and $R = \xi/L$.
As before, we have used polynomials for $f_1(x)$ and $f_2(x)$. 
Fifth-order polynomials allow us to obtain $\chi^2$/DOF $\approx 1$ 
even for $L_{\rm min} = 20$. We obtain 
$\alpha/\nu = -0.10(6)$, $0.01(9)$, and $0.16(25)$ for 
$L_{\rm min} = 20$, 40, 60 respectively. The results for the largest 
$L_{\rm min}$ are compatible with the estimates reported above. 
However, the very large errors indicate that our data are not precise enough
and that our set of values of $L$ is too small to disentangle the 
analytic background from the singular behavior. This probably means that the 
error reported in Eq.~(\ref{est-asun}) should not be taken too seriously and 
is most likely underestimated.

\section{Conclusions} \label{Conclusions}

In this paper we investigated the critical behavior of three-dimensional models 
with symmetry O(2)$\otimes$O($N$) described by the FT Hamiltonian
(\ref{LGWH}) in the case $v_0 > 0$, which corresponds to the 
symmetry-breaking pattern O(2)$\otimes$O($N$)$\to$O(2)$\otimes$O($N-2$) 
(the case $v_0 < 0$ has been discussed in detail in Ref.~\cite{DPV-03}). 

First, we considered the FT perturbative approach. The analysis of the
five-loop series in the $\overline{\rm MS}$ scheme without $\epsilon$
expansion provides strong evidence for the presence of a stable FP
with $v > 0$ for $N=2$ and $N=3$, and therefore for the existence of
the corresponding three-dimensional O(2)$\otimes$O($N$) universality
classes.  This result confirms the conclusions of Ref.~\cite{PRV-01},
in which a stable FP with $v > 0$ was found in the three-dimensional
MZM scheme.  Note that these FT perturbative analyses disagree with
the conclusions of Refs.~\cite{TDM-00,TDM-03,DMT-03}, in which no FP
was found by using a nonperturbative RG approach.  Moreover, the
$\overline{\rm MS}$ scheme without $\epsilon$ expansion allowed us to
obtain fixed-dimension results at any $d$. We recovered the
$\epsilon$-expansion results sufficiently close to four dimensions and
obtained a full picture of the fate of the different FPs as $d$ varies
from four to three dimensions.

In order to confirm the existence of a new universality class we have
performed a MC study of a lattice discretization of Hamiltonian
(\ref{LGWH}) for $N=2$.  The purpose is to identify a parameter region
in which the transition is of second order with the expect
symmetry-breaking pattern, O(2)$\otimes$O(2)$\to$O(2).  A detailed
analysis of the critical behavior of the model (\ref{HLi}) for $A_4 =
1$ and $A_{22} > A_4$ shows the following phase diagram. For $A_{22} >
A_{22}^* = 1.52(6)$ the model has a first-order transition. Such a
transition is easily identified for $A_{22} \gtrsim 3$---the energy and the 
magnetization show two peaks already for $L\approx 10$ with reduced latent heat
$\Delta E/T_c$ larger than 0.1.  The first-order transition becomes
weaker as $A_{22}$ decreases, the latent heat vanishing at the
tricritical point $A_{22} = A_{22}^*$. For $1 = A_4 < A_{22} <
A_{22}^*$ the transition is continuous. Its critical behavior is
expected to belong to the three-dimensional O(2)$\otimes$O(2)
universality class and therefore to be controlled by the stable FP of
the O(2)$\otimes$O(2) theory found within the FT methods.  We have
also performed simulations of the lattice model for $A_{22} =
7/5$ in order to identify the critical behavior. All results are
definitely compatible with the expected behavior at a second-order
transition. We are however unable to provide precise estimates of the
critical exponents since we observe strong crossover effects, probably
due to the presence of the nearby O(4) FP.  The effective exponents
computed in the MC simulation rensemble those observed in the FT model
for small $s\equiv v_0/u_0$, see Fig.~\ref{Fig:effexp}, indicating
that crossover effects play an important role in these systems and
make difficult, both numerically and experimentally, a precise
determination of the asymptotic critical behavior.

The FT Hamiltonian (\ref{LGWH}) is supposed to describe the critical
behavior of STA's and of helimagnets \cite{Kawamura-98}. Inelastic
neutron-scattering experiments show that STA's can be modeled by
three-component spin variables $\vec{s}$ associated with each site of
a stacked triangular lattice and by the Hamiltonian
\begin{equation}
{\cal H}_{\rm STA} = 
      J_\|\sum_{\langle vw\rangle_{xy}}  \vec{s}(v) \cdot \vec{s}(w) +
       J_\bot \,\sum_{\langle vw\rangle_z}  \vec{s}(v) \cdot \vec{s}(w) 
+ D \sum_v s_3(v)^2.
\label{latticeSTA}
\end{equation}
The first sum is over nearest-neighbor pairs within the triangular
layers ($xy$ planes) with an antiferromagnetic coupling $J_\|<0$, the
second one is over orthogonal interlayer nearest neighbors.  If the
uniaxial term is positive, one has an effective two-component theory.
Numerically, Hamiltonian (\ref{latticeSTA}) has been much studied in
the limiting cases $D=+\infty$ and $D = 0$. In the first case the
spins are confined to a plane, i.e., one is effectively considering XY
spins.  There is now evidence that XY STA's undergo first-order
transitions, at least for $|J_\bot/J_\| |$ not too small. Indeed,
first-order behavior has been observed for $J_\bot/J_\| = -3/4$
(Ref.~\cite{Itakura-03}), $J_\bot/J_\| = -1$ (Ref.~\cite{PSDMT-04}),
and $J_\bot/J_\| = -10$ (Ref.~\cite{PM-97}). The numerical results
indicate that the first-order transition becomes stronger as
$|J_\bot/J_\| |$ increases, and thus we can conclude that all these
models have first-order transitions, at least for $|J_\bot/J_\| | \gtrsim
3/4$.  It should however be noted that the latent heat is very small.
For $J_\bot/J_\| = -3/4$ and $J_\bot/J_\| = -1$, $\Delta E/T_c \approx
7\cdot 10^{-3}$ \cite{Itakura-03,PSDMT-04}. This means that small
modifications of the lattice model may turn the first-order transition
into a second-order one. In particular, it is not clear whether, on
the basis of these numerical simulations of the XY STA's, we should
expect first-order transitions also for experimental easy-plane
systems, which do not satisfy the condition $D \gg J_\|,
|J_\bot|$. For instance, in the case of CsMnBr$_3$ we have
\cite{CP-97} $J_\| \approx 0.0018$ meV, $J_\bot = 0.88$ meV, and
$D=0.013$ meV, while in other compounds like CsVX$_3$ (X = Cl, Br, I)
one observes \cite{CP-97} $D \approx J_\| \ll J_\bot$.
O(2)$\otimes$O(2) critical behavior is also expected in easy-axis
materials ($D < 0$) in the presence of a (large) magnetic field along
the easy direction.

Experiments on STA's favor a second-order transition, although the
estimates of $\beta$ do not satisfy the inequality $\beta \ge \nu/2$
that is expected on the basis of unitarity \cite{TDM-03,DMT-03}.  Of
course, this could be explained by the presence of a weak first-order
transition. A second possibility is that the experimental systems are
in the basin of attraction of the stable FP, but close to the boundary
of the stability region.  If this is the case, on the basis of the FT
crossover analysis, one expects strong crossover effects; see, for
example, the results presented in Fig.~\ref{Fig:effexp} for $s\approx
1$.  Thus, the exponents $\beta$ and $\nu$ that are measured
experimentally may well differ from their asymptotic value.  For what
concerns helimagnets, their critical behavior is somewhat different
from that observed in STA's. The exponent $\beta$ is always very close
to the O(4) value, while $\alpha$ varies between 0.1 and 0.3.  These
results strongly remind our MC ones with $A_{22} = 7/5$. In that case,
$\beta/\nu$ was close to the O(4) value and $\alpha \approx 0.10$.
Thus, the helimagnetic results can be explained by the presence of the
nearby O(4) FP that controls the critical behavior for $|t| \gtrsim
10^{-3}$.

Finally, we would like to conclude with some remarks on the
experimental relevance of the O(2)$\otimes$O(3) universality class.
Such a critical behavior is expected in some easy-axis materials that
have a small uniaxial anisotropy \cite{CP-97}, for instance in
RbNiCl$_3$, VCl$_2$, and VBr$_2$. However, the reduced-temperature
region in which O(2)$\otimes$O(3) behavior might be observed is
usually very small, i.e. for $t\gtrsim 10^{-2}$, because these systems
are expected to crossover to an XY critical behavior for $t\lesssim
10^{-2}$ \cite{CP-97,Kawamura-98}.  Therefore, the asymptotic
O(2)$\otimes$O(3) critical behavior can be hardly observed in these
materials and significant differences between theoretical predictions
and experimental results should not be unexpected.  As argued in
Ref.~\cite{KCP-90}, and usually assumed in the literature, the
O(2)$\otimes$O(3) critical behavior should also be experimentally
realized in easy-axis STA's, such as CsNiCl$_3$ and CsNiBr$_3$, at the
multicritical point observed in the presence of an external magnetic
field along the easy axis, or at the critical concentration of
mixtures of easy-axis and easy-plane materials, for instance in
CsMn(Br$_x$I$_{1-x})_3$ \cite{BWLOT-01}.  We note that the
identification of the multicritical point with the O(2)$\otimes$O(3)
universality class is not obvious and should be theoretically
analyzed.  The critical behavior at the multicritical point in a
magnetic field should be described by the stable FP of the most
general LGW theory with symmetry
O(2)$\otimes$[${\mathbb Z}_2 \oplus$O(2)] \cite{CPV-04}.  In
Ref.~\cite{KCP-90} only the quadratic terms have been considered and
discussed, but the relevant LGW Hamiltonian has also additional
quartic terms beside those appearing in the O(2)$\otimes$O(3)
Hamiltonian.  As a consequence, the O(2)$\otimes$O(3) FP, describing a
critical behavior with an enlarged O(2)$\otimes$O(3) symmetry,
determines the asymptotic behavior at the multicritical point only if
it remains stable with respect to the additional quartic terms
breaking O(2)$\otimes$O(3) to O(2)$\otimes$[${\mathbb Z}_2
\oplus$O(2)].  The critical behavior at the multicritical point is
determined by the stable FP of the RG flow of the complete LGW theory
with symmetry O(2)$\otimes$[${\mathbb Z}_2 \oplus$O(2)].  This issue
was recently investigated in Ref.~\cite{CPV-04} by a FT analysis
based on five-loop calculations within the $\overline{\rm MS}$ and MZM
schemes. Unfortunately, this study was unable to establish the 
stability properties of the O(2)$\otimes$O(3) FP. In any case, 
it did not provide evidence for any 
other stable FP.  Thus, on the basis of these
FT results, the transition at the multicritical point
is expected to be either continuous and controlled by the
O(2)$\otimes$O(3) fixed point or to be of first order.  Similar
arguments can be applied to the multicritical point in mixtures of
easy-axis and easy-plane materials, such as CsMn(Br$_x$I$_{1-x})_3$.
We believe that the identification of the multicritical behavior with
the O(2)$\otimes$O(3) universality class is even more questionable in this case,
since, beside the additional quartic terms considered above, there are other
perturbations related to the quenched randomness.

\section*{Acknowledgments}

We thank Maurizio Davini for his indispensable technical assistance to
manage the computer cluster where the MC simulations have been done.
PC acknowlegdes financial support from EPSRC Grant No. GR/R83712/01

\appendix

\section{The five-loop series of the $\overline{\rm MS}$ scheme}
\label{appmsb}

In the $\overline{\rm MS}$ scheme \cite{tHV-72}
one sets
\begin{eqnarray}
\phi &=& [Z_\phi(u,v)]^{1/2} \phi_R, \\
u_0 &=& A_d \mu^\epsilon Z_u(u,v) , \nonumber \\
v_0 &=& A_d \mu^\epsilon Z_v(u,v) , \nonumber
\end{eqnarray}
where the renormalization functions $Z_\phi$, $Z_u$, and $Z_v$ are
determined from the divergent part of the two- and four-point
one-particle irreducible correlation functions computed in dimensional
regularization.  They are normalized so that $Z_\phi(u,v) \approx 1$,
$Z_u(u,v) \approx u$, and $Z_v(u,v) \approx v$ at tree level.  Here
$A_d$ is a $d$-dependent constant given by $A_d= 2^{d-1} \pi^{d/2}
\Gamma(d/2)$.  Moreover, one defines a mass renormalization constant
$Z_t(f,g)$ by requiring $Z_t \Gamma^{(1,2)}$ to be finite when
expressed in terms of $u$ and $v$.  Here $\Gamma^{(1,2)}$ is the
one-particle irreducible two-point function with an insertion of
$\case{1}{2}\phi^2$.  The $\beta$ functions are computed from
\begin{equation}
\beta_u (u,v) = \mu \left. {\partial u \over \partial \mu} \right|_{u_0,v_0},
\qquad\qquad
\beta_v (u,v) = \mu \left. {\partial v \over \partial \mu} \right|_{u_0,v_0},
\end{equation}
while the RG functions $\eta_\phi$ and $\eta_t$ associated with the 
critical exponents are obtained from 
\begin{equation}
\eta_{\phi,t}(f,g) 
=  \left. {\partial \log Z_{\phi,t} \over \partial \log \mu} \right|_{u_0,v_0}.
\end{equation}
The $\beta$-functions have a simple dependence on $d$, indeed 
\begin{equation}
\beta_u = (d-4) u + B_u(u,v),\qquad
\beta_v = (d-4) v + B_v(u,v),
\label{Bdef}
\end{equation}
where the functions $B_u(u,v)$ and $B_v(u,v)$ are independent of $d$.
Also the RG functions $\eta_{\phi,t}$ are independent of $d$.  The
standard critical exponents are related to $\eta_{\phi,t}$ by
\begin{equation}
\eta = \eta_\phi(u^*,v^*),\qquad
\nu = \left[ 2 + \eta_t(u^*,v^*) - \eta_\phi(u^*,v^*) \right] ^{-1}.
\label{exponents} 
\end{equation}
We report the five-loop series \cite{CP-03} for the cases
$N=2$ and 3. The series for $N=2$ are:
\begin{eqnarray}
&&B_u(u,v)=
2\,u^2 - \case{1}{3} uv + \case{1}{6} v^2-\case{13}{6} u^3 + 
\case{11}{18} u^2 v - \case{13}{24} u v^2 + 
\case{5}{36} v^3+
6.95758\,u^{4} - 2.95970\,u^{3}\,v \nonumber\\ &&
+ 3.49036\,u^{2}\,v^{2} - 1.45080\,u\,v^{3} + 0.0939364\,v^{4} -33.3119\,u^{5} + 
18.9022\,u^{4}\,v - 
   25.3312\,u^{3}\,v^{2} \nonumber\\ &&
+ 14.5844\,u^{2}\,v^{3} -    3.31596\,u\,v^{4} + 0.260717\,v^{5}+
197.427\,u^{6} - 140.525\,u^{5}\,v + 211.453\,u^{4}\,v^{2}\nonumber\\ && 
- 152.787\,u^{3}\,v^{3} +  56.2903\,u^{2}\,v^{4}- 10.8790\,u\,v^{5} + 
   0.767937\,v^{6},
\nonumber\\[3mm] 
&& B_v(u,v)= 2 u v - \case{2}{3} v^2- \case{17}{6} u^2 v +  \case{29}{18} u v^2
- \case{11}{72} v^3
+10.0721\,u^{3}\,v - 8.38186\,u^{2}\,v^{2}+ 2.73239\,u\,v^{3}\nonumber\\ &&
 - 0.272799\,v^{4} -53.1466\,u^{4}\,v + 56.7468\,u^{3}\,v^{2} -  
29.2643\,u^{2}\,v^{3} + 7.15431\,u\,v^{4} -    0.598922\,v^{5}
\nonumber\\ &&
 + 341.414\,u^{5}\,v - 444.234\,u^{4}\,v^{2} + 
   311.112\,u^{3}\,v^{3} - 120.254\,u^{2}\,v^{4} + 
   22.4575\,u\,v^{5} - 1.52571\,v^{6},
\nonumber \\[3mm]
&&\eta_\phi(u,v)= \case{1}{12} u^2 - \case{1}{36} u v + \case{1}{48} v^2 
-\case{1}{24} u^3 + \case{1}{48} u^2 v 
- \case{5}{288} u v^2 + \case{7}{1728} v^3
\nonumber \\ &&
+ 0.112847 \,u^4 - 0.0752315\,u^3 v + 0.0998264\,u^2 v^2 
-    0.041956\,u v^3 + 0.00126591\,v^4 -0.41016\,u^5 
\nonumber \\ &&
+ 0.3418\,u^4 v  - 0.466487\,u^3 v^2 
+ 0.271787\,u^2 v^3 -    0.0630416\,u v^4 + 0.00479535\,v^5,\nonumber\\[3mm]
&&\eta_t(u,v) = -u + \case{1}{6} v+\case{1}{2} u^2 - \case{1}{6} u v + \case{1}{8} v^2
-\case{59}{48} u^3 + \case{59}{96} u^2 v 
- \case{97}{192} uv^2 + \case{401}{3456} v^3
\nonumber \\ &&
+4.13719\,u^4 - 2.75813\,u^3 v + 2.95416\,u^2 v^2 - 1.18534\,u v^3 
+ 0.121022\,v^4-18.2814\,u^5 
\nonumber \\ &&
+ 15.2345\,u^4 v - 19.0871\,u^3 v^2 + 10.7601\,u^2 v^3 -    2.69595\,u v^4 
+ 0.244579\,v^5.
\nonumber
\end{eqnarray}
The series for $N=3$ are:
\begin{eqnarray}
&&B_u(u,v) =
\case{7}{3} u^2 - \case{2}{3} u v + \case{1}{3} v^2-\case{8}{3} u^3 + \case{11}{9} u^2 v - 
\case{13}{12} u v^2 + \case{5}{18} v^3 
+ 9.07446\,u^{4} - 6.28514\,u^{3}\,v \nonumber \\ &&
 + 7.35688\,u^{2}\,v^{2} - 2.98029\,u\,v^{3} +  0.149100\,v^{4}- 46.7683\,u^5 +
 42.8264 \,u^4 v - 56.8289\,u^3 v^2 \nonumber \\ 
&&+ 32.0271\,u^2 v^3 -6.92247\, u v^4 + 0.503549\,v^5+296.166\,u^6 
- 338.69 \,u^5 v 
+ 506.708 \,u^4 v^2 \nonumber \\ &&
- 362.366 \,u^3 v^3 + 
131.029 \,u^2 v^4 - 24.7085 \,u v^5 + 1.71145 \,v^6,\nonumber\\[3mm]
&&B_v(u,v) = 2 u v - \case{1}{2} v^2- \case{28}{9} u^2 v + \case{14}{9} u v^2 
- \case{1}{36} v^3
+ 11.1573\,u^{3}\,v - 8.36797\,u^{2}\,v^{2}+2.43934\,u\,v^{3} \nonumber \\ &&
- 0.264816\,v^{4}
-62.5357\,u^4 v + 62.5357 \,u^3 v^2 - 30.8196\,u^2 v^3 
+ 7.59503 \,u v^4 - 0.660444 \,v^5\nonumber \\ 
&&+ 422.235 \,u^5 v - 527.794\,u^4 v^2 + 
364.748 \,u^3 v^3 -  141.636 \,u^2 v^4 + 25.7555 \,u v^5 - 1.63935\,v^6,
\nonumber\\[3mm]
&&\eta_\phi(u,v)=\case{1}{9} u^2 - \case{1}{18}u v + \case{1}{24} v^2-\case{7}{108} u^3 + \case{7}{144}
u^2 v - \case{11}{288} u v^2 + \case{13}{1728} v^3 \nonumber \\
&&+ 0.165895\,u^4 - 0.165895\,u^3 v + 0.228588\,u^2 v^2 
- 0.0935571\,u v^3 - 
   0.00108507\,v^4-0.709208\,u^5 \nonumber \\
&&+ 0.88651\,u^4 v - 1.19506\,u^3 v^2 
+ 0.674671\,u^2 v^3 -  0.144646\,u v^4 + 0.00961789\,v^5,\nonumber\\[3mm]
&&\eta_t(u,v) =-\case{4}{3}u + \case{1}{3}v+\case{2}{3} u^2 - 
\case{1}{3}uv + \case{1}{4} v^2-\case{52}{27} u^3 + \case{13}{9} u^2 v 
- \case{9}{8} u v^2 + \case{191}{864} v^3\nonumber \\
&&+7.00217\,u^4 - 7.00217\,u^3 v + 7.32279\,u^2 v^2 - 2.78612\,u v^3 
+ 0.227593\,v^4-33.6900\,u^5 \nonumber \\
&&+ 42.1125\,u^4 v 
- 51.8428\,u^3 v^2 + 28.3540\,u^2 v^3 -    6.77118\,u v^4 
+ 0.599424\,v^5\; .
\nonumber
\end{eqnarray}
The critical exponents associated with the chiral degrees of freedom
can be determined from the RG dimension of the chiral operator
$C_{ckdl}(x) = \phi_{ck}(x) \phi_{dl}(x) - \phi_{cl}(x) \phi_{dk}(x)$.
For this purpose, we computed the renormalization function $Z_c(u,v)$
by requiring $Z_c \Gamma^{(c,2)}$ to be finite when expressed in terms
of $u$ and $v$.  Here $\Gamma^{(c,2)}$ is the one-particle irreducible
two-point function $\Gamma^{(c,2)}$ with an insertion of the operator
$C_{ckdl}$. Then, one defines the RG function
\begin{equation}
\eta_{c}(u,v) 
=  \left. {\partial \log Z_{c} \over \partial \log \mu} \right|_{u_0,v_0}.
\end{equation}
The resulting series for $N=2$ and $N=3$ cases are respectively
\begin{eqnarray}
&&\eta_c(u,v) = - \case{1}{3} u + \case{1}{2} v+\case{5}{18} u^2 - 
\case{7}{18} u v + \case{5}{72} v^2
-\case{193}{432} u^3 + \case{307}{288} u^2 v 
- \case{353}{576} uv^2 + \case{163}{3456} v^3
\nonumber \\ &&
+ 1.51005 \,u^4 - 3.99379 \,u^3 v + 3.36615 \,u^2 v^2 - 1.05138 \,u v^3 
+  0.0947443 \,v^4- 6.32935\,u^5 
\nonumber \\ &&
+ 19.2654\,u^4 v - 21.2447 \,u^3 v^2 + 10.7984 \,u^2 v^3 -   2.47513 \,u v^4 
+ 0.197296 \,v^5, 
\nonumber
\end{eqnarray}
and
\begin{eqnarray}
&&\eta_c(u,v) = - \case{1}{3} u + \case{1}{2} v+\case{1}{3} u^2 - \case{4}{9} u v + \case{1}{18} v^2
-\case{55}{108} u^3 + \case{5}{4} u^2 v 
- \case{205}{288} uv^2 + \case{55}{1728} v^3
\nonumber \\ &&
+ 1.93436 \,u^4 - 5.12306 \,u^3 v +   4.24956 \,u^2 v^2 - 1.26059 \,u v^3 
+  0.105233 \,v^4- 8.73201 \,u^5 
\nonumber \\ &&
+ 26.6393 \,u^4 v - 29.198 \,u^3 v^2 + 14.5119 \,u^2 v^3 - 3.18347 \,u v^4 
+ 0.246754 \,v^5\ .
\nonumber
\end{eqnarray}
The chiral crossover exponent $\phi_c$ can be determined by 
using the RG scaling relation
\begin{equation}
\phi_c = \nu \left[ 2 + \eta_c(u^*,v^*) - \eta_\phi(u^*,v^*)\right].
\end{equation}
We have computed $\eta_c(u,v)$ for generic values of $N$. Thus, we have 
been able to compute the expansion of $\phi_c$ in powers of $\epsilon$
for any $N$. We have compared the result with the large-$N$ 
expression of Ref.~\cite{Gracey-02}, finding full agreement.

\section{Medium-range models} \label{App:mediumrange}

In this Appendix we consider a $d$-dimensional theory on a hypercubic lattice
with general Hamiltonian 
\begin{equation}
{\cal H} = - {\beta\over2} 
    \sum_{x,y} J_\rho(x - y) \sum_{ab} 
    \varphi_{x}^{ab} \varphi_{y}^{ab} + 
    \sum_x V(\varphi_x),
\label{H-MR}
\end{equation}
where $\varphi_{x2}^{ab}$ are $M\times N$ matrices, $V(\varphi_x)$ is
an O($M$)$\otimes$O($N$) invariant function, and the sums over $x$ and
$y$ are extended over all lattice points.  The coupling $J_\rho(x)$
depends on a parameter $\rho$. For instance, one may take the explicit
form (\ref{Coupling-J}), but this is not necessary for the discussion
that will be presented below. Indeed, one can consider more general
families of couplings, as discussed in Sec. 3 of
Ref.~\cite{PRV-lr}. The relevant property is that $J_\rho(x)$ couples
all lattice points for $\rho\to \infty$, i.e., that for $\rho\to
\infty$ one recovers a mean-field theory.

Models like (\ref{H-MR}) are called medium-range models and admit an
interesting scaling limit called critical crossover limit.  If $R$
parametrizes the range of the interactions, cf. Eq.~(\ref{def-R2}),
and $\beta_c(R)$ is the critical temperature as a function of $R$,
then for $R\to \infty$, $t\equiv (\beta_c(R) - \beta)/\beta_c(R) \to
0$ at fixed $\widetilde{t} \equiv R^{2d/(4 - d)} t$, critical
quantities show a scaling behavior. For instance, the susceptibility
$\chi(\beta,R)$ and the correlation length $\xi(\beta,R)$ scale
\cite{crossover1} according to Eq.~(\ref{MR-criticalscaling}).  The
functions $f_\chi(\widetilde{t})$ and $f_\xi(\widetilde{t})$ are
directly related to the crossover functions $F_\chi(t,s)$ and
$F_\xi(t,s)$ computed in field theory, cf.~Eq.~(\ref{relations-CCL})
\cite{PRV-lr,crossover2}.
The purpose of this Section is the computation of the nonuniversal
constants $\mu_\chi$, $\mu_\xi$, $a$, and $s$.  Interestingly enough,
if the range $R$ is defined according to Eq.~(\ref{def-R2}), they do
not depend on the explicit form of the coupling $J(x)$, but only on
the potential $V(\varphi)$. The dependence on $J(x)$ is effectively
encoded in the variable $R$.

The calculation can be done by a straightforward generalization of the
results of Ref.~\cite{PRV-lr}. Following Sec.~4.1 of
Ref.~\cite{PRV-lr} we first perform a transformation
\cite{Baker_62,Hubbard_72}---we use matrix notation and drop the subscript 
$\rho$ from $J_\rho(x)$ to simplify the notation---rewriting
\cite{footnote-HS}
\begin{equation}
\exp\left( {\beta\over 2} \varphi {J} \varphi \right)
=\left( {\rm det} \beta {J}\right)^{-NM/2}
\int {d^{NM}\phi\over (2\pi)^{NM/2}}
\exp \left( -{1\over 2\beta}
\phi {J}^{-1} \phi + \phi \varphi \right),
\label{fmf}
\end{equation}
where $\phi$ is an $N\times M$ matrix.
Then, we define a function $A(\phi)$ by requiring 
\begin{equation}
z e^{A(\phi)} \equiv
\int d^{NM}\varphi \;e^{- V(\varphi) + \phi \varphi} ,
\label{Adef1}
\end{equation}
where $z$ is a normalization factor ensuring $A(0)= 0$.    
We need to  compute the small-$\phi$ expansion of $A(\phi)$. For this 
purpose we define the integrals
\begin{eqnarray}
   I_{2n,1} &=& \int d^{NM}\varphi \; e^{- V(\varphi)} (\varphi^2)^n, \\
   I_{2n,2} &=& \int d^{NM}\varphi \; e^{- V(\varphi)} (\varphi^2)^{n -2}
  \sum_{ab}^N \sum_{cd}^M \varphi^{ac} \varphi^{ad} \varphi^{bc} \varphi^{bd},
\end{eqnarray}
and $f_{n,m} \equiv I_{n,m}/I_{0,1}$. Then, a straightforward calculation
gives 
\begin{equation} 
A(\phi) = {\bar{a}_{2,1}\over2} \phi^2 + 
          {\bar{a}_{4,1}\over 4!} (\phi^2)^2 + 
          {\bar{a}_{4,2}\over 4!} \left[(\sum_{ab}^N \sum_{cd}^M 
            \phi^{ac} \phi^{ad} \phi^{bc} \phi^{bd}) - (\phi^2)^2\right]
            + O(\phi^6),
\end{equation}
where
\begin{eqnarray}
  \bar{a}_{2,1} &=& {1\over NM} f_{2,1}, \\
  \bar{a}_{4,1} &=& {3 (f_{4,1} + 2 f_{4,2})\over M N (M+2) (N+2)} - 
              {3\over N^2M^2} f_{2,1}^2, \\
  \bar{a}_{4,2} &=& 6 {(MN+2) f_{4,2} - (M+N+1) f_{4,1} \over 
                 MN(M-1)(N-1)(M+2)(N+2)}.
\end{eqnarray}
The expansion of $A(\phi)$ is all we need to compute the critical crossover 
limit. Indeed, it is possible to show that $\varphi$ correlations are 
directly related to $\phi$ correlations \cite{PRV-lr}. For instance, 
\begin{eqnarray}
\langle \varphi_x\cdot\varphi_y\rangle\, =\,
 - {1\over\beta}\left( \widehat{J}^{-1}\right)_{xy} +
   {1\over\beta^2} \sum_{wz}
   \left( \widehat{J}^{-1}\right)_{xw}
   \left( \widehat{J}^{-1}\right)_{yz}
   \langle \phi_w \cdot \phi_z \rangle.
\end{eqnarray}
In the critical crossover limit, the first term in the right-hand side 
represents a subleading correction and can be neglected. 
This equation implies that $\chi_\varphi = \bar{a}_{2,1}^2\chi_\phi$, 
where we have used the fact that 
$\beta_c (\sum_x J(x)) = 1/\bar{a}_{2,1} + O(R^{-d})$. 

Moreover, $\phi$ correlations can be computed by using the following 
continuum theory
\begin{eqnarray}
{\cal H } &=& \int d^dx\, \left\{
   {1\over 2} \bar{a}_{2,1} R^2 \sum_\mu(\partial_\mu\phi)^2 + 
       {1\over 2}  \bar{a}_{2,1} t \phi^2 
   \right. \nonumber \\ 
   && \left. \qquad -
       {\bar{a}_{4,1}\over 4!} (\phi^2)^2 -
       {\bar{a}_{4,2}\over 4!} \left[(\sum_{ab}^N \sum_{cd}^M 
            \phi^{ac} \phi^{ad} \phi^{bc} \phi^{bd}) - (\phi^2)^2\right]
       \right\},
\end{eqnarray}
with a proper mass renormalization that is discussed in detail in
Ref.~\cite{PRV-lr}. This identification gives necessary conditions in
order to observe a second-order O($M$)$\otimes$O($N$) transition.  The
bare parameters should belong to the stability region, which implies
$\bar{a}_{4,1} < 0$ and $\bar{a}_{4,1} - {1\over2} \bar{a}_{4,2} < 0$,
and should be in the attraction domain of the O($M$)$\otimes$O($N$) FP
(assuming it exists), which implies (at least) $\bar{a}_{4,2} < 0$.

The above-reported results allow us to compute the nonuniversal constants 
appearing in Eq.~(\ref{relations-CCL}) \cite{footnote-error}. We obtain
\begin{eqnarray}
   \mu_\chi &=& \bar{a}_{2,1}^2 \mu_\xi^2 = 
      {\bar{a}_{2,1}^5\over \bar{a}_{4,1}^2},\\
   a &=& {\bar{a}_{2,1}^4\over \bar{a}_{4,1}^2},\\
   s &=& {\bar{a}_{4,2} \over \bar{a}_{4,1}}.
\end{eqnarray}
In Sec.~\ref{secmc} we will be interested in the specific potential 
(\ref{lattice-potential}) for $N=M=2$. 
In this case  
\begin{equation}
I_{2n,1} = {\pi^2} \int_{0}^\infty y^{n+1} dy 
      \int_0^1 dx\; \exp\left(-ry - {U_0\over24} y^2 + 
            {V_0\over 48} x^2 y^2\right),
\end{equation}
while $I_{2n,2}$ can be obtained by taking derivatives of $I_{2n,1}$
with respect to $V_0$ and $U_0$.  Exact results can be obtained for
$V_0$, $U_0\to 0$ at fixed $s_L \equiv V_0/U_0$. It is easy to verify
that $s \approx f(s_L)/U_0 \to 0$ if $r < 0$ and $s\to s_L$ if $r >
0$, as expected in a weakly coupled system.  We can also compute the
nonuniversal constants and, in particular, $s$ for the parameter
values used in the MC simulation.  For instance, we can consider two
cases: (a) $A_{22} = 2$ that corresponds to $r = -1$, $U_0 = 36$, and
$V_0 = 24$ (cf. Eq.~\ref{parHL-parMR}); (b) $A_{22} = 7/5$, i.e. $r =
-1$, $U_0 = 144/5$, and $V_0 = 48/5$.  For case (a) we obtain
$\bar{a}_{4,1} = -0.0370438$ and $\bar{a}_{4,2} = -0.0336483$, that
satisfy the necessary conditions reported above. Correspondingly, $s =
0.908337$, $a = 2.51497$, $\mu_\chi = 0.609569$, and $\mu_\xi =
3.22122$.  For case (b) we obtain $\bar{a}_{4,1} = -0.0412887$,
$\bar{a}_{4,2} = -0.0175521$, $s = 0.425106$, $a = 2.82353$, $\mu_\chi
= 0.743715$, and $\mu_\xi = 3.27408$.  It is interesting to note that
for the family of Hamiltonians considered in Sec.~\ref{secmc}---those
with $A_4=1$---we always have $s>s_L$, the difference $s-s_L$
decreasing as $s_L \to 0$.  The exact mapping between the lattice
model and the FT one also allows us to determine bounds on $A_{22}^*$
for $A_4 = 1$. Since $s = 1$ corresponds to $A_{22} \approx 2.13$
($s_L \approx 0.72$) and $s = 2$ to $A_{22} \approx 4.17$ ($s_L
\approx 1.23$), the FT bound $1 \lesssim s^* < 2$ implies $2.13
\lesssim A_{22}^* < 4.17$ or, equivalently, $0.72 \lesssim s^*_L <
1.23$.


\begin{references}
\bibitem{SBP} 
If we write the order parameter as a matrix $\phi^{aA}$, $a=1,2$, 
$A=1,\ldots,N$, the symmetries are 
$\phi \to L \phi R_1$ in the high-temperature phase and 
$\phi \to L^T \phi (L\oplus R_2)$ in the low-temperature phase. Here 
$L\in$ O(2), $R_1 \in$ O($N$) and $R_2 \in$ O($N-2$).
Note that the O(2) subgroup is different in the two phases.
\bibitem{footnotesp4}
Actually, using the same arguments,
this result can be extended to any dimension
$2<d<4$.
\bibitem{norm-U-V}
The quantities $u$ and $v$ are related to $\bar{u}$ and $\bar{v}$
defined in Ref.~\cite{PRV-01} by $u = 4 \pi \overline{u}$ and 
$v = 16\pi\overline{v}/3$.
\bibitem{microcanonical}
Explicitly, if we define a 4-dimensional vector $\Psi = (\varphi,\psi)$,
we generate a new vector $\Psi'_x$ according to the rule
$\Psi'_x = - \Psi_x + 2 s_x (s_x\cdot \Psi_x)/(s_x \cdot s_x)$, 
where $s_x = \sum_y \Psi_y$ and the sum runs over all nearest neighbors
$y$ of $x$. The update $\Psi_x \to \Psi'_x$ is accepted/rejected 
on the basis of a Metropolis test with Hamiltonian 
$2 (A_{22} -A_4)\varphi_x^2 \psi^2_x$.
\bibitem{nota-Sarbach}
Note that if $\phi > 1$
the exponent controlling the behavior of $\Delta_h$ 
close to the endpoint is $(2-\alpha - \phi)$ instead of $(1-\alpha)/\phi$,
since $y_1 = \phi/\nu$ and $y_2 = 1/\nu$.
\bibitem{footnote-error}
We mention that the general expressions for $\mu_\chi$ 
[Eqs. (4.38) and (4.68)] in Ref.~\cite{PRV-lr} are incorrect. 
The correct expressions are: $\mu_\chi = - {N \bar{a}_2^5/ \bar{a}_4}$ 
in two dimensions and $\mu_\chi = {N^2 \bar{a}_2^5/ \bar{a}_4^2}$ 
in three dimensions. Since $\bar{a}_2=1$ in the $N$-vector model, 
Eqs. (4.41) and (4.71) in Ref.~\cite{PRV-lr} are correct.
\end{references}
\end{document}